\documentclass[onecolumn,usenatbib]{mn2e}
\usepackage{graphicx}

\input{buote.defs}

\newcommand\lcdm{\hbox{{$\Lambda$CDM}}}

\title[Spherically Averaging Clusters: II.\ Biases]{Spherically Averaging Ellipsoidal Galaxy Clusters
in X-Ray and Sunyaev-Zel'dovich Studies: II.\ Biases}
\author[D.\ A.\ Buote and P.\ J.\ Humphrey]{David A. Buote\thanks{E-mail: buote@uci.edu} and
Philip J. Humphrey \\ Department of Physics and Astronomy, University
of California, Irvine, 4129 Frederick Reines Hall, Irvine, CA
92697-4575, USA}

\date{Accepted 2011 December 15. Received 2011 December 14; in original form 2011 September 29}

\begin{document} 
\maketitle

\begin{abstract}
  This is the second of two papers investigating the spherical
  averaging of ellipsoidal galaxy clusters in the context of X-ray and
  Sunyaev-Zel'dovich (SZ) observations. In the present study we
  quantify the orientation-average bias and scatter in observables
  that result from spherically averaging clusters described by
  ellipsoidal generalizations of the NFW profile or a nearly
  scale-free logarithmic potential.  Although the mean biases are
  small and mostly $<1\%$, the flattest cluster models generally have
  a significant mean bias; i.e., averaging over all orientations does
  not always eliminate projection biases. Substantial biases can
  result from different viewing orientations, where the integrated
  Compton-y parameter $(Y_{\rm SZ})$ and the concentration have the
  largest scatter (as large as $\sigma\sim 10\%$ for $Y_{\rm SZ}$),
  and the emission-weighted temperature $(T_{\rm X})$ has the smallest
  $(\sigma\la 0.5\%)$.  The very small scatter for $T_{\rm X}$ leads
  to $Y_{\rm X}$ and $M_{\rm gas}$ having virtually the same
  orientation biases. Substantial scatter is expected for individual
  clusters (up to $\sigma\sim 8\%$) in the correlation between $Y_{\rm
  SZ}$ and $Y_{\rm X}$ in comparison to the small mean bias
  ($\sigma\la 1\%$) applicable to a random sample of clusters of
  sufficient size. For ellipsoidal NFW models we show that the
  orientation bias for the total cluster mass attains a minimum near
  the radius $r_{2500}$ so that the spherically averaged mass computed
  at this radius is always within $\approx 0.5\%$ of the true value
  for any orientation.  Finally, to facilitate the accounting for
  orientation bias in X-ray and SZ cluster studies, we provide cubic
  polynomial approximations to the mean orientation bias and $1\sigma$
  scatter for each cluster observable as a function of axial ratio for
  the ellipsoidal NFW models.

\end{abstract}

\begin{keywords}{X-rays: galaxies: clusters ---  X-rays: galaxies --- dark
matter --- cosmological parameters --- cosmology:observations}
\end{keywords}

\section{Introduction}
\label{intro}

Galaxy clusters are potent tools for cosmological studies~\citep[e.g.,
see reviews by][]{henr03a,schu05a,arna05b,voit05b,tozz07a,alle11a}.
The redshift evolution of cluster baryon fractions supplements
well-known geometrical cosmological constraints obtained by the cosmic
microwave background radiation, supernova distances, and baryon
acoustic oscillations. Important cosmological information
complementary to these probes is provided by the growth of cosmic
structure, which is manifested in a variety of ways for clusters,
notably via the evolution of the cluster mass function. Since the hot
intracluster medium (ICM) dominates the cluster baryons, X-ray
observations are essential for measuring baryon fractions. They also
provide one of the most effective means to measure the total
gravitating mass for a cluster that obeys approximate hydrostatic
equilibrium.

To obtain the most accurate constraints on the baryon fraction and
total gravitating mass with X-ray observations requires obtaining
accurate measurements of both the radial ICM density and temperature
profiles out to a large fraction of the virial radius. These demanding
requirements impose rather strict limits on the number of clusters
which can be usefully analyzed with current data and data anticipated
in the near future.  Consequently, for cosmological studies with
X-rays there is much interest in using global scaling relations of
quantities that act as mass proxies; e.g., the ICM temperature and gas
mass measured from X-ray observations and the ICM pressure measured
from either X-ray or Sunyaev-Zel'dovich (SZ) observations. Useful
measurements of such proxies are possible for data of much lower
quality and therefore may be applicable to much larger cluster samples
than if detailed radial ICM profiles are needed. Moreover, even if the
measurements of such mass proxies are of low quality, by averaging
over large samples the impact of large statistical errors on
measurements for individual systems is mitigated leading to precise
cosmological constraints, which are accurate provided systematic
errors also ``average out'' or are understood and controlled.

The hot ICM also traces feedback processes during cluster formation
and evolution. For purely gravitational evolution the radial ICM
entropy profile behaves as approximately
$r^{1.1}$~\citep[e.g.,][]{tozz01a,voit05a}, whereas feedback from an
AGN and supernovas flattens the entropy profile mostly in the central
regions~\citep[e.g.,][]{tozz01a,brig01a,voit02a,borg05a,rome06a,youn07a,mcca08a,mcca10a}.
Seemingly at odds with these theoretical expectations, recent X-ray
studies of the Perseus and Virgo clusters measured entropy profiles
that flatten near the virial radius~\citep{simi11a,urba11a}. In
addition, \citet{simi11a} measure a gas fraction for Perseus that rises
well above (over 50\%) the cosmic mean baryon fraction. To reconcile
these unanticipated results with the theoretical picture mentioned
above, the authors of these studies postulate that near the virial
radius the cluster ICM is substantially multiphase due to clumping
from incompletely mixed material currently falling into the cluster --
a hypothesis that subsequently has been given support by a
cosmological simulation~\citep{naga11a}. In light of these results it
is interesting that our recent \chandra\ and \suzaku\ study of the
fossil group/cluster RXJ~1159+5531 out to the virial radius does not
find a flat entropy profile and also yields a measurement of the
baryon fraction that is very consistent with the cosmic mean
value~\citep{hump11b}. A possible interpretation of these different
results is that the magnitude of the gas clumping varies between
clusters such that the most relaxed systems (like fossil groups)
display the least amount of clumping. Of course, verifying this
interpretation requires an accurate accounting of the important
systematic errors involved with such measurements.

Deviations from hydrostatic equilibrium likely account for the largest
systematic errors in X-ray measurements of cluster masses. Theoretical
studies of clusters formed in hydrodynamical cosmological simulations
find that such deviations can lead to mass errors ranging from a few
percent up to $\sim 30\%$ even for nominally relaxed
clusters~\citep[e.g.,][]{tsai94a,evra96a,rasi06a,naga07a,piff08a,vazz11a}. The
simulations typically find that non-thermal pressure generated by
turbulent motions in the ICM is primarily responsible for the
systematic underestimates of the masses. While the modest turbulent
velocities associated with these mass biases are not strongly
constrained by current spectral data~\citep[e.g.,][]{schu04a,sand11a},
the calorimeter to be operated on {\sl ASTRO-H}~\citep{taka11a} should
measure the amount of turbulent motions with novel precision, thereby
providing a direct constraint on the accuracy of the hydrostatic
equilibrium approximation and substantially reducing this source of
systematic error for many clusters.

Arguably next in importance is the systematic error associated with
the ubiquitous assumption of spherical symmetry, since it is
well-known that clusters are not spherical. While this assumption is
not required, it is the default standard for several reasons.  First,
the intrinsic shape and orientation of a cluster, even if it is
assumed to be a triaxial ellipsoid, are not directly
observable. Second, although promising approaches exist to constrain
these properties indirectly though various combinations of X-ray, SZ,
and lensing data~\citep[e.g.,][]{zaro98a}, they necessarily require
more data and involve more complex modeling than if spherical symmetry
is assumed. Finally, while it is generally recognized that for an
individual cluster substantial error might be expected by assuming
spherical symmetry (e.g., when viewing a highly flattened spheroidal
cluster face-on), it is widely assumed (or hoped) that if it were
possible to average such measurements over all possible cluster
orientations, or rather average results over a statistically large
cluster sample, that the effects of random projection orientations on
cluster measurements should be eliminated entirely.

Over the years several theoretical studies have mentioned that
observed ICM properties should not vary dramatically as a result of
different intrinsic shapes and orientations.  The pioneering study by
\citet{binn78a} showed that even for spheroidal clusters highly
inclined to the line-of-sight, deprojection assuming that the symmetry
axis lies in the plane of the sky (effectively an assumption of
spherical symmetry for large inclination angles) did not produce very
large errors in the deprojected X-ray emissivity profile.  It has been
known since the first hydrodynamical cosmological simulation of a
cluster that also simulated X-ray observations to quantify errors in
the recovered mass, that significant ($\sim 25\%$) errors could be
expected from a combination of deviations from hydrostatic equilibrium
and spherical symmetry~\citep{tsai94a}.  More recently, cluster
simulations have shown that significant scatter in the measurement of
the integrated Compton-y parameter $(Y_{\rm SZ})$ (proportional to the
integrated ICM pressure) arises from effects due to non-spherical
cluster morphology~\citep[e.g.,][]{whit02a,shaw08a,krau11a}. But
precisely how much scatter arises from morphology alone is difficult
to extract from such simulations; e.g., the large scatter in the
relationship between measures of cluster morphology and deviations in
the correlation between $Y_{\rm SZ}$ and mass~\citep{krau11a}.

It is therefore worth examining comparatively simple non-spherical
models that effectively isolate the impact of assuming spherical
symmetry on cluster measurements in X-rays or the SZ effect.  Such
examinations are quite uncommon in the literature, and when they do
occur, they most often take the form of ancillary systematic error
checks on X-ray cluster measurements of the total mass and gas mass.
For example, to examine the error from assuming spherical symmetry on
their measurements of the gas mass in the cluster A~478, \citet{daw94}
divided up the radial surface brightness profile from \rosat\ into
different azimuthal sectors, concluding that the masses were
consistent between the different sectors within the estimated
$3\sigma$ statistical errors. This approach has also been applied to
other systems, including the more recent study by \citet{chur08a} of
\chandra\ observations of the low-mass clusters, M~87 and NGC~1399,
in which they conclude errors of a few percent in their total mass
measurements can be attributed to the assumption of spherical
symmetry. \citet{frg} and \citet{buot96c} noted in their studies of
the intrinsic shapes of some clusters with \einstein\ and
\rosat\ that the spherically averaged total mass profiles obtained
using both oblate and prolate spheroidal models were consistent,
within the statistical errors, with the results obtained from assuming
spherical symmetry. \citet{buot96c} also mentioned small
$(\sim 2\%)$, but significant, errors on the gas mass if spherical
symmetry is assumed. Finally, using spheroidal models \citet{greg00a}
estimate $\sim 20\%$ systematic errors from assuming spherical
symmetry on their SZ-derived estimates of the gas mass in A~370.

\citet{piff03a} and \citet{gava05a} conducted the first studies
dedicated to quantifying biases owing to the assumption of spherical
symmetry in measurements of total cluster masses, in particular with
respect to X-ray observations. \citet{piff03a} presented results for
ten clusters observed with \rosat. They modeled each cluster ICM as a
triaxial ellipsoid projected down its intermediate principal axis,
with the major-axis profile described by the isothermal $\beta$
model~\citep{cava76a,sara86a}. Their models indicated that the biases
from assuming spherical symmetry were $\la 4\%$ for the total mass and
$\la 5\%$ for the gas fraction. \citet{gava05a} conducted a
predominantly theoretical study of clusters modeled as spheroidal
NFW~\citep{nfw} mass distributions and isothermal ICM. To emphasize
projection effects he focused on the special case of a cluster viewed
down its symmetry axis (i.e., face-on). Like \citet{piff03a},
generally he found mass biases of only a few percent owing to the
assumption of spherical symmetry.

While significant biases from assuming spherical symmetry have been
clearly demonstrated for particular intrinsic shapes and viewing
orientations, the critical question of whether such biases completely
``average out'' when viewed from all directions remains unanswered. To
address this important issue we compute theoretical
orientation-average biases and scatter for ideal X-ray and SZ
measurements as a function of the intrinsic shape of a cluster. We
also consider several observables of interest to X-ray and SZ studies
-- concentration, total mass, gas mass, gas fraction,
emission-weighted temperature, $Y_{\rm X}$, and $Y_{\rm SZ}$ -- and a
wider range of models, including models with ICM temperature
gradients. Finally, for a subset of models we provide cubic polynomial
approximations to the mean orientation bias and $1\sigma$ scatter of
each observable that may be convenient for assessing the errors from
assuming spherical symmetry in X-ray and SZ cluster studies

The paper is organized as follows. We define the ellipsoidal models in
\S \ref{mass} and specify the default parameter values in \S
\ref{fiducial}. In \S \ref{method} we describe the procedures to
produce theoretical and observed spherically averaged observables. We
present the results in \S \ref{results} and our conclusions in \S
\ref{conc}. Our computations employ several of the analytical results
for the deprojection and spherical averaging of ellipsoidal potentials
derived in our companion paper~\citep[][hereafter Paper~1]{buot11c}.

\section{Mass Models}
\label{mass}

It is customary to approximate the isodensity contours of cluster dark
matter halos formed in cosmological simulations by ellipsoids of
constant shape and
orientation~\citep[e.g.,][]{jing02a,bail05a,allg06a,muno11a}.  While
this is often a satisfactory approximation, more generally the
isodensity contours of dark matter halos exhibit variations of shape
and orientation with radius~\citep[e.g.,][]{bail05a,allg06a}. To
explore a variation of shape with radius in the mass distribution we
also consider models where the isopotential contours have constant
shape and orientation; e.g., the cored logarithmic
potential~\citep[e.g.,][]{binn81a,evan93a}. Here we extend the
nomenclature of \citet{kass93a} to three-dimensional systems by
referring to models having ellipsoidal isodensity surfaces of constant
shape and orientation as ``Ellipsoidal Mass Distributions'' and models
with ellipsoidal isopotential surfaces of constant shape and
orientation as ``Ellipsoidal Potentials'', the latter of which are the
focus of Paper~1. 

%For our present investigation it is noteworthy that these two types of
%models have proved useful for modeling X-ray and SZ observations of
%clusters~\citep[e.g.,][]{binn78a,frg,buot92a,buot95a,buot96c,piff03a,wang04a,lee04a,defi05a,gava05a,flor07a,sere07a,kawa10a,mora10a,saye11a,mora11a,mora11b}.

Since each of these model types determines only how the flattening of
the mass distribution varies with radius, the radial profile must also
be specified.  We will consider two different models: NFW~\citep{nfw}
and the cored logarithmic potential mentioned above. The NFW profile
provides a good overall description of dark matter halos formed in
cosmological simulations and also of the total mass profile inferred
from X-ray observations of massive elliptical galaxies and
clusters~\citep[e.g., see reviews by][and references
therein]{arna05b,buot12a}. Although the centers of dark matter halos
are better represented by a Sersic/Einasto model~\citep{nava04a,merr05a}, the
additional accuracy at small radii is unimportant for our present
study which focuses on global halo structure.  Unlike the NFW model,
the cored logarithmic potential is not motivated by cosmological
simulations. Nevertheless, it also corresponds to the widely used
isothermal $\beta$-model~\citep{cava76a,sara86a} which is well-known
to provide a good description of the radial X-ray surface
brightness profiles of many galaxies and clusters.

Finally, we take the ICM to be a tracer population within a cluster,
which is a reasonable approximation given that the observed cluster
gas fraction is $\sim 10\%$ within the radius ($r_{500}$) we
investigate~\citep[e.g.,][]{prat10a}. Although we ignore the
self-gravity of the ICM we do examine how the gas mass (and gas
fraction) implied by hydrostatic equilibrium in the single-component
potential are affected by spherical averaging (see \S \ref{method}).

\subsection{Spherical Models}

We begin by presenting the density, enclosed mass, and gravitational
potential for a spherically symmetric system for both the NFW and
cored logarithmic potential models. This provides an opportunity to
define some useful quantities that will be used throughout the paper
and will serve as a convenient point of comparison for the
ellipsoidal generalizations defined below in \S~\ref{ep} and
\S~\ref{emd}.

\subsubsection{NFW}
\label{nfw}

The NFW mass density profile, $\rho(r)\propto r^{-1}(r_s + r)^{-2}$,
has two free parameters: (1) a normalization and (2) a scale radius
$(r_s)$ denoting where the logarithmic slope of the density profile
equals -2. It is conventional to replace these with the following related
parameters. Define the concentration parameter, $c\equiv
r_{\Delta}/r_s$, where $r_{\Delta}$ is defined so that the average
density of the halo with that radius equals $\Delta$ times the
critical density of the universe $(\rho_{\rm crit})$. Typically quoted
values for $\Delta$ range from 100-2500. For the other parameter,
define $M_{\Delta}$ to be the mass enclosed within $r_{\Delta}$. With
these definitions, the NFW density, enclosed mass, and gravitational
potential take the form,
\begin{equation}
\rho(\tilde{r}) = \frac{M_{\Delta}}{4\pi r_{\Delta}^3}\frac{c^3}{A(c)}
\frac{1}{c\tilde{r}\left(1 + c\tilde{r}\right)^2}, \label{eqn.nfw.rho}
\end{equation}
\begin{equation}
M(<\tilde{r})  = M_{\Delta}\frac{1}{A(c)}\left[\ln\left(1 +
    c\tilde{r}\right) - \frac{c\tilde{r}}{1 + c\tilde{r}}\right], \label{eqn.nfw.mass}
\end{equation}
\begin{equation}
\Phi(\tilde{r}) = -\frac{GM_{\Delta}}{r_{\Delta}}\frac{1}{A(c)}
\frac{\ln (1+c\tilde{r})}{\tilde{r}},  \label{eqn.nfw.phi}
\end{equation}
where,
\begin{equation}
\tilde{r} \equiv \frac{r}{r_{\Delta}}, \hskip 0.5cm A(c) \equiv
\ln\left(1+c\right) - \frac{c}{1+c}, \hskip 0.5cm
\frac{3M_{\Delta}}{4\pi r_{\Delta}^3} \equiv \Delta\rho_{\rm crit}. \label{eqn.sp.extra}
\end{equation}

\subsubsection{CORELOG (also Isothermal $\beta$-model)}
\label{corelog}

As befits its name, the cored logarithmic potential is defined by the
form of its potential, $\Phi = - (v_0^2/2)\ln(r_c^2 +
r^2)$~\citep[e.g.,][]{binn81a,evan93a,bt}. Like the NFW model it has
two free parameters: a normalization $(v_o)$ and a scale radius
$(r_c)$, usually called a ``core radius.'' For consistency of
presentation we redefine these parameters in analogy with that done
for the NFW model above. That is, we define a concentration parameter,
$c\equiv r_{\Delta}/r_c$, and mass, $M_{\Delta}$, where $r_{\Delta}$
is again defined so that the average density within a sphere of that
radius equals $\Delta$ times the critical density. When expressed in
terms of $c$ and $M_{\Delta}$ the cored logarithmic potential and
derived quantities take the form,
\begin{equation}
\rho(\tilde{r}) = \frac{M_{\Delta}}{4\pi r_{\Delta}^3}(1+c^2)
\frac{3 + (c\tilde{r})^2}{\left[1 +
    \left(c\tilde{r}\right)^2\right]^2}, \label{eqn.corelog.rho}
\end{equation}
\begin{equation}
M(<\tilde{r})  = M_{\Delta}(1+c^2)\frac{\tilde{r}^3}{1+\left(c\tilde{r}\right)^2},
\end{equation}
\begin{equation}
\Phi(\tilde{r}) = -\frac{GM_{\Delta}}{2r_{\Delta}}\frac{(1+c^2)}{c^2}\ln
\left(\frac{1+c^2}{1+\left(c\tilde{r}\right)^2}\right), \label{eqn.corelog.phi}
\end{equation}
where, $\tilde{r}$ and $M_{\Delta}$, are defined as above (eqn.\
\ref{eqn.sp.extra}). 
We have chosen the arbitrary constant term in the potential so that
$\Phi=0$ when $r=r_{\Delta}$. Henceforth we will refer to this model
as the ``CORELOG'' model. In order to provide a sharper contrast with
the NFW model, we will focus on the case $c\gg 1$ so that $\Phi\sim
\ln(r)$, corresponding to the singular isothermal sphere. It is
noteworthy that eqn.\ (\ref{eqn.corelog.phi}) also corresponds to the
potential of the isothermal $\beta$-model~\citep{cava76a,trin86},
where the ICM density is assumed to obey, $\rho_{\rm gas} \propto (1 +
(r/r_c)^2)^{-3\beta/2}$, and the factor outside the logarithm equals,
$-(3\beta/2)(k_BT/\mu m_a)$, where $T$ is the ICM temperature, $k_B$
is Boltzmann's constant, $\mu$ is the mean atomic weight of the ICM,
and $m_a$ is the atomic mass unit.

\subsection{Ellipsoidal Models -- General Considerations}
\label{gen}

As in Paper~1 we consider an ellipsoid of principal axes $a,b,c$ with axis
ratios $p_v \equiv b/a$ and $q_v \equiv c/a$ satisfying $0 < q_v \le
p_v \le 1$. When the ellipsoid is oriented so that $a$ lies along the
$x$-axis, $b$ along the $y$-axis, and $c$ along the $z$-axis, the
ellipsoidal radius is defined by,
\begin{equation}
a_v^2 = x^2 + \frac{y^2}{p_v^2} + \frac{z^2}{q_v^2}. \label{eqn.av}
\end{equation}
For the ellipsoidal generalizations of the spherical NFW and CORELOG
models we consider below,  we will make use of the following quantities, 
\begin{equation}
\tilde{a} \equiv \frac{a_v}{a_{\Delta}}, \hskip 0.5cm 
\frac{3M_{\Delta}}{4\pi p_vq_va_{\Delta}^3} \equiv \Delta\rho_{\rm crit},
\end{equation}
where, in analogy to the spherical case, $a_{\Delta}$ is defined so
that the average density within an ellipsoid $a_v = a_{\Delta}$ equals
$\Delta \rho_{\rm crit}$, and $M_{\Delta}$ refers to the mass
contained within that ellipsoid.

\subsection{Ellipsoidal Potentials -- EPs}
\label{ep}

For a detailed discussion of this model type we refer the reader to
Paper~1. Briefly, we define an ellipsoidal potential (EP) model such
that the gravitational potential is an ellipsoid of constant shape
$(p_v,q_v)$ and orientation; i.e., $\Phi$ depends only
on $a_v$. For this model type one begins by specifying $\Phi(a_v)$ and
then deduces the mass distribution from it. The key advantage of an EP
is that a simple, analytic form for $\Phi$ can be adopted, enabling
much faster computational evaluation than for ellipsoidal mass
distributions (\S \ref{emd}). A disadvantage is that when the
potential is sufficiently flattened (the amount depending on how steep
is the radial potential profile), the model can become unphysical in
regions near the shortest principal axis; e.g., as manifested by
negative values for the mass density~\citep[e.g.,][]{bt}. However,
since we primarily make use of the integrated mass profile, which is
as well-behaved as for spherical analogs of $\Phi$ (see Paper~1),
localized peculiar features in the density profiles of EP models are
not important for our investigation.

In Theorem~1 of Paper~1 we show that the mass enclosed within $a_v$,
$M(<a_v)$, is related to the the potential $\Phi(a_v)$ in the same
manner as the spherical $M(<r)$ is related to $\Phi(r)$, with the
introduction of a proportionality constant,
\begin{equation}
\eta (p_v,q_v) \equiv \frac{p_v q_v}{3}\left[1 + \frac{1}{p_v^2} +
\frac{1}{q_v^2} \right], \label{eqn.eta}
\end{equation}
which is of order unity. We emphasize that $p_v$ and $q_v$ define the
shape of $\Phi$ and the bounding surface for the calculation of
$M(<a_v)$ (and thus definition of $M_{\Delta}$ below in \S
\ref{nfw.ep} and \S \ref{corelog.ep}), but the mass density distribution of an
EP generally has smaller axis ratios (see \S \ref{fiducial}) and is
not strictly ellipsoidal.

We now define versions of the NFW and CORELOG models appropriate for
EPs.

\subsubsection{NFW-EP}
\label{nfw.ep}

We convert the spherical NFW model to an EP starting with the
potential defined by eqn.\ (\ref{eqn.nfw.phi}) and letting
$r\rightarrow a_v$ and $r_{\Delta}\rightarrow a_{\Delta}$,
\begin{equation}
\Phi(\tilde{a}) = -\frac{GM_{\Delta}}{a_{\Delta}}\frac{1}{A(c)}
\frac{\ln (1+c\tilde{a})}{\tilde{a}},
\end{equation}
where $c\equiv a_{\Delta}/a_{\rm scale}$ and $a_{\rm scale}$ is an
ellipsoidal analog of $r_s$.  From this modified potential the mass
profile is computed using eqn.\ (7) of Paper~1 to give,
\begin{equation}
M(<\tilde{a})  =  M_{\Delta}\frac{\eta (p_v,q_v)}{A(c)}\left[\ln\left(1 +
    c\tilde{a}\right) - \frac{c\tilde{a}}{1 + c\tilde{a}}\right], \label{eqn.nfw.mass.ep}
\end{equation}
and the density profile is computed from Poisson's equation which yields,
\begin{eqnarray}
\lefteqn{\rho(x,y,z) =  \frac{M_{\Delta}}{4\pi a_{\Delta}^3}\frac{c^3}{A(c)}
\frac{1}{(c\tilde{a})^2\left(1 + c\tilde{a}\right)} \, \times} \nonumber\\
& & \left[ \frac{3+4c\tilde{a}}{1+c\tilde{a}} \frac{1}{a_v^2} \left(x^2 +
    \frac{y^2}{p_v^4} + \frac{z^2}{q_v^4}\right) 
  -\left(1 + \frac{1}{p_v^2} + \frac{1}{q_v^2} \right) +
  \frac{\left(1+c\tilde{a}\right)\ln\left(1+c\tilde{a}\right)}{c\tilde{a}}
\left[\left(1 + \frac{1}{p_v^2} + \frac{1}{q_v^2} \right) -
\frac{3}{a_v^2} \left(x^2 + \frac{y^2}{p_v^4} +
    \frac{z^2}{q_v^4}\right)   \right]  \right].  \label{eqn.nfw.density.ep}
\end{eqnarray}

\subsubsection{CORELOG-EP (also EP analog of Isothermal $\beta$ model)}
\label{corelog.ep}

The procedure to construct the EP analog of the CORELOG model
follows the same procedure as for the NFW model above but starting
from the potential given by eqn.\ (\ref{eqn.corelog.phi}). We obtain
for the potential,
\begin{equation}
\Phi(\tilde{a}) = -\frac{GM_{\Delta}}{2a_{\Delta}}\frac{(1+c^2)}{c^2}\ln
\left(\frac{1+c^2}{1+\left(c\tilde{a}\right)^2}\right), \label{eqn.corelog.phi.ep}
\end{equation}
where $c\equiv a_{\Delta}/a_c$ and $a_c$ is an ellipsoidal analog of
$r_c$. The mass and density profiles derived from this potential are respectively, 
\begin{equation}
M(<\tilde{a})  = \eta (p_v,q_v)
M_{\Delta}(1+c^2)\frac{\tilde{a}^3}{1+\left(c\tilde{a}\right)^2}, 
\end{equation}
and,
\begin{equation}
\rho(x,y,z) =  \frac{M_{\Delta}}{4\pi a_{\Delta}^3}
  \left(1+c^2\right) \frac{1}{\left(1+\left(c\tilde{a}\right)^2\right)^2}
  \left[ \left(1 + \frac{1}{p_v^2} + \frac{1}{q_v^2} \right)
    \left(1+\left(c\tilde{a}\right)^2\right) 
  - \frac{2c^2}{a_{\Delta}^2} \left(x^2 + \frac{y^2}{p_v^4} + \frac{z^2}{q_v^4}\right)\right].  \label{eqn.corelog.density.ep}
\end{equation}
This expression for the density is the triaxial generalization of the
well-known spheroidal form~\citep[e.g.,][]{bt}.

\subsection{Ellipsoidal Mass Distributions -- EMDs}
\label{emd}

We define an ellipsoidal mass distribution (EMD) such that the mass
density is an ellipsoid of constant shape and orientation; i.e.,
$\rho$ depends only on the ellipsoidal radius $a_v$, where now the
axial ratios $p_v$ and $q_v$ refer to $\rho$ rather than $\Phi$. The
gravitational potential generated by an EMD is not an ellipsoid but
instead has the form \citep[e.g.,][]{chan87,bt},
\begin{equation}
\Phi(\vec{x}) = -\pi G p_vq_va_{\Delta}^3\int_0^{\infty}\frac{du}{\Delta (u)}
\int_{\tilde{a}^2(\vec{x},u)}^{\infty} d\tilde{a}^{\prime 2} \rho (
\tilde{a}^{\prime 2} ),
\end{equation}
where,
\begin{equation}
\tilde{a}^2(\vec{x},u) \equiv \frac{x^2}{a_{\Delta}^2 + u} + \frac{y^2}{(p_va_{\Delta})^2 +
  u} + \frac{z^2}{(q_va_{\Delta})^2 + u},  \hskip 0.5cm
\Delta(u) \equiv \sqrt{(a_{\Delta}^2 + u)((p_va_{\Delta})^2 + u)((q_va_{\Delta})^2 + u)},
\end{equation}
and the function $\Delta(u)$ should not be confused with the
over-density value as signified by $a_{\Delta}$. Because the improper
integral over $du$ with upper limit $\infty$ is inconvenient for
numerical evaluation, we follow
\citet{merr96a} and define a new variable,
\begin{equation}
s \equiv \frac{1}{\sqrt{1+u}},
\end{equation}
so that,
\begin{equation}
\tilde{a}^2(\vec{x},s) \equiv s^2\left[ \frac{x^2}{1 + (a_{\Delta}^2-1)s^2} +
\frac{y^2}{ 1 + [(p_va_{\Delta})^2 - 1]s^2} + \frac{z^2}{1 +
  [(q_va_{\Delta})^2 -1]s^2} \right],
\end{equation}
\begin{equation}
\Delta(s) = \frac{\delta(s)}{s^3},  \hskip 0.5cm
\delta(s) \equiv \sqrt{(1 + (a_{\Delta}^2-1)s^2)(1 + [(p_va_{\Delta})^2 - 1]s^2) (1 +
  [(q_va_{\Delta})^2 -1]s^2)},
\end{equation}
\begin{equation}
\Phi(\vec{x}) = -2\pi G p_vq_va_{\Delta}^3\int_0^{1}\frac{ds}{\delta (s)}
\int_{\tilde{a}^2(\vec{x},s)}^{\infty} d\tilde{a}^{\prime 2} \rho (
\tilde{a}^{\prime 2} ). \label{eqn.phi.emd}
\end{equation}
We use eqn.\ (\ref{eqn.phi.emd}) for all numerical evaluations of EMD
potentials in this paper.

\subsubsection{NFW-EMD}
\label{nfw.emd}

We convert the spherical NFW model to an EMD starting with the mass
density defined by eqn.\ (\ref{eqn.nfw.rho}) and letting $r\rightarrow
a_v$, $r_{\Delta}\rightarrow a_{\Delta}$, and $4\pi\rightarrow 4\pi p_vq_v$,
\begin{equation}
\rho(\tilde{a}) = \frac{M_{\Delta}}{4\pi p_vq_va_{\Delta}^3}\frac{c^3}{A(c)}
\frac{1}{c\tilde{a}\left(1 + c\tilde{a}\right)^2} \label{eqn.nfw.rho.emd},
\end{equation}
where $c\equiv a_{\Delta}/a_{\rm scale}$, and $a_{\rm scale}$ is an
ellipsoidal analog of $r_s$. From this modified density we compute the
mass enclosed within $a_v$ by direct integration of $\rho(a_v)$ which
gives,
\begin{equation}
M(<\tilde{a})  = M_{\Delta}\frac{1}{A(c)}\left[\ln\left(1 +
    c\tilde{a}\right) - \frac{c\tilde{a}}{1 + c\tilde{a}}\right].
\end{equation}
Finally, from the expression for the density we construct the
potential,
\begin{equation}
\Phi(\vec{x}) = -\frac{GM_{\Delta}}{2}\frac{c}{A(c)}
\int_0^{\infty}\frac{du}{\Delta (u)} \frac{1}{1+c\tilde{a}
  (\vec{x},u)} = -GM_{\Delta}\frac{c}{A(c)}
\int_0^{1}\frac{ds}{\delta (s)} \frac{1}{1+c\tilde{a}
  (\vec{x},s)}.
 \label{eqn.nfw.phi.emd}
\end{equation}
For $p_v=q_v=1$ this expression for $\Phi$ reduces exactly to the
spherical case (eqn.\ \ref{eqn.nfw.phi}).

\subsubsection{CORELOG-EMD}
\label{corelog.emd}

The procedure to construct the EMD analog of the spherical CORELOG
model is fully analogous to that used for the NFW-EMD
model. Converting eqn.\ (\ref{eqn.corelog.rho}) to an EMD gives the
mass density,
\begin{equation}
\rho(\tilde{a}) = \frac{M_{\Delta}}{4\pi p_vq_va_{\Delta}^3}(1+c^2)
\frac{3 + (c\tilde{a})^2}{\left[1 +
\left(c\tilde{a}\right)^2\right]^2}, \label{eqn.corelog.rho.emd}
\end{equation}
where $c\equiv a_{\Delta}/a_c$, and $a_c$ is an ellipsoidal analog of
$r_c$. From this expression for the density follow the equations for
the enclosed mass and the potential as a function of ellipsoidal
radius,
\begin{equation}
M(<\tilde{a})  =
M_{\Delta}(1+c^2)\frac{\tilde{a}^3}{1+\left(c\tilde{a}\right)^2},
\end{equation}
and,
\begin{eqnarray}
\Phi(\vec{x}) & = & -\frac{GM_{\Delta}}{2}\frac{1+c^2}{c^2}
\int_0^{\infty}\frac{du}{\Delta (u)} \left[ \frac{1}{2} \ln\left(
    \frac{1+c^2}{1+[c\tilde{a}(\vec{x},u)]^2} \right) +
  \frac{1}{1+[c\tilde{a}(\vec{x},u)]^2} - \frac{1}{1+c^2}\right], \\
& = & -GM_{\Delta}\frac{1+c^2}{c^2}
\int_0^{1}\frac{ds}{\delta (s)} \left[ \frac{1}{2} \ln\left(
    \frac{1+c^2}{1+[c\tilde{a}(\vec{x},s)]^2} \right) +
  \frac{1}{1+[c\tilde{a}(\vec{x},s)]^2} - \frac{1}{1+c^2}\right]
 \label{eqn.corelog.phi.emd}.
\end{eqnarray}
Because the integral over $\tilde{a}$ diverges at $\infty$, we
replaced the upper limit with $\tilde{a}=1$. With this choice, we
have for $p_v=q_v=1$ that $\Phi$ reduces to the spherical case (eqn.\
\ref{eqn.corelog.phi}) but offset by a constant so that,
$\Phi(r_{\Delta}) = -GM_{\Delta}/r_{\Delta}$, rather than
$\Phi(r_{\Delta}) = 0$.

\section{Fiducial Mass Model Parameters and Axial Ratio Profiles}
\label{fiducial}

The ellipsoidal NFW and CORELOG models have four free parameters that
need to be specified: concentration $(c)$, mass $(M_{\Delta})$, and
axis ratios ($p_v$,$q_v$). The axis ratios refer to the potential for
an EP and to the mass distribution for an EMD.  As we show below in \S
\ref{results}, our results are not very sensitive to the degree of
triaxiality of the ellipsoid as quantified by the triaxiality
parameter $T$~\citep{fran91a}. Consequently, unless stated otherwise,
throughout this paper we set,
\begin{equation}
p_v = \sqrt{1 - T\left(1 - q_v^2\right)} = \sqrt{\frac{1+q_v^2}{2}},
\label{eqn.triax} 
\end{equation}
where we adopt $T=0.5$ for the ``maximally triaxial'' ellipsoid that
lies midway between the oblate $(T=0)$ and prolate $(T=1)$ spheroids.

Since we also find that our results are very insensitive to the halo
mass scale over the range of interest $(10^{12}-10^{15}\, M_{\odot})$,
we adopt a fiducial mass, $M_{\Delta}=10^{14}\, \msun$, which lies
near the middle of this range. With this value for the mass, we set
the NFW concentration to $c=9$ (assuming $\Delta\approx 100$)
appropriate for the results we obtained from X-ray measurements
spanning the same large mass range~\citep{buot07a}. Although this
normalization of the concentration-mass relation is 30-60\% larger
than that obtained from dissipationless cosmological
simulations~\citep[e.g.,][]{macc08a}, the results we present in this
paper are unchanged if instead we use such a smaller concentration
value.  These values for $c$ and $M_{\Delta}$ are used for both the EP
and EMD variations of the NFW model.  To provide a clear contrast with
the NFW models, we adopt a large value of the concentration parameter
$(c=100)$ for all CORELOG models which insures nearly scale-free
behavior (e.g, $M\sim a_v$ for EP) over most of the system.

\begin{figure*}
\parbox{0.49\textwidth}{
\centerline{\includegraphics[scale=0.42,angle=0]{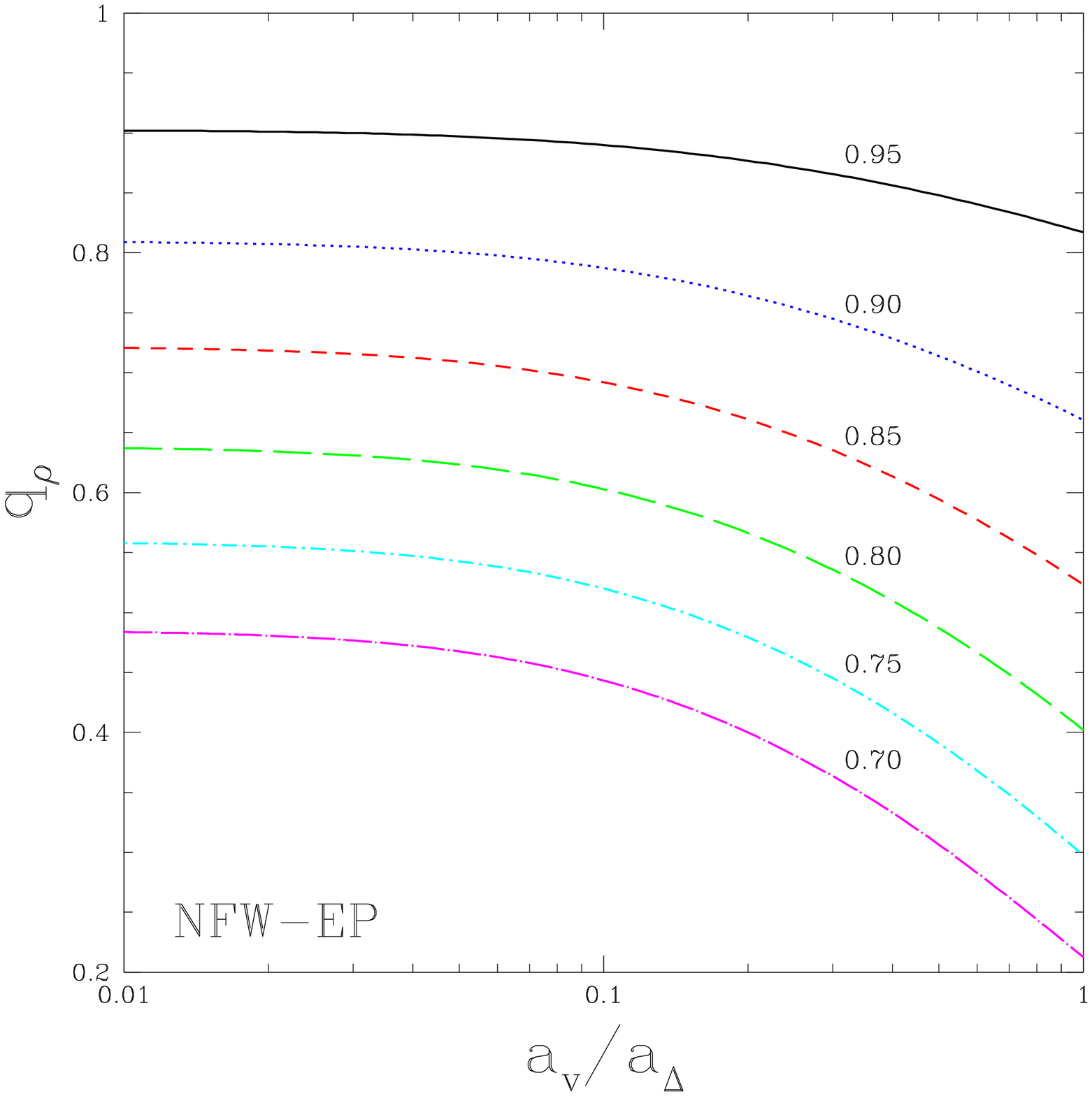}}}
\parbox{0.49\textwidth}{
\centerline{\includegraphics[scale=0.42,angle=0]{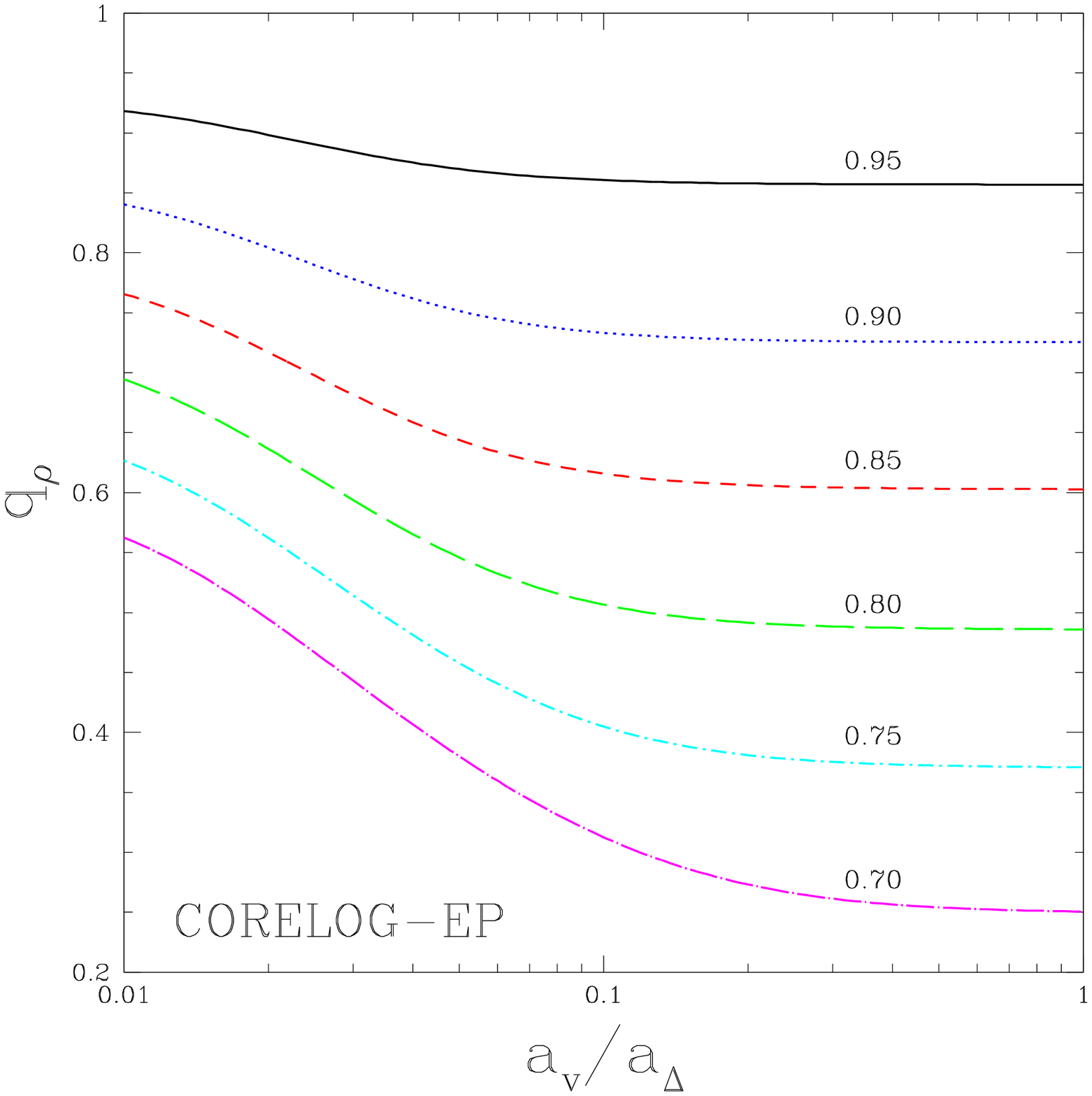}}}
\caption{\label{fig.ep} Axial ratio profiles of the mass density for
  the fiducial NFW-EP ({\sl Left Panel}) and CORELOG-EP ({\sl Right
  Panel}) models (\S \ref{fiducial}). Shown are results for six values
  of the axial ratio in the potential $q_{\Phi}$. Note that for
  hydrostatic equilibrium the axial ratio of the X-ray emissivity is
  the same as the potential, independent of the temperature profile
  (see \S \ref{he}).}
\end{figure*}

Using these fiducial parameters for the EP models, we display in
Figure~\ref{fig.ep} the (short-to-long) axial ratios $q_{\rho}$ of the
isodensity contours plotted as a function of $a_v/a_{\Delta}$ (with
$\Delta = 100$) for different values of $q_{\Phi}\equiv q_v$, the
axial ratio of the potential ellipsoid. We consider the range,
$q_{\Phi}=0.7-1$, corresponding to approximately, $q_{\rho}=0.4-1$,
appropriate for \lcdm\ halos~\citep[e.g.,][and references
therein]{bail05a}.  The $q_{\rho}$ were computed for a given $a_v$ by
solving the equation, $\rho(a_v,0,0) = \rho(0,0,z)$, for $z$ and then
setting $q_{\rho} = z/a_v$. We find that for both models, as is
typical for EPs, that $q_{\rho}$ decreases (i.e., the density
distribution becomes flatter) with increasing $a_v$.  As $a_v$
approaches $a_{\Delta}$, $q_{\rho}$ approaches a constant value for
each $q_{\Phi}$ for CORELOG-EP, whereas $q_{\rho}$ continues to
decrease for NFW-EP; i.e., the NFW-EP mass distribution is flatter
than CORELOG-EP for $a_v \sim a_{\Delta}$ and is slightly rounder for
$a_v \sim 0.01a_{\Delta}$.  (Note that for $q_{\Phi}<0.8$ the density
of the fiducial NFW-EP model takes negative values on the $z$ axis for
some $z<a_{\Delta}$.)
%; i.e., $a_v\sim (0.4,0.6)a_{\Delta}$ for $q_{\Phi}<(0.70,0.75)$.)
We mention that the decrease of $q_{\rho}$
with radius for the NFW-EP model disagrees with the generally
increasing trend found in dissipationless cosmological
simulations~\citep[e.g.,][]{bail05a,allg06a}, although $q_{\rho}$
profiles that decrease with radius tend to result when dissipation is
considered~\citep[e.g.,][]{kaza04a,deba08a}.

\begin{figure*}
\parbox{0.49\textwidth}{
\centerline{\includegraphics[scale=0.42,angle=0]{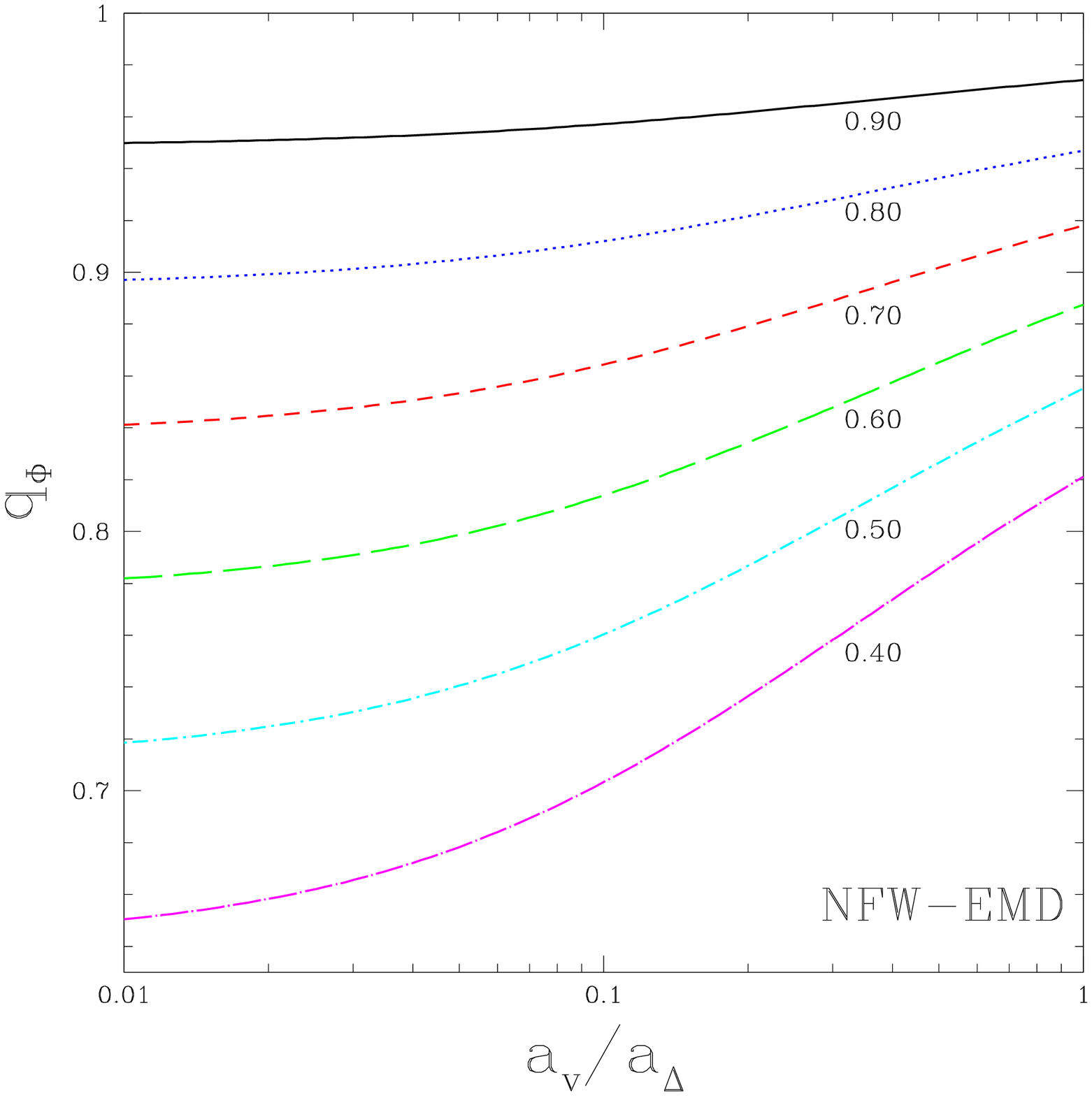}}}
\parbox{0.49\textwidth}{
\centerline{\includegraphics[scale=0.42,angle=0]{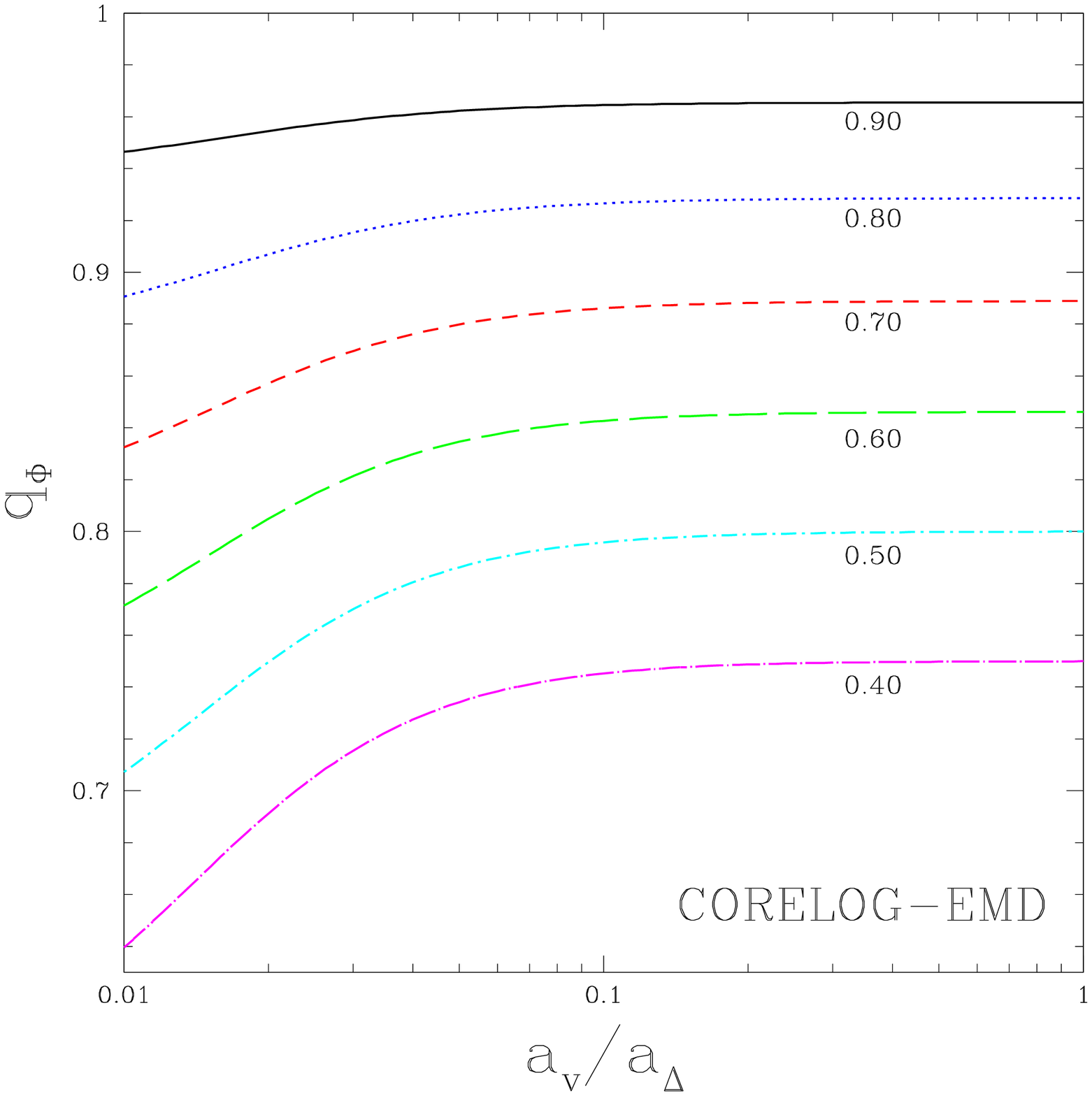}}}
\caption{\label{fig.emd} Axial ratio profiles of the potential for the
  fiducial NFW-EMD ({\sl Left Panel}) and CORELOG-EMD ({\sl Right
  Panel}) models (\S \ref{fiducial}). Shown are results for six values
  of the axial ratio in the mass density $q_{\rho}$. Note that for
  hydrostatic equilibrium the axial ratio of the X-ray emissivity is
  the same as the potential, independent of the temperature profile
  (see \S \ref{he}).}
\end{figure*}

For EMD models with the fiducial parameters, we display in
Figure~\ref{fig.emd} the axial ratios $(q_{\Phi})$ of the isopotential
contours plotted as a function of $a_v/a_{\Delta}$ for different
values of $q_{\rho}\equiv q_v$, the axial ratio of the mass (density)
ellipsoid. The $q_{\Phi}$ were computed for a given $a_v$ by solving
the equation, $\Phi(a_v,0,0) = \Phi(0,0,z)$, for $z$ and then setting
$q_{\Phi} = z/a_v$. For both the NFW-EMD and CORELOG-EMD models
$q_{\Phi}$ increases (i.e., the potential becomes rounder) with
increasing radius, though for CORELOG-EMD $q_{\Phi}$ attains a
constant value for $a_v\ga 0.1a_{\Delta}$ for all $q_{\rho}$
explored. Generally, for $a_v \sim a_{\Delta}$ the potential is
rounder for NFW-EMD compared to a CORELOG-EMD model with the same
$q_{\rho}$. The models have very similar values of $q_{\Phi}$ for $a_v
\sim 0.01a_{\Delta}$.

\section{Method to Test Spherical Averaging}
\label{method}

Our objective is to quantify average biases and scatter in cluster
properties arising from inconsistent spherical averaging procedures
employed by observers and theorists, with particular attention devoted
to properties measured by X-ray and SZ studies. It is most common for
the theorist to spherically average the three-dimensional cluster
available in their models, from which ``true'' spherically averaged
quantities are derived. The observer, however, has access only to the
cluster projected onto the sky. Generally, the observer circularly
averages the cluster data projected onto the sky and then performs a
deprojection assuming spherical symmetry. We refer to cluster
properties inferred from such a deprojection as ``observed''
spherically averaged properties. By comparing the ``observed'' and
``true'' values for clusters of different shapes and projection
orientations we quantify any bias due to spherical averaging in this
way. Below we provide more details on how we make this comparison.

\subsection{Hydrostatic Equilibrium -- Gas Mass and X-ray Emissivity}
\label{he}

To associate X-ray emission with a cluster entails filling the
gravitational potential well with hot plasma, which is approximated
very well by a non-rotating ideal gas in hydrostatic equilibrium,
\begin{equation}
\nabla P_{\rm gas} = -\rho_{\rm gas}\nabla\Phi, \label{eqn.he}
\end{equation}
where $P_{\rm gas}$ is the thermal pressure and $\rho_{\rm gas}$ is
the gas density.  For an isothermal gas, this equation can be solved
for the gas density to yield,
\begin{equation}
\rho_{\rm gas}(\vec{x}) = \rho_{\rm gas, 0} 
\exp\left(\frac{\mu m_{\rm a}}{k_B
    T}\left[\Phi_0 - \Phi(\vec{x}) \right]\right),
\end{equation}
where $\rho_{\rm gas, 0}$ and $\Phi_0$ are, respectively, the central
values of the gas density and the potential, and $T$ is the (constant)
temperature of the gas. Our study focuses on the isothermal solution
because temperature gradients have little impact on the flattening of
X-ray images of galaxies and clusters. This is a consequence of the
isopotential surfaces having the same shapes as the constant surfaces
of gas density, temperature, and X-ray emissivity; i.e., the X-ray
Shape Theorem~\citep{buot94,buot96a,buot12a}. In fact, in Theorem~7 of
Paper~1 we show that the observed deprojected spherically averaged
mass profile for any EP model is independent of the temperature
profile. Nevertheless, because other quantities may be more affected,
we also explore models with temperature gradients using a polytropic
equation of state $P_{\rm gas}\propto \rho_{\rm
gas}^{\gamma}$. Inserting this relationship between the pressure and
density into eqn.\ (\ref{eqn.he}) gives~\citep[e.g.,][]{sara86a},
\begin{equation}
T(\vec{x}) = T_0 + \frac{\gamma - 1}{\gamma} \frac{\mu m_{\rm a}}{k_B}
\left( \Phi_0 - \Phi(\vec{x})\right),
\end{equation}
\begin{equation}
\rho_{\rm gas}(\vec{x}) = \rho_{\rm gas,0}
\left(\frac{T(\vec{x})}{T_0}\right)^{\frac{1}{\gamma - 1}}.
\end{equation}
We focus our attention on models with $\gamma = 1.2$ which best represent the
falling temperature profiles (outside of any cool core) of observed
clusters and those produced in cosmological
simulations~\citep[e.g.,][]{ostr05a}.

To define completely these models we need to specify the
normalizations of the temperature and density profiles of the hot gas.
We normalize the temperature using the $M_{500}-T_{\rm X}$ relation
given by~\citet{arna05a,arna07a}.  For isothermal models, we simply
set $T=T_{\rm X}$. For the polytropic models we determine the
normalization constant $T_0$ by computing the gas-mass-weighted
temperature within a radius of $r_{500}$ and then set the result equal
to $T_{\rm X}$. The results presented below in \S \ref{results} are
very insensitive to the adopted form of the $M_{500}-T_{\rm X}$
relation.

Similarly, we normalize the gas density using the $M_{500}-M_{\rm gas,
\, 500}$ relation given by~\citet{arna07a}. That is, we determine $\rho_{\rm
gas, 0}$ by computing the total gas mass within a radius of $r_{500}$
and then set it equal to $M_{\rm gas, \, 500}$. While this
normalization is convenient, it is not strictly necessary because we
choose to neglect the self-gravity of the gas in our models
$(\Phi_{\rm gas} \ll \Phi)$ -- see beginning of \S \ref{mass}.
%This is justified by the small cluster gas fractions $(\sim 10\%)$
%observed within $r_{500}$~\citep[e.g.,][]{prat10a}.

Given the gas density and temperature we compute the X-ray emissivity,
$\epsilon_{\rm X} = \rho_{\rm gas}^2\Lambda(T)$, where $\Lambda(T)$ is the
plasma emissivity for a coronal plasma, which we take to be the APEC
model~\citep{apec} as implemented in \xspec~\citep{xspec} integrated
over energies 0.5-8~keV. We assume a constant metallicity of 0.5 solar
throughout the system, though our analysis is not very sensitive to
the assumed abundance profile. For isothermal models, we set $\Lambda
= 1$ so that the emissivity is simply, $\epsilon_{\rm X} = \rho_{\rm
gas}^2$.

\subsection{True Spherically Averaged Profiles of Mass and other Quantities}
\label{true}

For a given mass model (\S \ref{mass}) we compute the true spherically
averaged profile of the (total) mass as follows. For EP models the
spherical averaging of the mass profile is conveniently achieved by
applying Gauss's Law over a sphere of radius $r$ (Theorem~2 of
Paper~1). There is no such simplification for the EMD models, where
instead we directly integrate the density (i.e., either eqn.\
\ref{eqn.nfw.rho.emd} or \ref{eqn.corelog.rho.emd}) over a spherical
volume. We evaluate the spherically averaged mass profile in 500 (200)
logarithmically spaced bins for EP (EMD) models within a radius
$r=a_{\Delta}$, where $\Delta = 100$ (\S
\ref{fiducial}).  By fitting this profile with the spherical version
of the mass model (e.g., NFW -- eqn.\ \ref{eqn.nfw.mass}), we obtain
the ``true'' spherically averaged parameters for the mass model,
$c_{\rm sp}$ and $M_{\Delta, \, \rm sp}$. For consistency with the
observed mass analysis (see \S \ref{obs}), the fitting is performed
only within $r=(p_vq_v)^{1/3}a_{500}$ which is $\approx r_{500}$. Note
in most cases at least 80\% of the logarithmically spaced bins are
within this radius. Moreover, the inferred values of $c_{\rm sp}$ and
$M_{\Delta, \, \rm sp}$ are not very sensitive to small changes ($\sim
10\%$) in the outer radius used in the fits.

Having chosen both a mass model and a gas equation of state (\S
\ref{he}) we compute the spherically averaged profiles of quantities
directly associated with the hot gas.  For EMD models we compute the
``true'' spherically averaged gas mass $(M_{\rm gas, sp})$ by
integrating $\rho_{\rm gas}$ over a spherical volume of radius
$r=r_{500}$, where $r_{500}$ is evaluated using $M_{500, \, \rm sp}$
determined by fitting the mass profile as described above.  For EPs
the effective spherical average is used by setting $M_{\rm gas, sp} = M_{\rm
gas}(<(p_vq_v)^{-1/3}r_{500})$, which is the three-dimensional analog of Definition~1
of Paper~1. These procedures insure a consistent comparison of the gas
mass with the observations (\S \ref{obs}). It then follows that the
spherically averaged gas fraction is, $f_{\rm gas, sp} = M_{\rm gas,
sp}/ M_{500, \, \rm sp}$. For the emission weighted temperature
$(T_{\rm X, sp})$ we integrate $\epsilon_{\rm X}T$ within $r_{500}$
and then divide by the luminosity within that volume. Using the
temperature and gas mass we compute, $Y_{\rm X, sp} = M_{\rm gas,
sp}T_{\rm X, sp}$~\citep{krav06a}. Finally, since the integrated
Compton-y parameter is proportional to the volume integral of the gas
pressure (e.g., eqn.\ 48 of Paper~1), we set $Y_{\rm SZ, sp}$ equal to
the integral of $P_{\rm gas}$ within $r_{500}$, ignoring
proportionality constants which are unimportant for our study.

\subsection{Projection}
\label{proj}

We construct the X-ray image by projecting the X-ray emissivity onto
the sky for a given orientation.  In Paper~1 we present analytical
formulas for EPs relating the ``observed'' deprojected spherically
averaged quantities (e.g., mass, gas mass, etc.) to their
three-dimensional ellipsoidal counterparts. This means that for a
given shape and projection orientation one may, e.g., immediately
evaluate the ``observed'' deprojected spherically averaged mass
profile for NFW-EP using eqn.\ (\ref{eqn.nfw.mass.ep}) inserted into
the result of Theorem~7 of Paper~1. Nevertheless, in our present
investigation we find it useful to carry out the projection (and
subsequent deprojection) numerically in order to test the accuracy of
other aspects of our computer code, particularly via comparison to the
EMDs which cannot be evaluated with simple analytical formulas.

Since the X-ray emissivity of an EP is an ellipsoid, we follow the
procedure described in \S~5 of Paper~1 and carry out the projection by
partitioning the emissivity into a series of concentric, similar
triaxial ellipsoidal shells and by approximating the emissivity as a
constant within each shell. For such a system viewed with orientation
$(\theta, \phi)$, expressions for the surface brightness
$\Sigma_{{\rm X},i}$ and temperature map $\langle T\rangle_i$ are given by
eqns.\ (66) and (69) in Paper~1. Each quantity is defined within an
elliptical annulus $i$ on the sky with inner semi-major axis
$\alpha_{i-1}$, outer semi-major axis $\alpha_i$, and axial ratio
$q_s$ (eqn.\ 18 of Paper~1) which is same for all annuli.

We compute the surface brightness of an EMD,
$\Sigma_{\rm X}(x^{\prime},y^{\prime}) = \int \epsilon_{\rm X} (\vec{x}^{\prime})
dz^{\prime}$, via full integration of the X-ray emissivity along the
line-of-sight for a given $(\theta, \phi)$.  Using the same procedure
we compute the emission-weighted temperature map, $\langle T\rangle
(x^{\prime},y^{\prime}) = \int \epsilon_X (\vec{x}^{\prime}) T
(\vec{x}^{\prime}) dz^{\prime} / \Sigma_{\rm X}(x^{\prime},y^{\prime})$.
The integrations proceed interior to the isopotential surface that
passes through $a_{\Delta}$ on the major axis. Since the potential of
an EMD is not itself ellipsoidal in shape, but is very nearly so, we
approximate this boundary surface by an ellipsoid in our
computations.

We make no further modifications to the model surface brightness and
temperature map. In particular, we do not fold the image through an
image point spread function, nor do we fold the spectrum through a
response matrix. We assume the observer is able to account for such
issues perfectly, so our analysis is able to focus on potential biases
arising only from geometrical issues associated with projection and
spherical averaging of the galaxy or cluster.

(Note that the sky map of the Compton-y parameter is obtained in the
same manner as $\Sigma_{\rm X}$ by replacing the emissivity with the
gas pressure.)

\subsection{Spherical Deprojection}
\label{deproj}

Now turning our perspective to that of an observer, we begin with the
surface brightness, temperature, and Compton-y maps on the sky as
produced in the previous section. In general these maps have
non-circular shapes, but it is the goal of the fictitious observer to
infer spherically averaged profiles from these maps. To this end
we follow standard practice and bin the surface brightness into a
series of $N$ concentric, circular annuli defined so that $R_i$ is the
outer radius of annulus $i$. For consistency, we use the same bin
definitions adopted in \S \ref{true} for the three-dimensional radius
of the true mass profile; i.e., we associate each $R_i$ on the sky
with the equivalent $r_i$ in three dimensions defined in \S
\ref{true}.

The procedure to create the circular binning differs for the EP and
EMD models. For EPs the constructed image is already binned in
elliptical annuli, where the annuli definitions correspond to
elliptical versions of the spherical radii adopted in \S \ref{true} for
the true mass profile; i.e., for each bin we have $\alpha_i =
a_{v,i}=r_i$, where $\alpha_i$ is the outer semi-major axis of
elliptical annulus $i$ on the sky and $a_{v,i}$ is the outer
semi-major axis of the shell $i$ in three dimensions. To achieve the
circular binning on the sky, for $\Sigma_{X,i}$ defined in each
elliptical bin we associate an inner radius $R_{i-1} \equiv
\alpha_{i-1}\sqrt{q_s}$ and an outer radius $R_{i} \equiv
\alpha_{i}\sqrt{q_s}$ (see Definition~1 of Paper~1). In contrast, for the
circular binning of the EMDs we must resort to computing, $\int
\Sigma_X(x^{\prime},y^{\prime}) dx^{\prime} dy^{\prime}$, the full
integration of the surface brightness over each circular bin
$(R_{i-1},R_i)$, a much more computationally expensive procedure than
for the EPs. To improve computational speed for the EMDs we use only
ten points, equally spaced in azimuth, for each circular bin. We also
take advantage of the elliptical symmetry and perform the calculation
over half of the circle. Finally, as in previous studies (e.g.,
\citealt{lewi03a}, see also~\citealt{mcla99a}), we find it useful for
each annulus to choose an intermediate average radius to represent the
result for the entire bin.

We obtain the observed radial volume emissivity in each shell $i$
$(\epsilon_i^{\rm sp, obs})$ by deprojecting the circularly averaged
surface brightness $(\Sigma_{X,i}^{\rm circ})$ using the standard
spherical ``onion peeling'' method~\citep{deproj,kris83}. In
particular, we convert the binned surface brightness values into
luminosities, $L_i^{\rm circ} = \pi (R_{i}^2 -
R_{i-1}^2)\Sigma_{X,i}^{\rm circ}$ and then apply the onion peeling
method as given by eqn.\ (66) of Paper~1 but using the spherical
version of the projection matrix $V^{\rm int}_{ij}$.  This procedure
is also used for the SZ map where the pressure takes the place of the
emissivity, and $L_i^{\rm circ}$ is replaced by the Compton-y
parameter within bin $i$.

The observed spherically averaged gas density follows immediately from
the emissivity, $\rho_{\rm gas, i}^{\rm sp, obs} =
\sqrt{\epsilon_i^{\rm sp, obs}}$, for isothermal models. For the
models with temperature gradients, however, we require the temperature
profile in order to evaluate the density.  To deproject the
temperature map we follow the same procedure as for the surface
brightness with the following substitutions to eqn.\ (66) of Paper~1:
$L_i\rightarrow\langle T\rangle_i L_i$ and $\epsilon\rightarrow
\epsilon T$. The result is the radial profile of the product of the
emissivity and the temperature $(\epsilon T)_i^{\rm sp, obs}$ in three
dimensions. Using this, along with the deprojected emissivity, gives
the radial temperature profile, $T_i^{\rm sp, obs} = (\epsilon
T)_i^{\rm sp, obs} / \epsilon_i^{\rm sp, obs}$, which allows the gas
density to be computed, $\rho_{\rm gas, i}^{\rm sp, obs} =
\sqrt{\epsilon_i^{\rm sp, obs} / \Lambda(T_i^{\rm sp, obs})}$. (We
assume the observer has perfect knowledge of the constant metallicity
for evaluation of $\Lambda$.)

\subsection{Observed Spherically Averaged Mass Profile}
\label{obs}

From the observed spherically averaged density and temperature
profiles, we compute the mass profile assuming hydrostatic
equilibrium,
\begin{equation}
M(<r)_{\rm sp}^{\rm obs} = -\left[\frac{rk_{\rm B}T_{\rm sp}^{\rm obs}}{\mu m_{\rm a}G}\right] \left[
\frac{d\ln\rho_{\rm gas, sp}^{\rm obs}}{d\ln r} + \frac{d\ln T_{\rm
sp}^{\rm obs}}{d\ln r} \right], \label{eqn.trad}
\end{equation}
where we evaluate the derivatives by interpolating the binned profiles
with cubic splines (except for the outermost bin where we use
logarithmic interpolation).  We compute the mass profile in the same
radial bins as done for the true mass profile (\S \ref{true}), and
then fit this observed mass profile with the same spherical version of
the mass model to obtain the observed spherically averaged model
parameters, $c_{\rm sp}^{\rm obs}$ and $M_{\Delta, \, \rm sp}^{\rm
  obs}$.  In order to minimize any weak sensitivity of our results to
the chosen outer boundary defined by the elliptical radius $a_{100}$
in the potential (mass) for EPs (EMDs), we performed the fitting of
the observed mass profile within a smaller radius. As remarked already
in \S \ref{true}, we adopted $r_{500}$ for this purpose, although we
found little difference when using other conventional choices (e.g.,
$r_{2500},r_{200}$).  The procedure to determine the other gas
parameters is the same as described for the ``true'' values in \S
\ref{true}. For example, for both EP and EMD models we compute the
spherically averaged gas mass $(M_{\rm gas, sp}^{\rm obs})$ by
integrating $\rho_{\rm gas, sp}^{\rm obs}$ within a spherical volume
of radius $r=r_{500}$, where $r_{500}$ is evaluated using $M_{500, \,
  \rm sp}^{\rm obs}$, from which immediately follows the spherically
averaged gas fraction, $f_{\rm gas, sp}^{\rm obs} = M_{\rm gas,
  sp}^{\rm obs}/ M_{500, \, \rm sp}^{\rm obs}$. In the same way we
compute $Y_{\rm SZ, sp}^{\rm obs}$ within $r_{500}$, though we note
that in practice SZ observations without corresponding X-ray data must
estimate $r_{500}$ in a different manner.

\section{Results}
\label{results}

\subsection{Preliminaries}
\label{prelim}

For our comparison of the ``observed'' (\S \ref{obs}) with the
``true'' (\S \ref{true}) spherically averaged quantities, we will
focus on the percentage bias for each parameter; e.g., the total mass, 
\begin{equation}
100\%\left(\frac{M - M_{\rm true}}{M_{\rm true}} \right)\equiv
100\%\left(\frac{M_{500, \, \rm sp}^{\rm obs} - M_{500, \, \rm sp}^{\rm
true}}{M_{500, \, \rm sp}^{\rm true}} \right), \label{eqn.bias}
\end{equation}
where we suppress the notation for the spherical averaging,
over-density, and ``obs'' denoting the observed quantity.  With regard
to the numerical accuracy of the biases presented below, there are
differences between the EP and EMD models.  In Paper~1 we showed that
for EPs the deprojected spherical average of a parameter (e.g.,
$M_{500, \, \rm sp}^{\rm obs}$)\footnote{For an EP this is also
denoted by $\langle M(<r_{500})\rangle^{\rm d+}$ using the
notation of Paper~1.} can be evaluated directly from the ellipsoidal
distribution (e.g., $M(<a_v)$) without the need to perform explicitly
any projection and deprojection. Consequently, the biases for EPs can
be evaluated extremely accurately and efficiently. Recall in \S
\ref{proj} we chose, nevertheless, to numerically project and
deproject the EPs to assess the performance of our computer code. We
found no significant differences between the results obtained via
numerical projection and deprojection with those obtained using the
analytical formulas from Paper~1; i.e., the numerical uncertainty for
the EPs is negligible for our investigation.

However, the much higher computational expense of the EMD models,
owing to the need for numerical integration of their non-analytical
potentials, limits the number of radial and azimuthal grid points that
can be practically employed, rendering the EMDs less accurate than the
EPs. By experimenting with models having different numbers of radial
and azimuthal grid points and having different tolerance precisions
for the numerical integrations, we conclude that in most cases the
numerical uncertainty of the EMDs is negligible for our
investigation. The largest uncertainties usually occur for models
with the smallest axial ratios ($q_v<0.5$) which translate at most to
bias errors of $\approx 0.2\%$ for EMDs.

\subsection{Bias Distributions}
\label{dist}

\begin{figure*}
\parbox{0.49\textwidth}{
\centerline{\includegraphics[scale=0.32,angle=0]{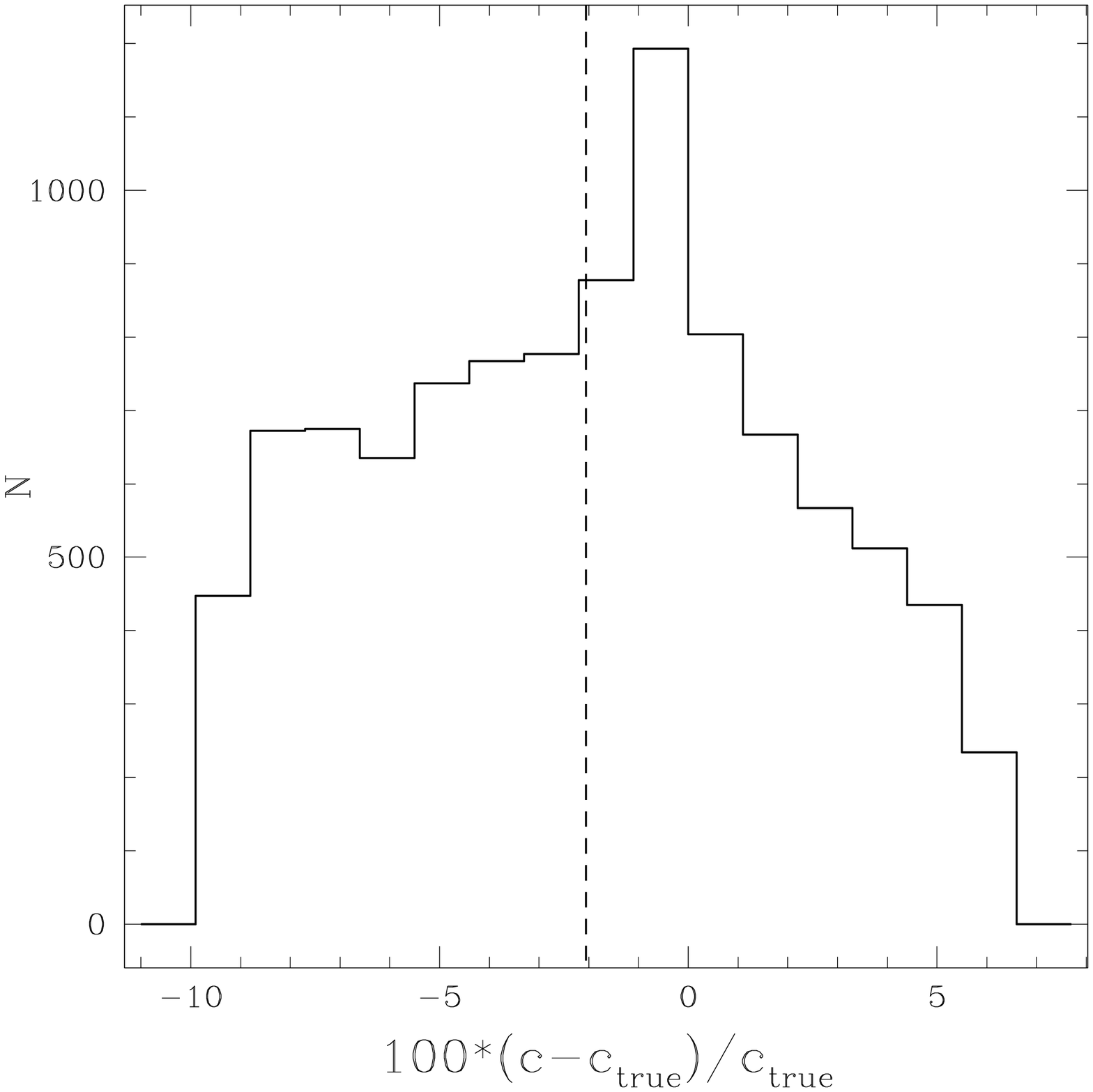}}}
\parbox{0.49\textwidth}{
\centerline{\includegraphics[scale=0.32,angle=0]{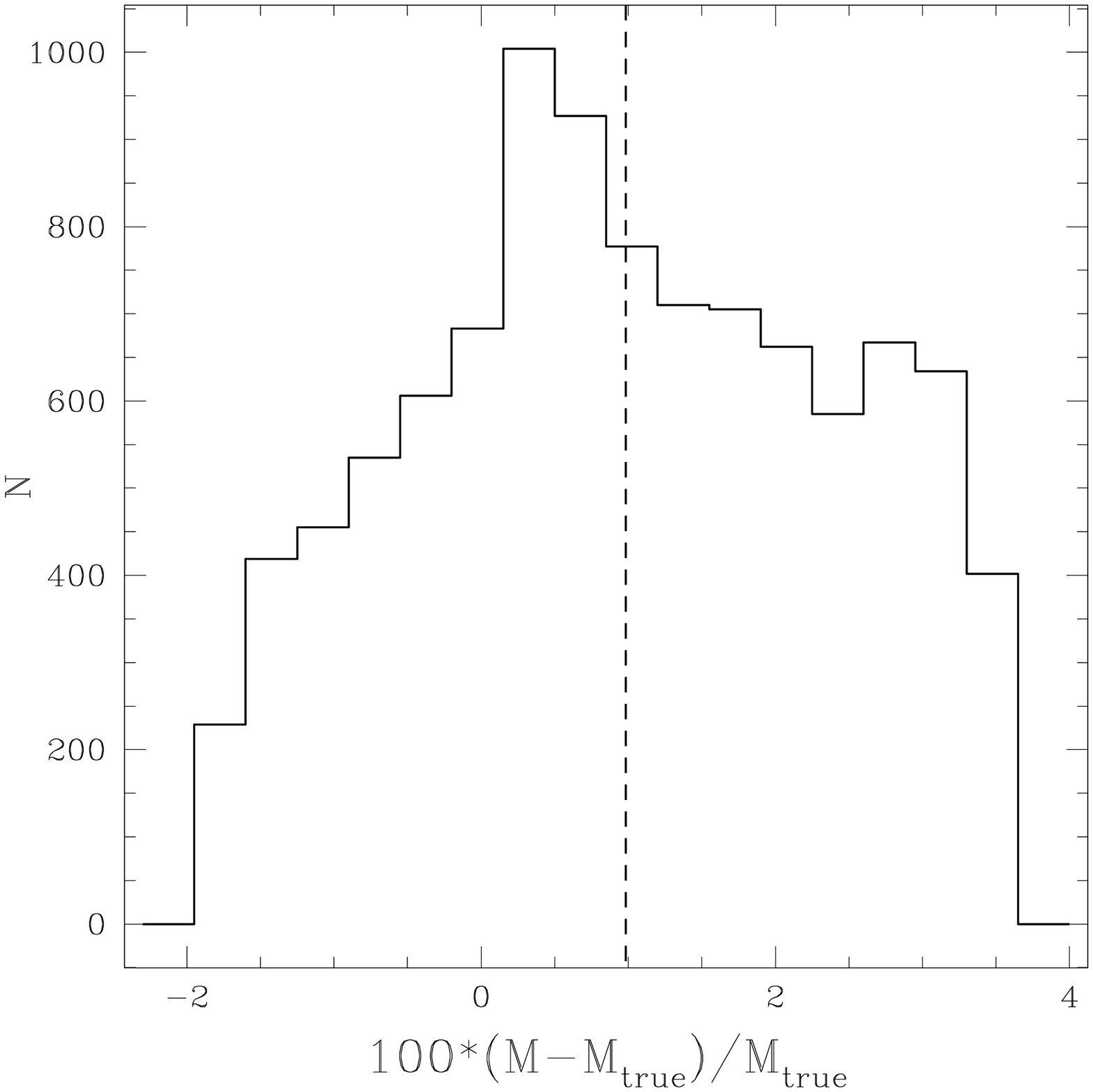}}}

\vskip 0.05cm

\parbox{0.49\textwidth}{
\centerline{\includegraphics[scale=0.32,angle=0]{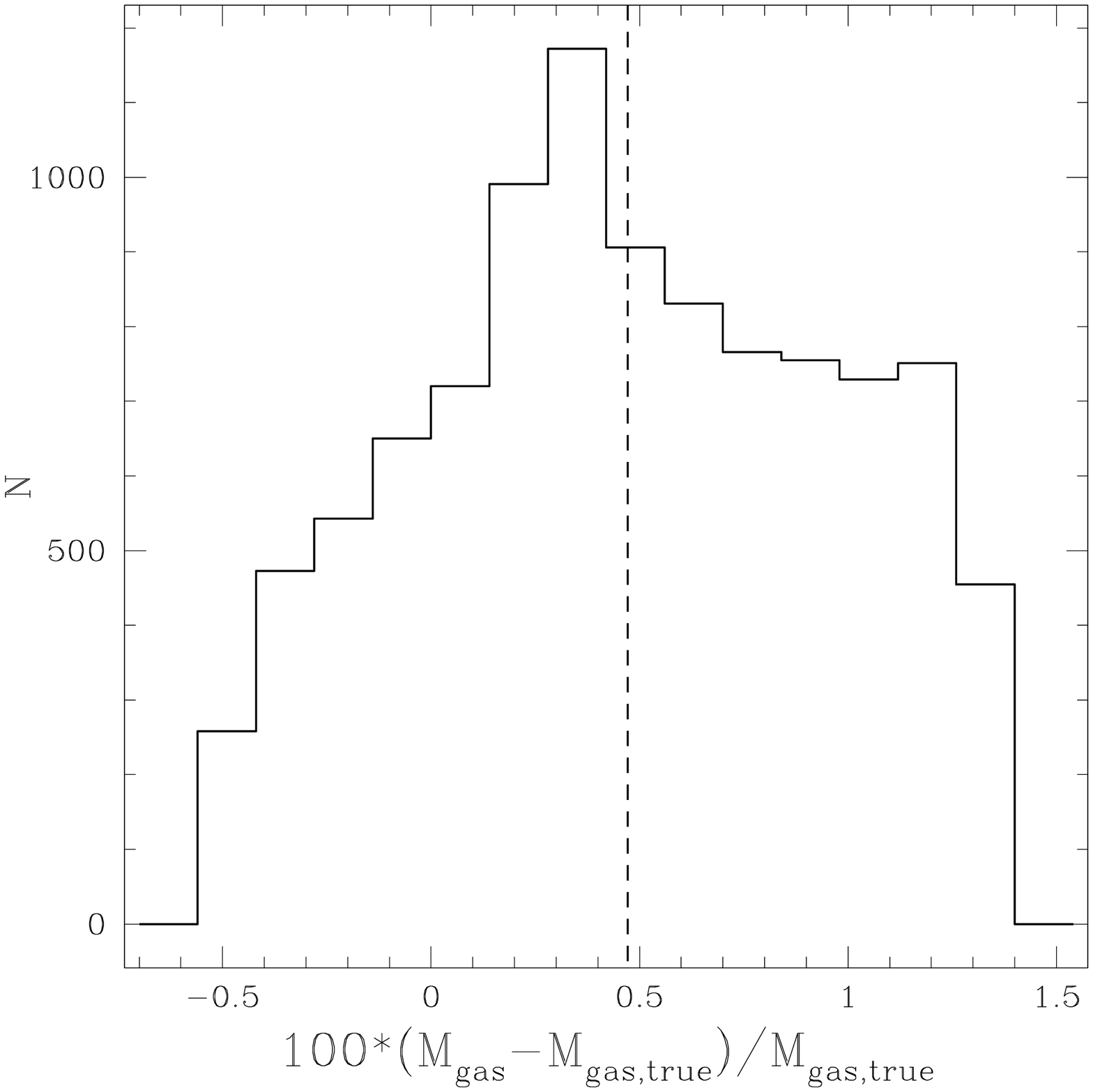}}}
\parbox{0.49\textwidth}{
\centerline{\includegraphics[scale=0.32,angle=0]{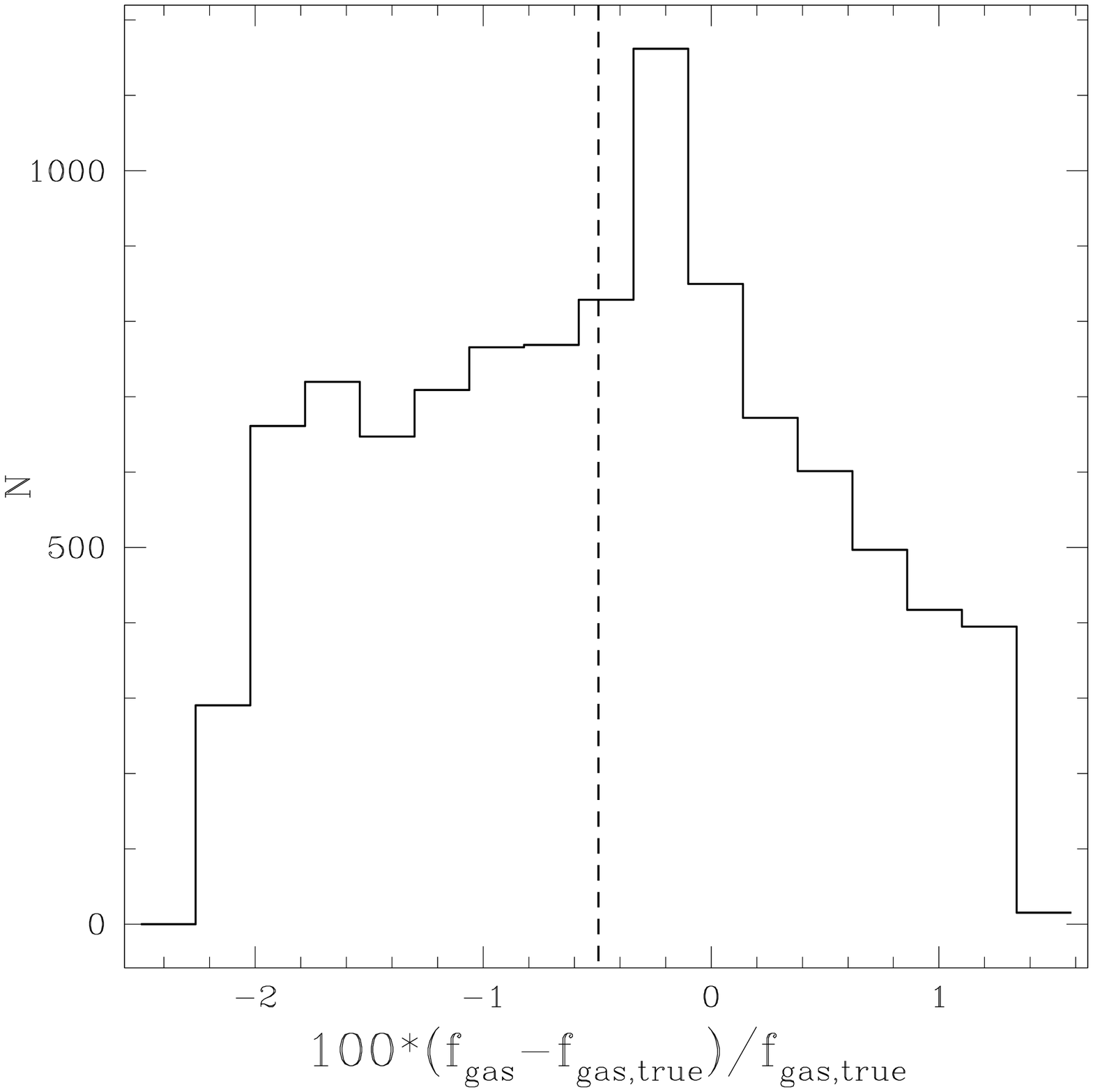}}}

\vskip 0.05cm

\parbox{0.49\textwidth}{
\centerline{\includegraphics[scale=0.32,angle=0]{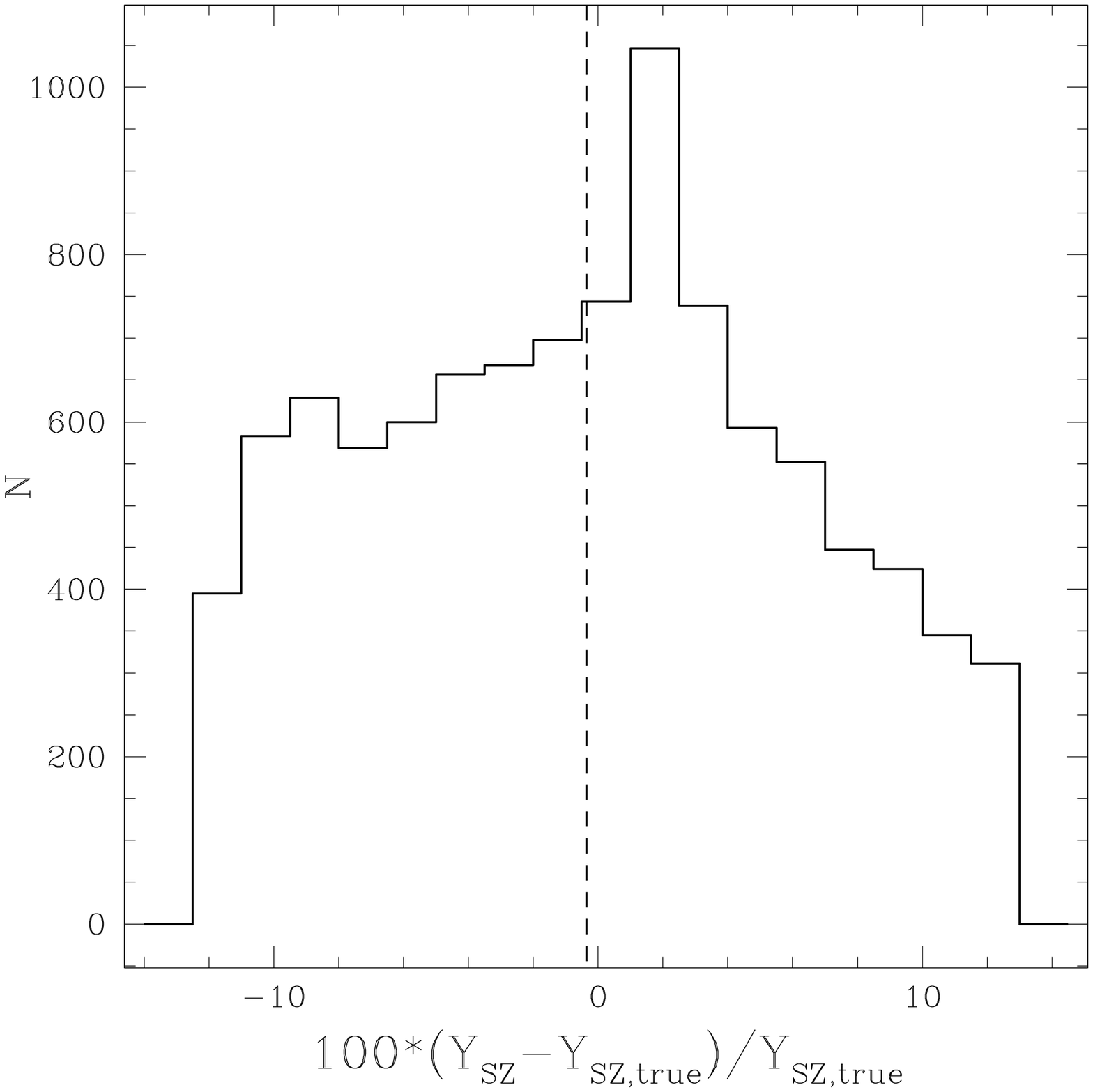}}}
\parbox{0.49\textwidth}{
\centerline{\includegraphics[scale=0.32,angle=0]{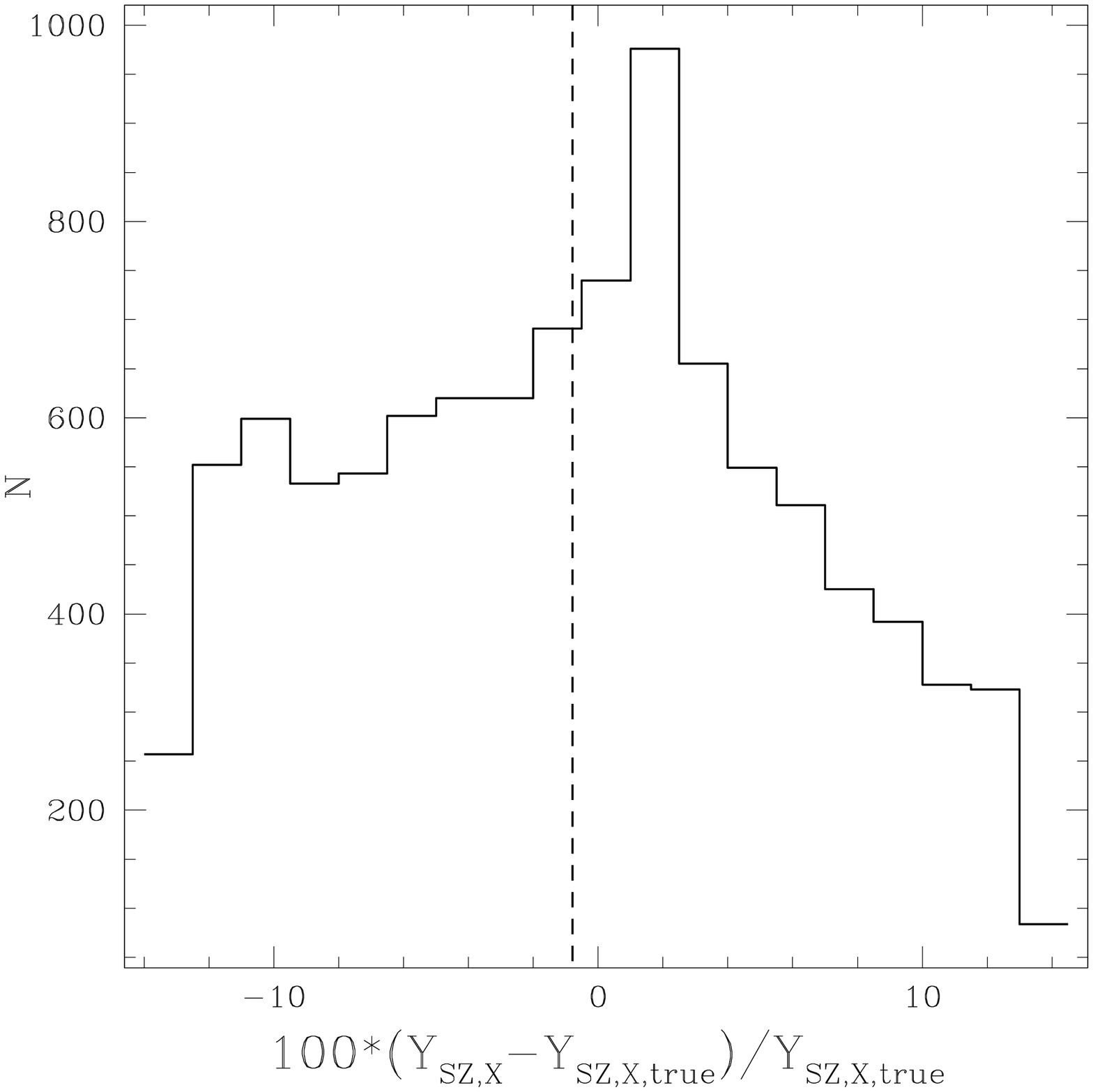}}}

\caption{\label{fig.hist} 
Bias distributions generated from $10^4$ random trials in viewing
orientation for the isothermal NFW-EP models with $q_v=0.70$ for the
gravitational potential.  Displayed are results for the concentration
({\sl top, left}), mass ({\sl top, right}), gas mass ({\sl middle,
left}), gas fraction ({\sl middle, right}), $Y_{\rm SZ}$ ({\sl bottom,
left}), and $Y_{\rm SZ,X}$ ({\sl bottom, right}).  The vertical dashed
lines show the mean of each distribution.}
\end{figure*}

We construct bias probability distributions by evaluating eqn.\
(\ref{eqn.bias}) (appropriate for each parameter) for a large number
of viewing orientations $(\theta,\phi)$ generated by randomly sampling
the solid angle over the orientation sphere. By taking advantage of
the ellipsoidal symmetry of the cluster models, we restrict the
orientation angles to the first octant of the sphere; i.e.,
$0\le\cos\theta\le 1$ and $0\le\phi\le\pi/2$. In Figure \ref{fig.hist}
we display the bias distributions generated by $10^4$ random
orientations of the isothermal NFW-EP model with $q_v=0.70$ for the
gravitational potential. All of the distributions are non-gaussian and
somewhat asymmetric, though they are peaked toward the center.
Interestingly, the mean bias for each parameter is significantly
different from zero; i.e., biases are not eliminated through
angle-averaging. As $q_v$ increases toward 1 the distributions of the
isothermal NFW-EP model become progressively narrower and more
symmetric (not shown). Despite the non-gaussian behavior for smaller
$q_v$, the mean and standard deviation provide a useful, convenient
characterization of each distribution, which we employ below to frame
our discussion of the systematic variations of the bias distributions
with $q_v$.

\subsection{Angle Averages and Standard Deviations}
\label{avg}

\begin{figure*}
\parbox{0.32\textwidth}{
\centerline{\includegraphics[scale=0.29,angle=0]{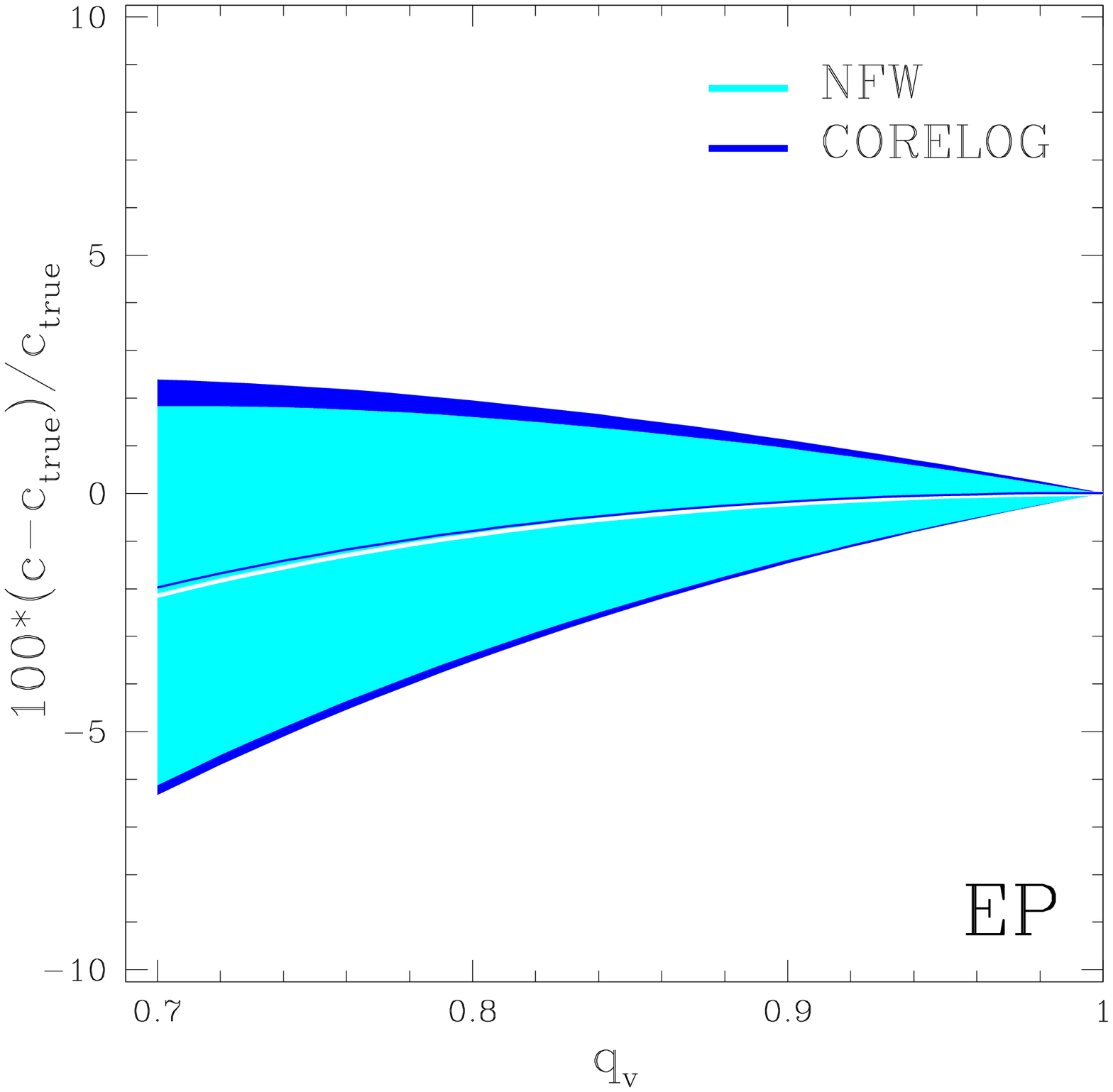}}}
\parbox{0.32\textwidth}{
\centerline{\includegraphics[scale=0.29,angle=0]{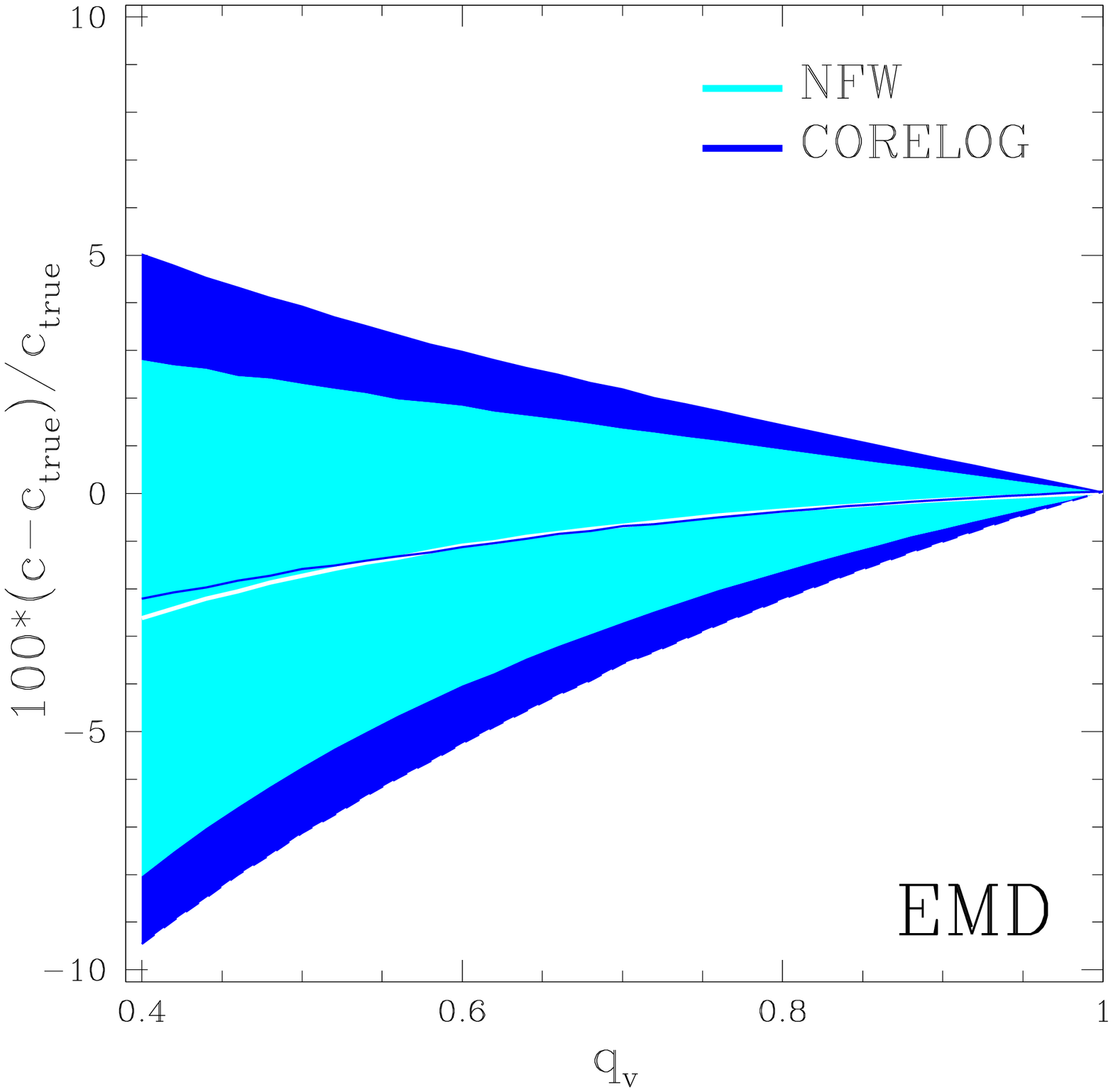}}}
\parbox{0.32\textwidth}{
\centerline{\includegraphics[scale=0.29,angle=0]{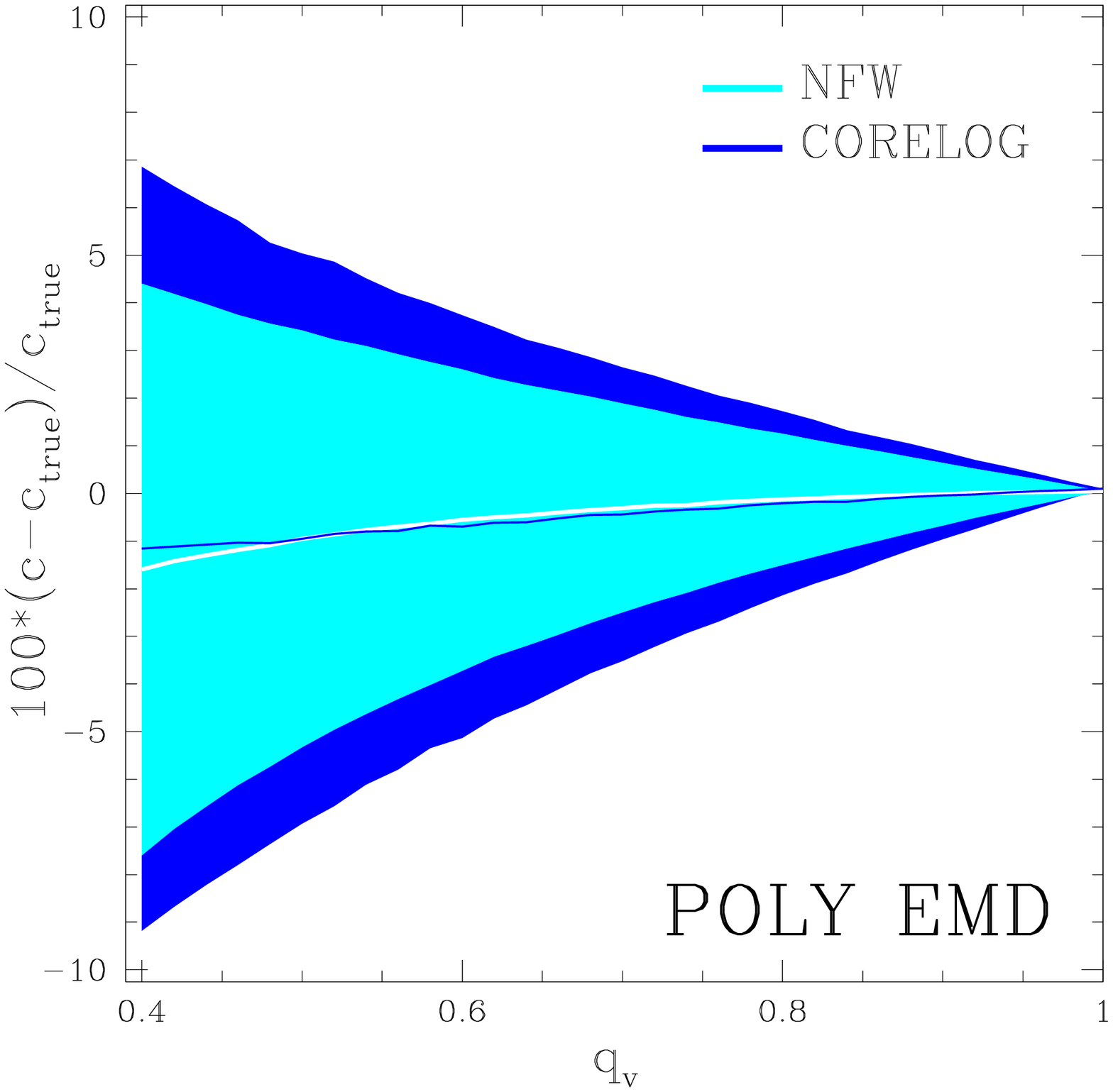}}}
\caption{\label{fig.error.c} 
Bias on the concentration parameter as a consequence of spherical
averaging. ({\sl Left Panel}) Results are shown for isothermal EP
models plotted as a function of $q_v$ in the gravitational
potential. (In hydrostatic equilibrium this is also the $q_v$ of the
X-ray emissivity.) The solid lines are orientation angle-averaged
values (NFW --white, CORELOG -- blue), while the shaded regions (NFW
-- light blue/cyan, CORELOG -- dark blue) are the $1\sigma$
ranges. ({\sl Middle Panel}) Same as the left panel except now the EMD
models are shown and, therefore, $q_v$ corresponds to the axial ratio
of the mass density profile.  ({\sl Right Panel}) Same as the middle
panel except now results for the $\gamma = 1.2$ polytrope with a
temperature gradient are shown.}
\end{figure*}

\begin{figure*}
\parbox{0.32\textwidth}{
\centerline{\includegraphics[scale=0.29,angle=0]{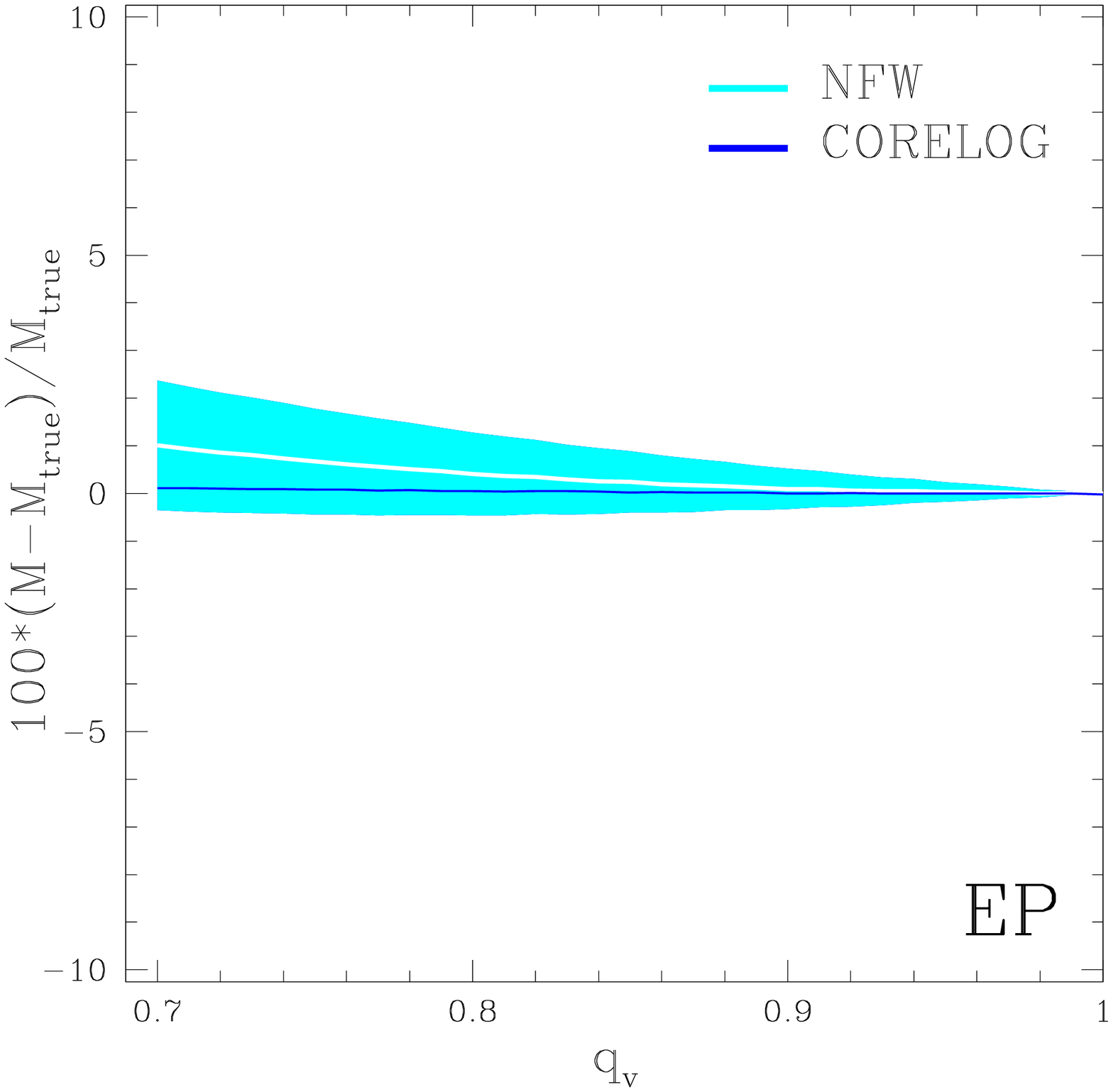}}}
\parbox{0.32\textwidth}{
\centerline{\includegraphics[scale=0.29,angle=0]{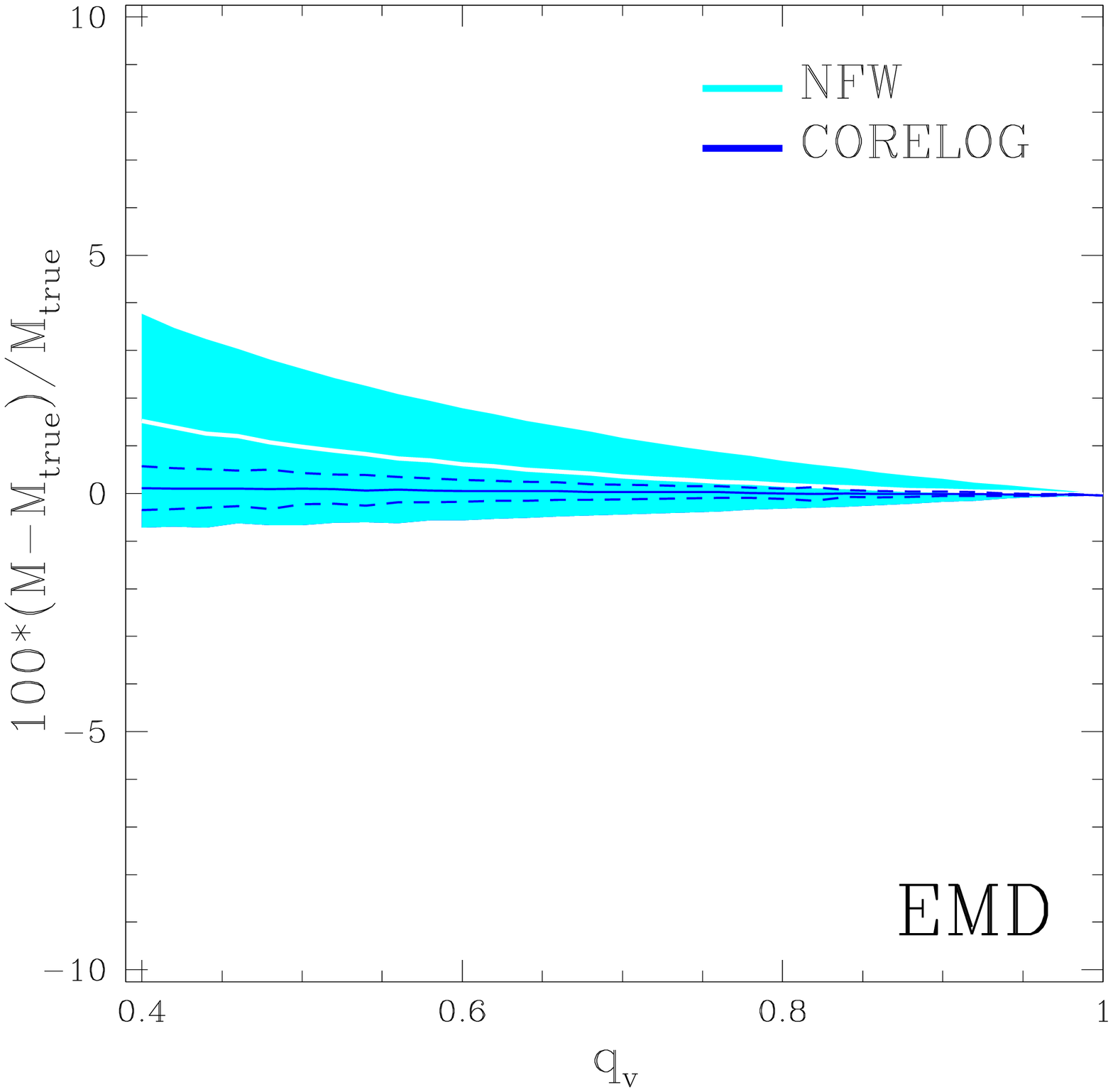}}}
\parbox{0.32\textwidth}{
\centerline{\includegraphics[scale=0.29,angle=0]{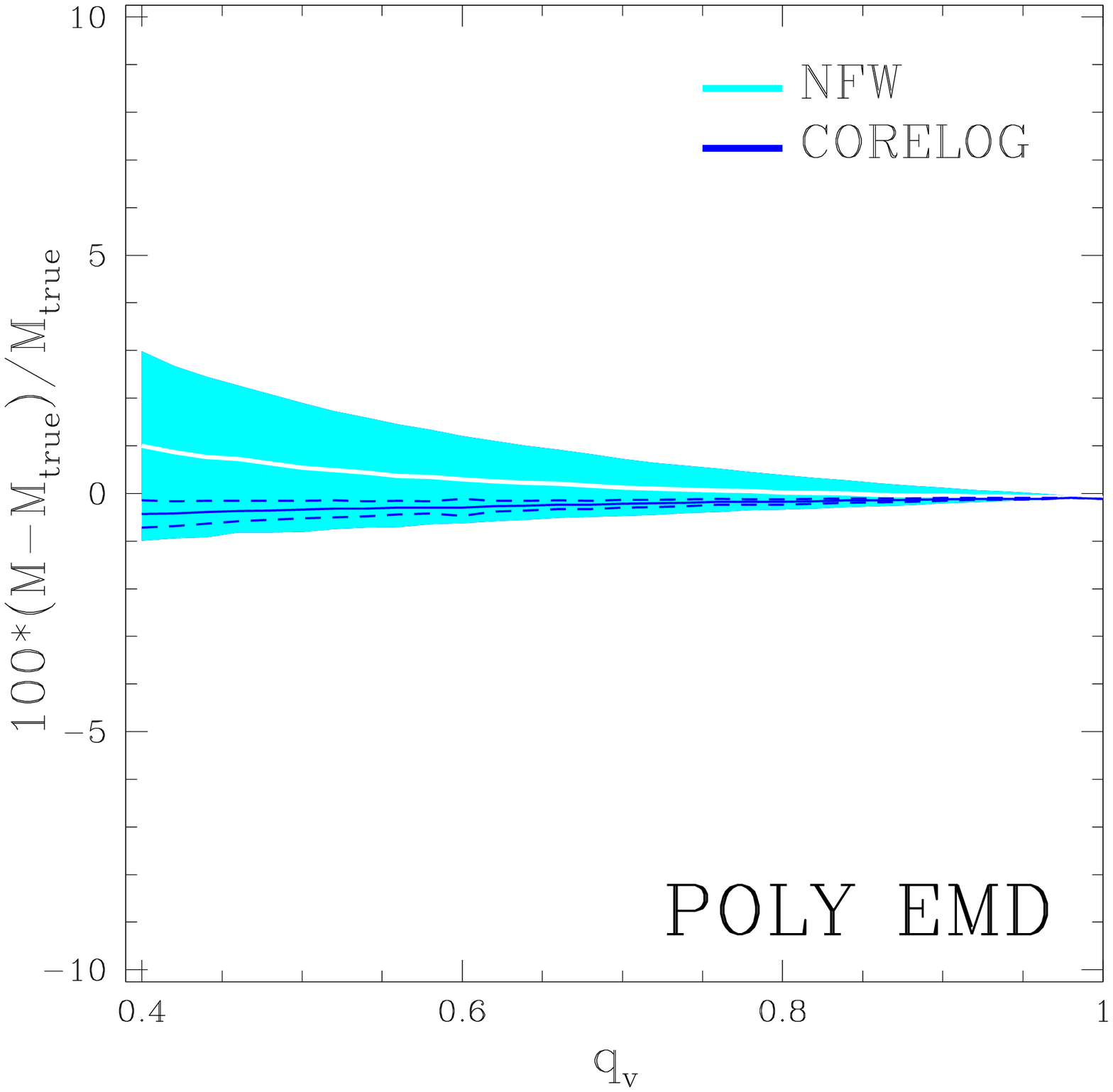}}}
\caption{\label{fig.error.m} Same as Figure \ref{fig.error.c} except
now the bias on the mass is shown. For clarity, the blue dashed lines
also represent the $1\sigma$ CORELOG region when it is contained
within the $1\sigma$ NFW region. Note that the $1\sigma$ region for
CORELOG-EP is negligible.}
\end{figure*}

\begin{figure*}
\parbox{0.49\textwidth}{
\centerline{\includegraphics[scale=0.32,angle=0]{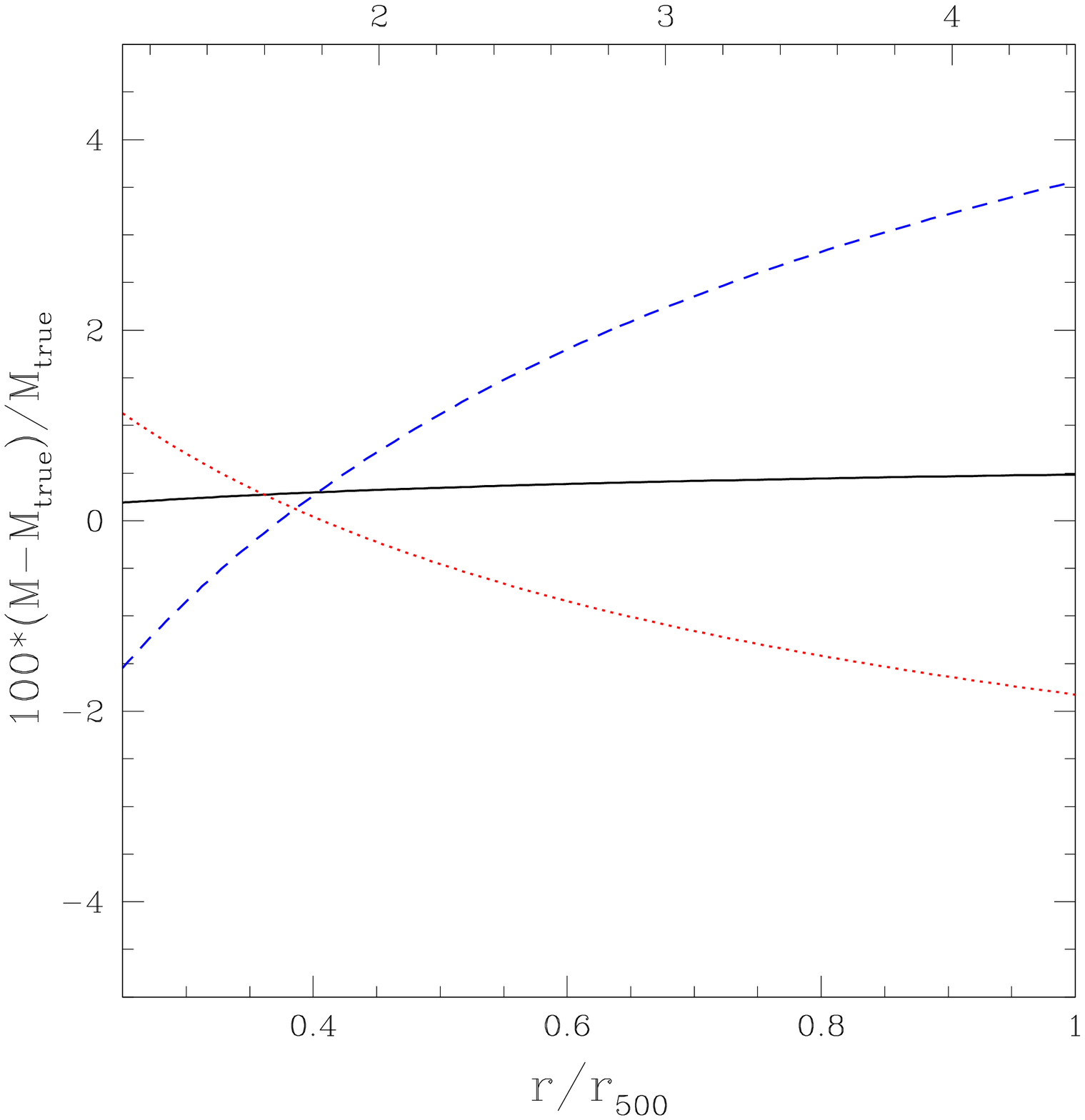}}}
\parbox{0.49\textwidth}{
\centerline{\includegraphics[scale=0.32,angle=0]{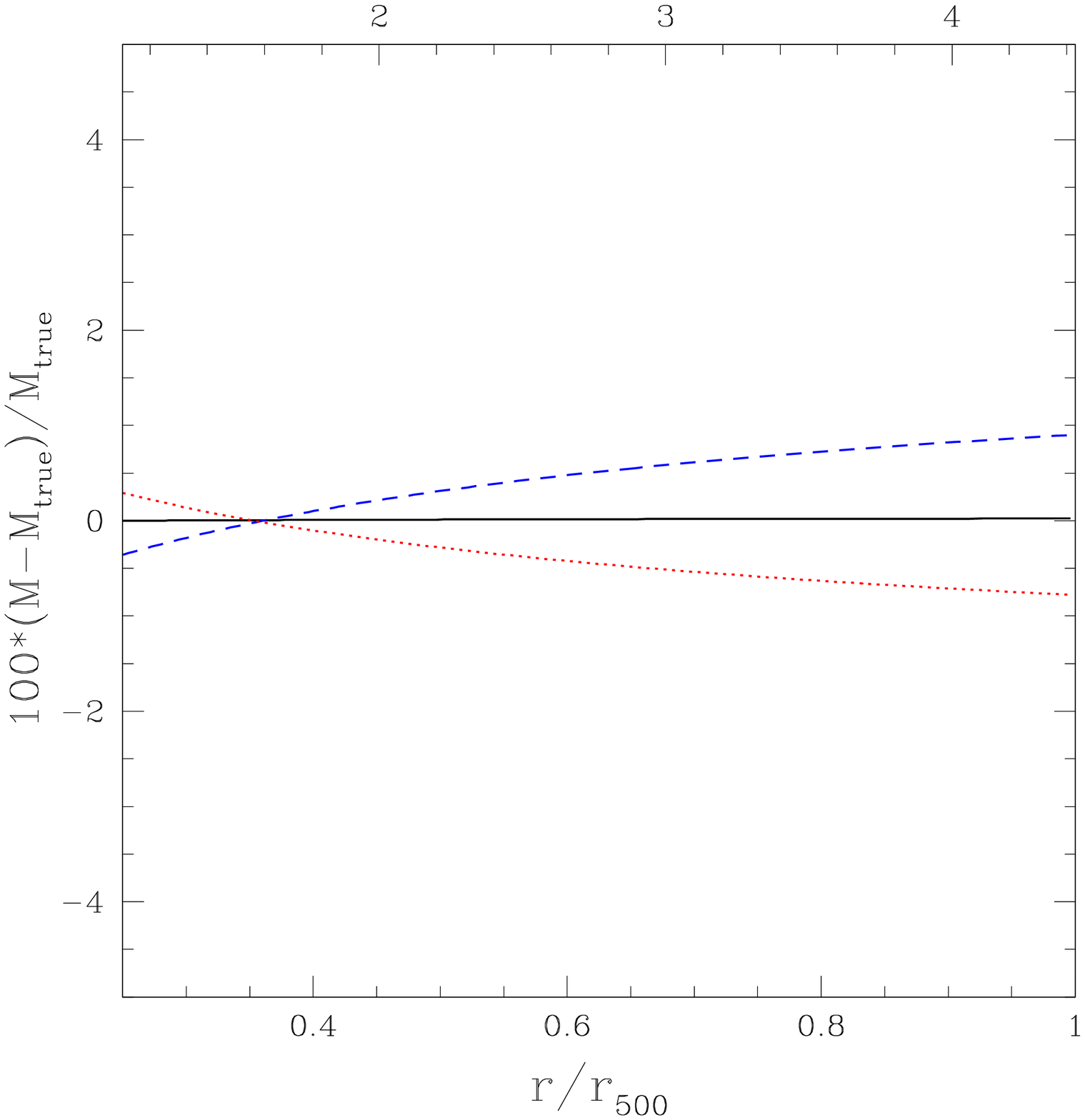}}}
\caption{\label{fig.radpro.ep} 
Bias of the mass plotted as a function of radius between $0.25-1\,
r_{500}$ for the isothermal NFW-EP model for $q_v=0.7$  ({\sl left
panel}) and $q_v=0.9$ ({\sl right panel}) in the gravitational
potential. Results are displayed for projections down the three
principal axes: short (dashed, blue), long (dotted, red), and intermediate
(solid, black). The top axis gives $r/r_s$, where $r_s$ is the NFW-EP
scale radius. }
\end{figure*}

\begin{figure*}
\parbox{0.49\textwidth}{
\centerline{\includegraphics[scale=0.32,angle=0]{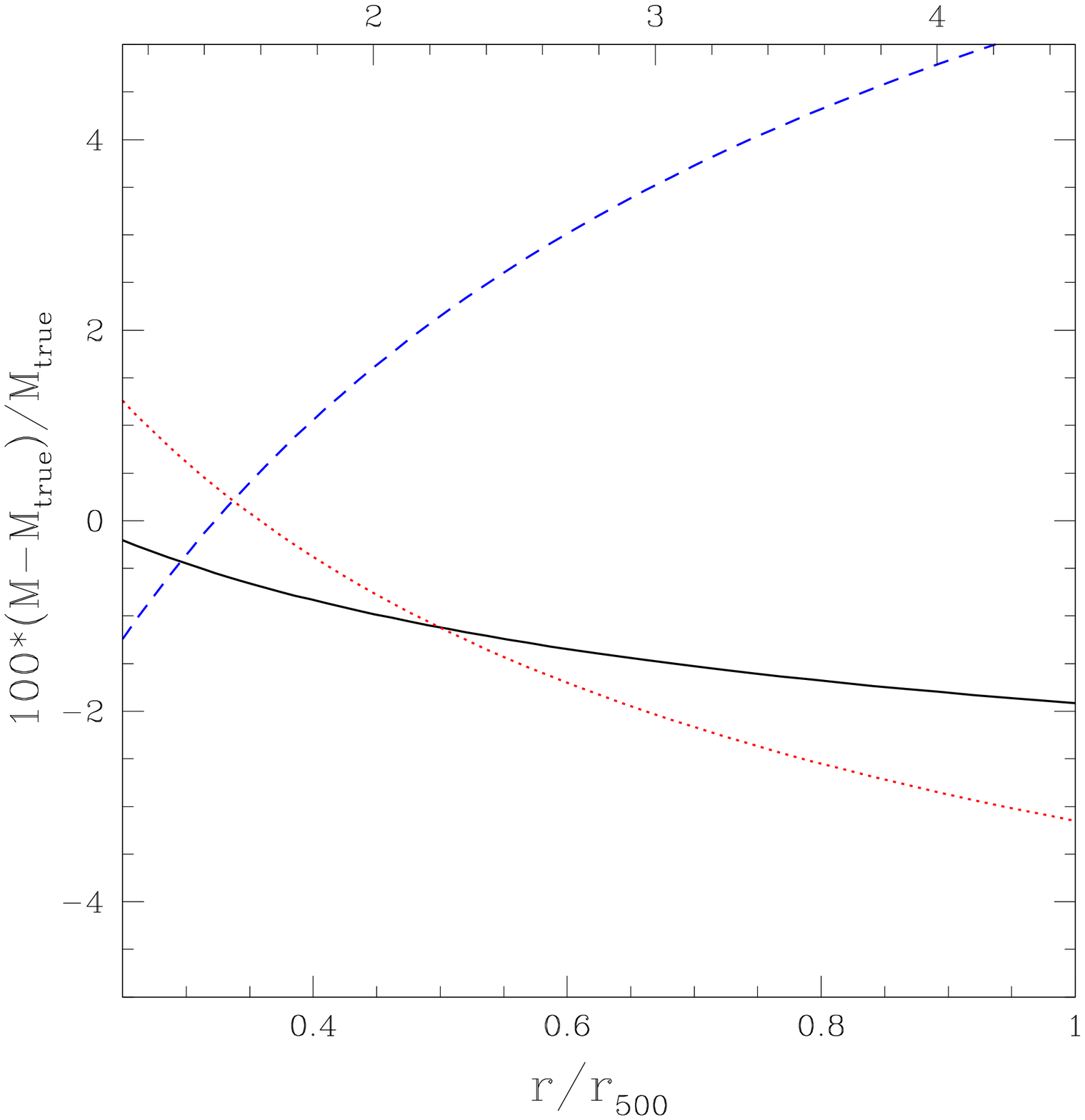}}}
\parbox{0.49\textwidth}{
\centerline{\includegraphics[scale=0.32,angle=0]{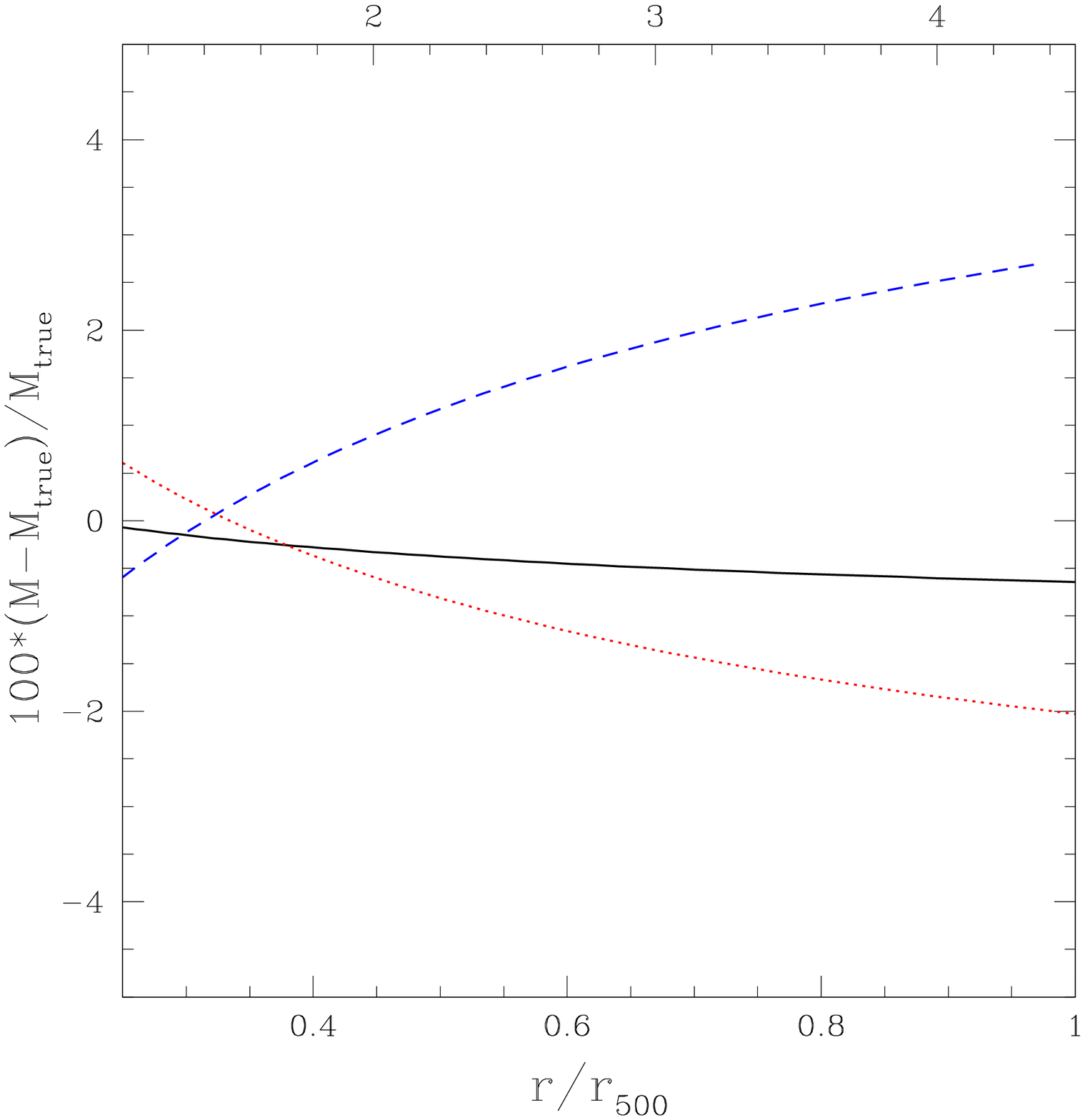}}}
\caption{\label{fig.radpro.emd} 
Bias of the mass plotted as a function of radius between $0.25-1\,
r_{500}$ for the isothermal NFW-EMD model for $q_v=0.4$ ({\sl left
panel}) and $q_v=0.6$ ({\sl right panel}) in the mass. Results are
displayed for projections down the three principal axes: short
(dashed, blue), long (dotted, red), and intermediate (solid,
black). The top axis gives $r/r_s$, where $r_s$ is the NFW-EMD scale
radius.}
\end{figure*}

\begin{figure*}
\parbox{0.32\textwidth}{
\centerline{\includegraphics[scale=0.29,angle=0]{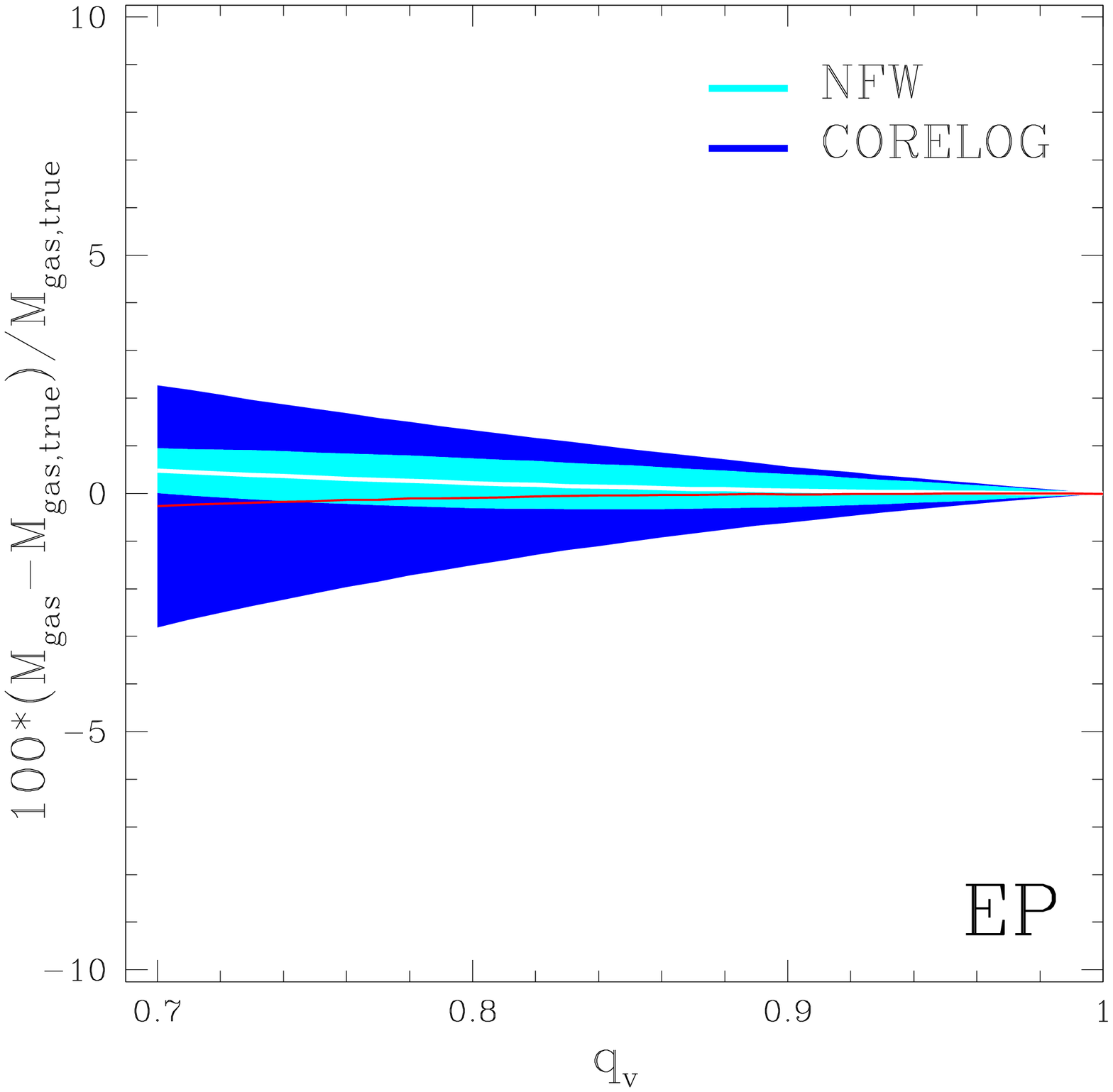}}}
\parbox{0.32\textwidth}{
\centerline{\includegraphics[scale=0.29,angle=0]{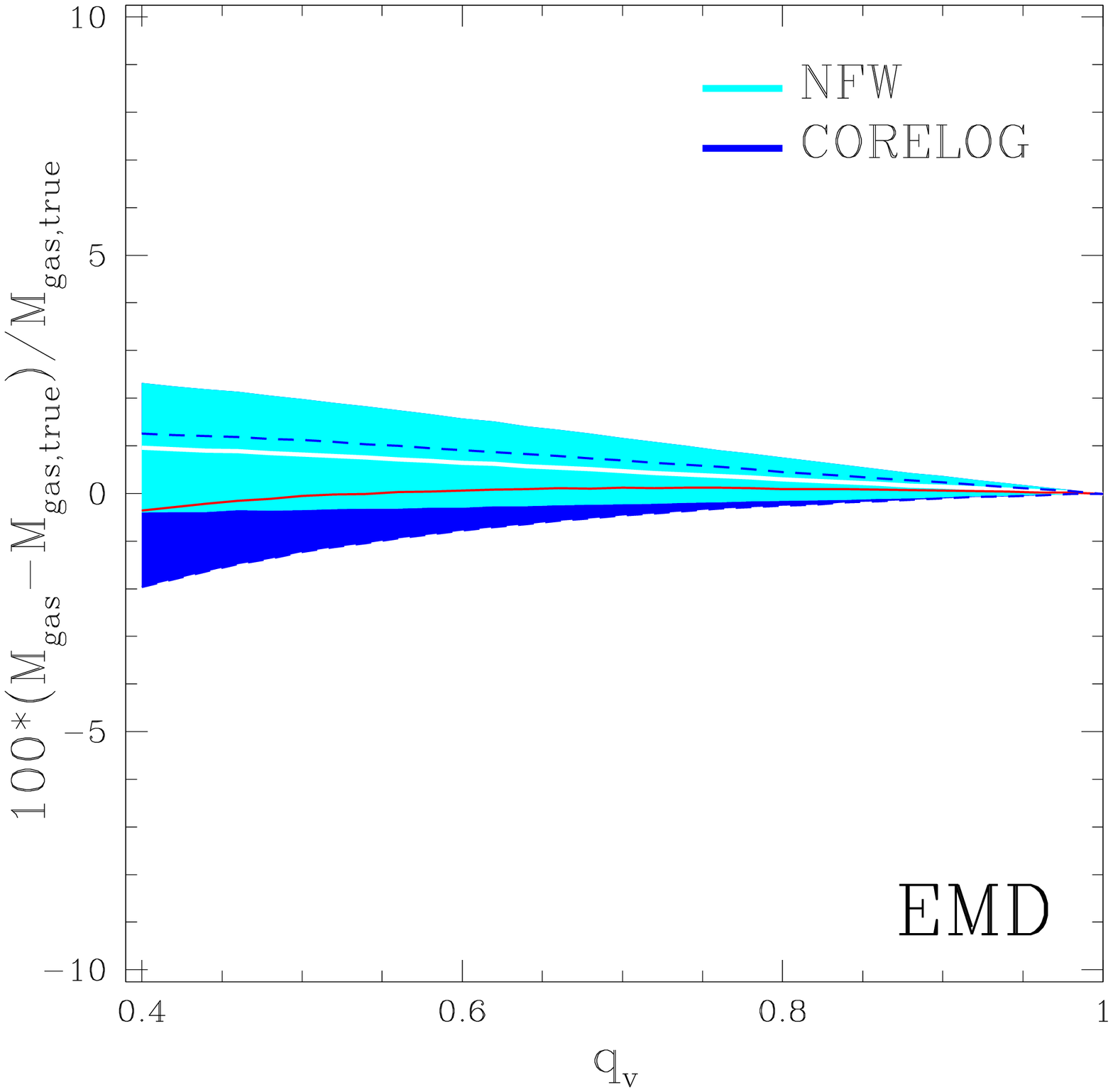}}}
\parbox{0.32\textwidth}{
\centerline{\includegraphics[scale=0.29,angle=0]{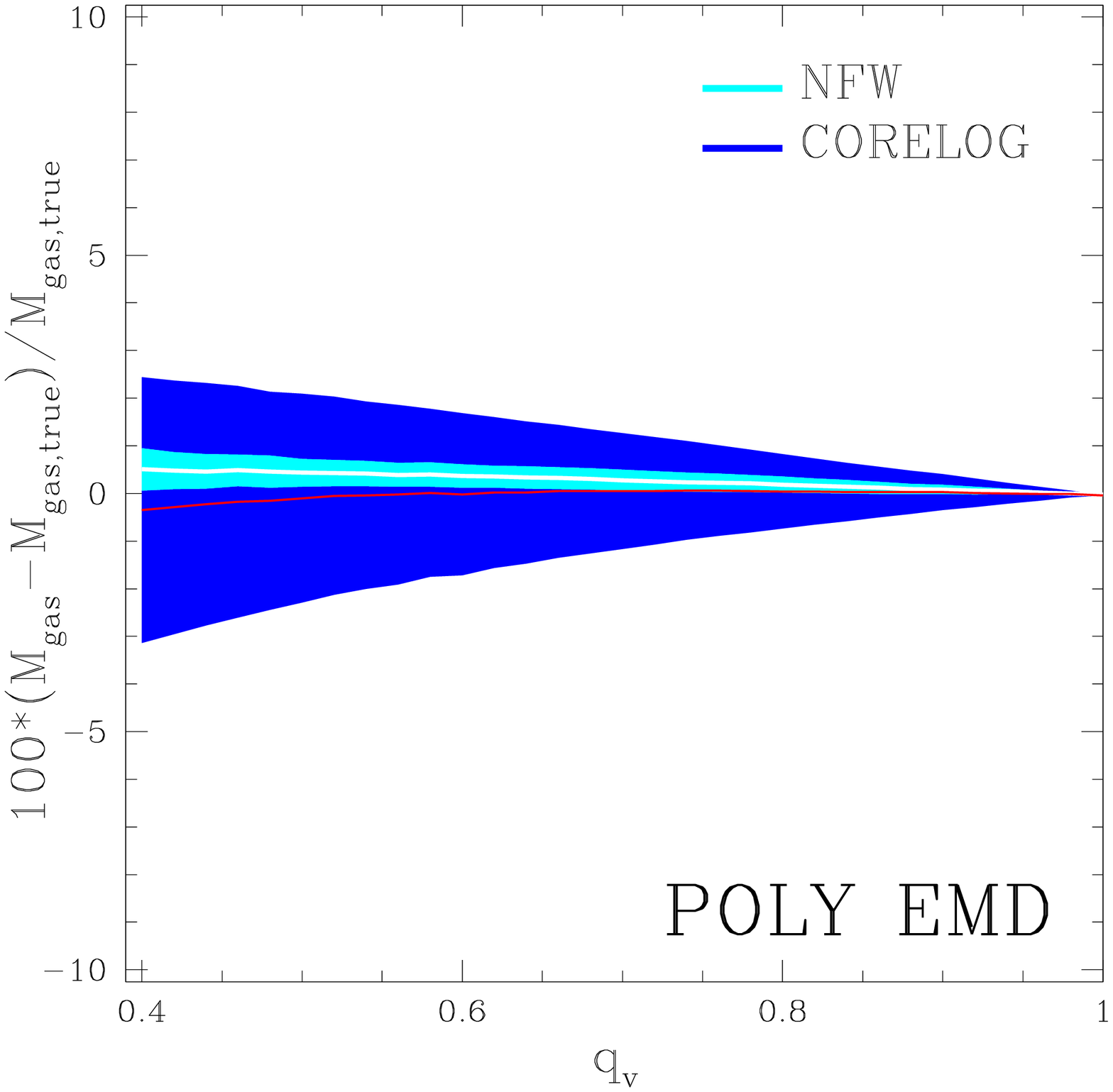}}}
\caption{\label{fig.error.mgas} Same as Figure \ref{fig.error.c} except
now the bias on the gas mass is shown. For clarity, the blue dashed lines
also represent the $1\sigma$ CORELOG region when it is contained
within the $1\sigma$ NFW region, and the average bias for CORELOG is
denoted by the red solid line.}
\end{figure*}

\begin{figure*}
\parbox{0.49\textwidth}{
\centerline{\includegraphics[scale=0.32,angle=0]{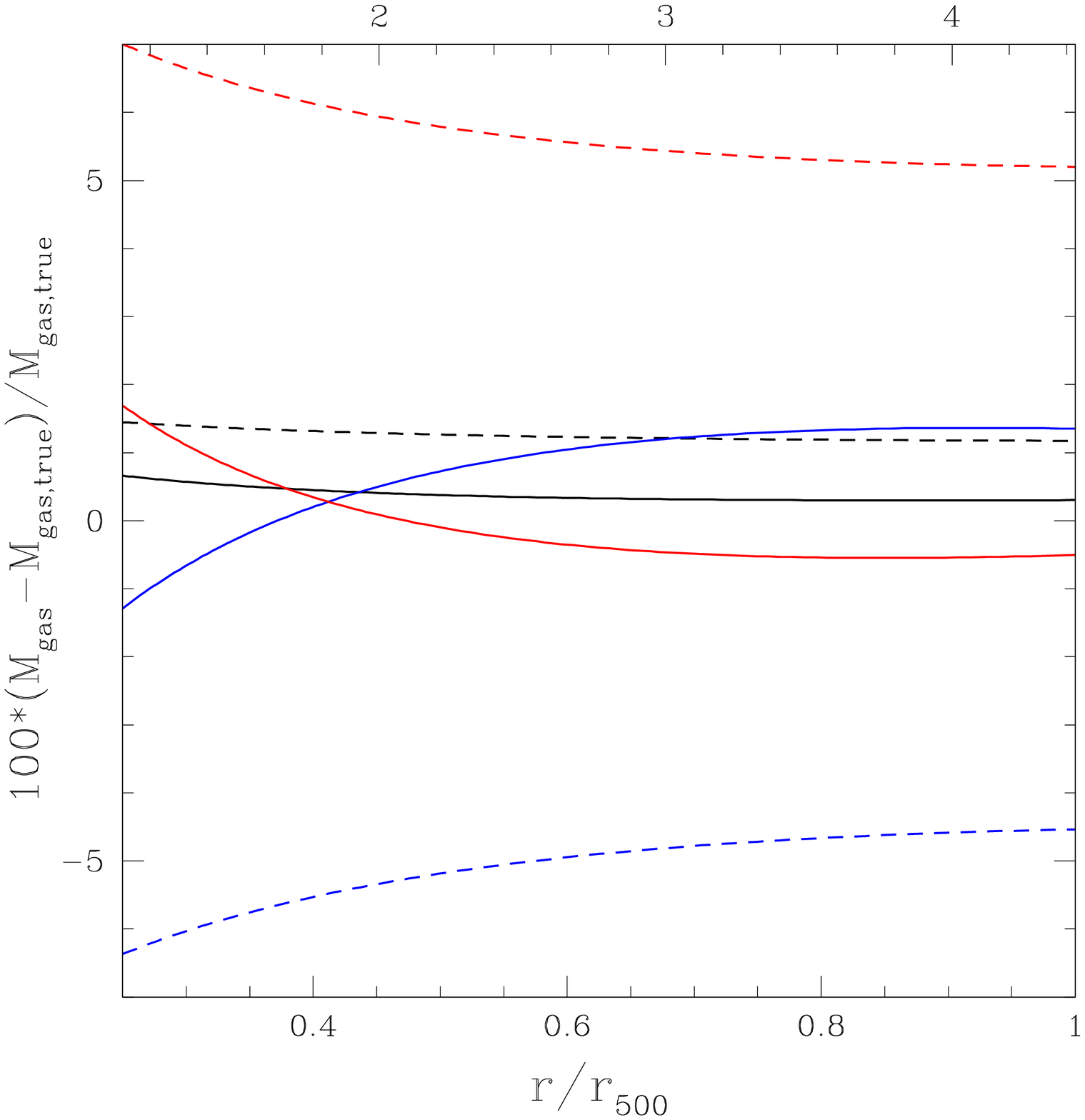}}}
\parbox{0.49\textwidth}{
\centerline{\includegraphics[scale=0.32,angle=0]{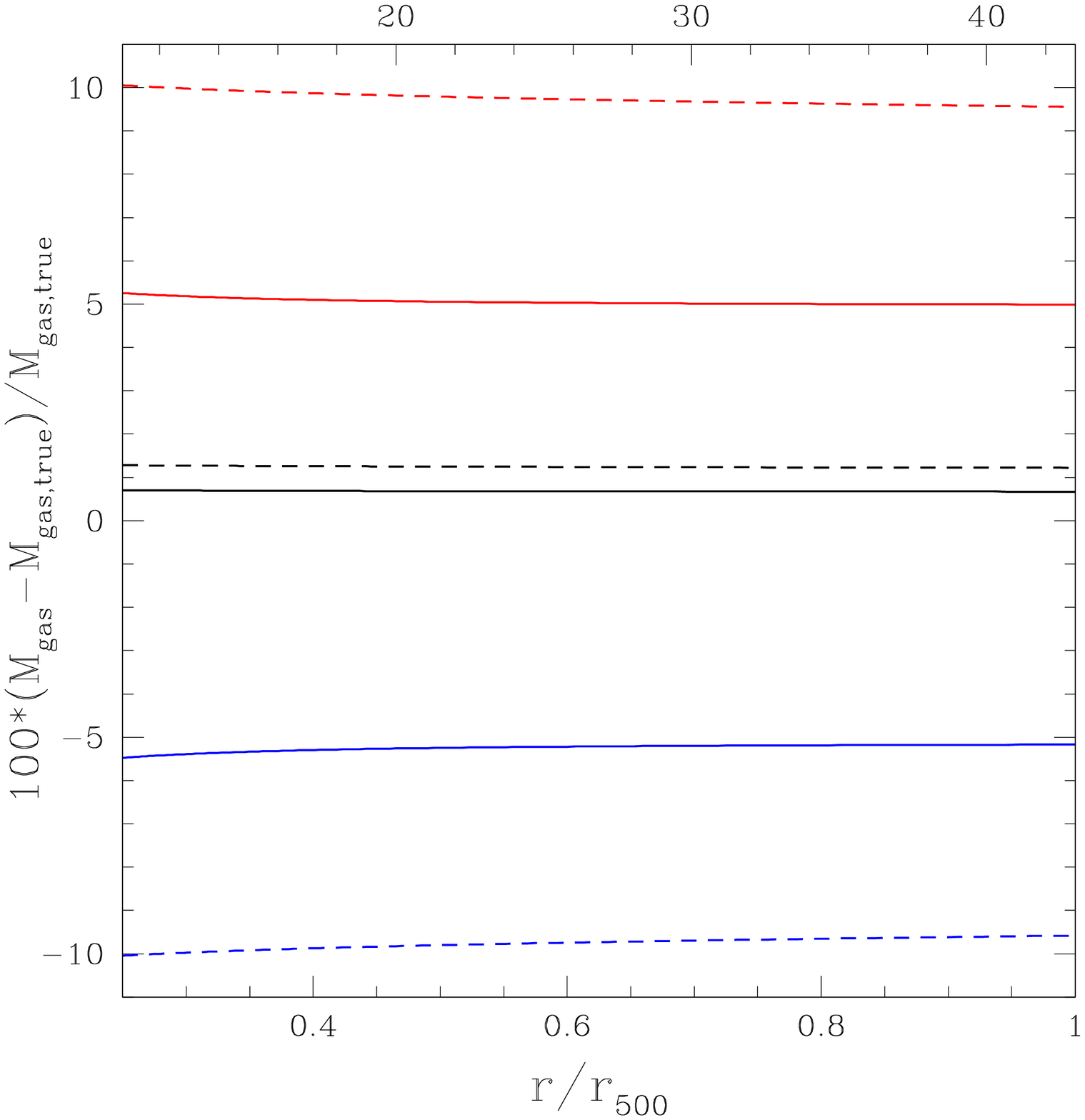}}}
\caption{\label{fig.radpro.mgas} 
Bias of the gas mass plotted as a function of radius between $0.25-1\,
r_{500}$ for the isothermal (solid lines) and polytropic (dashed
lines) NFW-EP ({\sl left panel}) and CORELOG-EP ({\sl
right panel}) models for $q_v=0.7$ in the potential. Results are displayed
for projections down the three principal axes: short (blue), long
(red), and intermediate (black). The top axis gives the radius
expressed in units of the scale radius  for NFW-EP and in units
of the core radius for CORELOG-EP.}
\end{figure*}

\begin{figure*}
\parbox{0.32\textwidth}{
\centerline{\includegraphics[scale=0.29,angle=0]{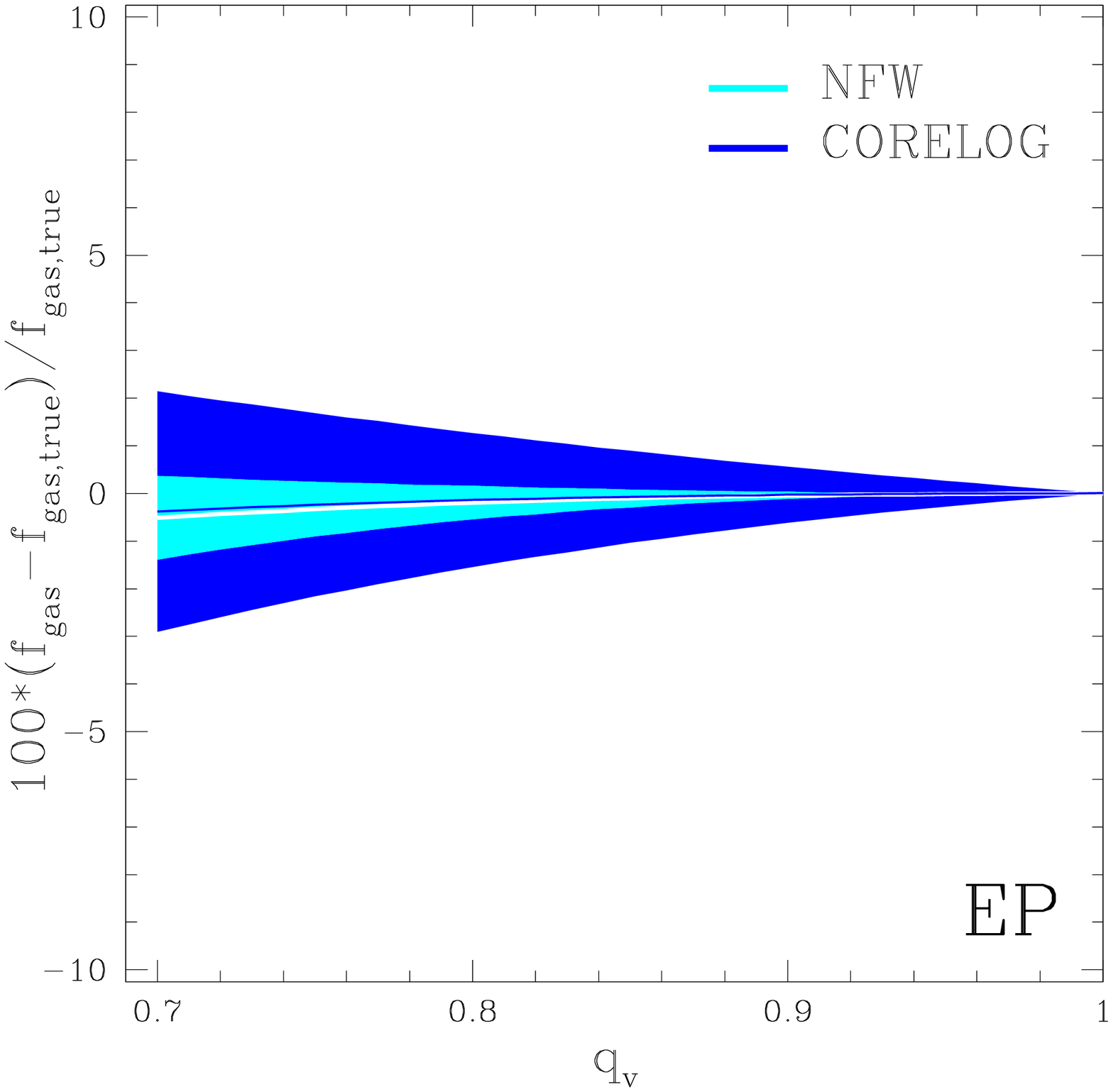}}}
\parbox{0.32\textwidth}{
\centerline{\includegraphics[scale=0.29,angle=0]{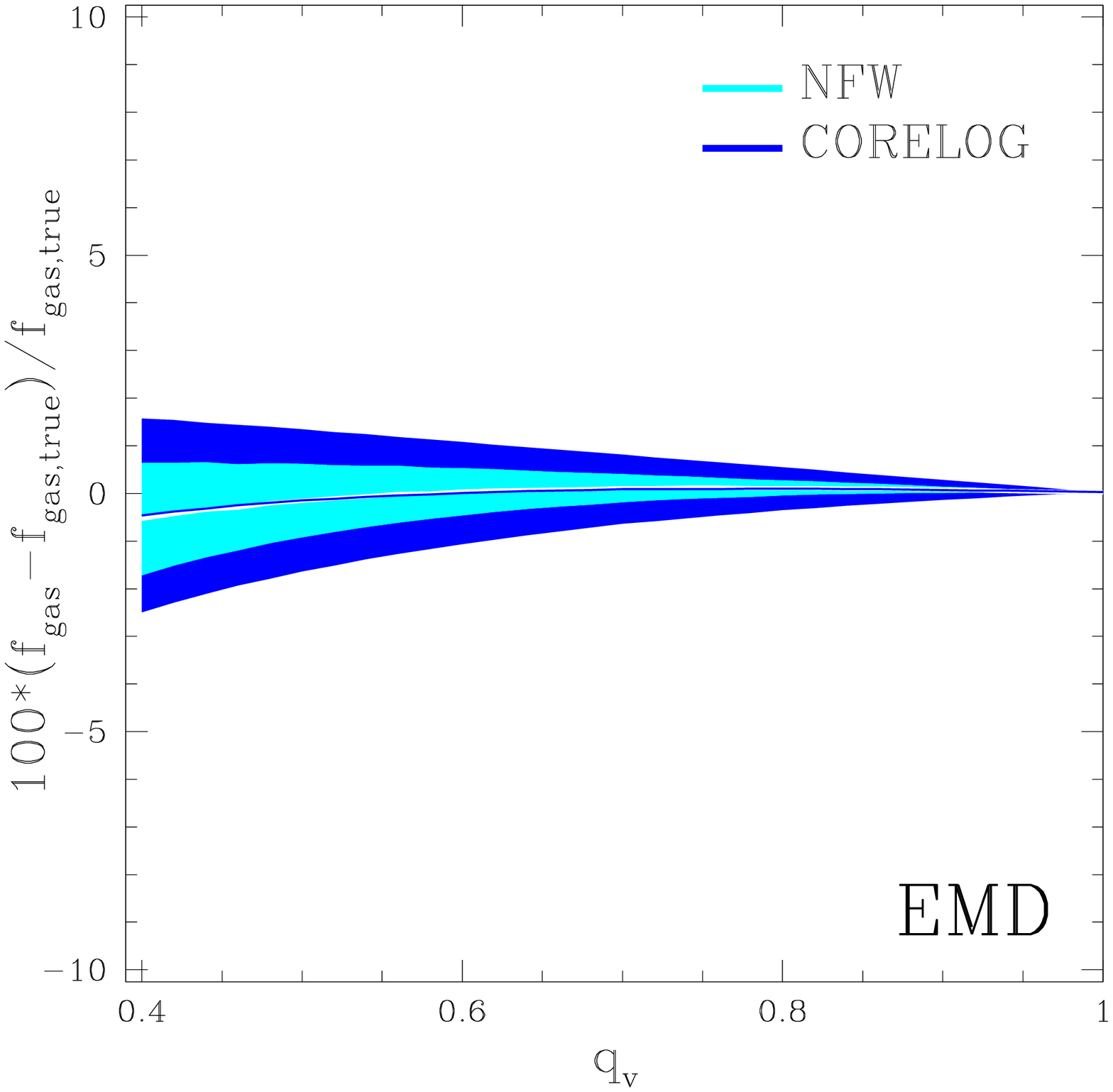}}}
\parbox{0.32\textwidth}{
\centerline{\includegraphics[scale=0.29,angle=0]{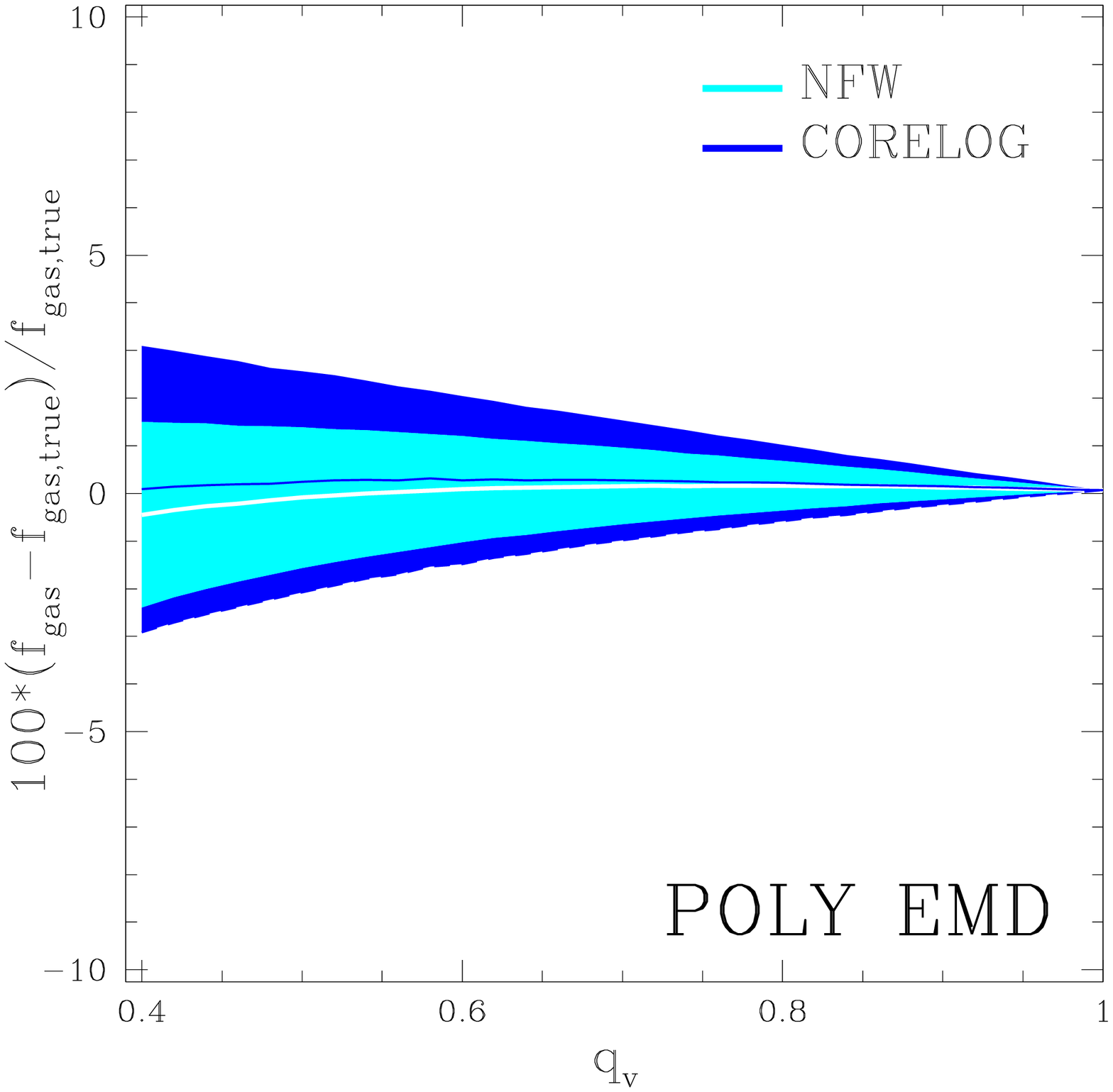}}}
\caption{\label{fig.error.fgas} Same as Figure \ref{fig.error.c} except
now the bias on the gas fraction is shown.}
\end{figure*}

\begin{figure*}
\parbox{0.49\textwidth}{
\centerline{\includegraphics[scale=0.29,angle=0]{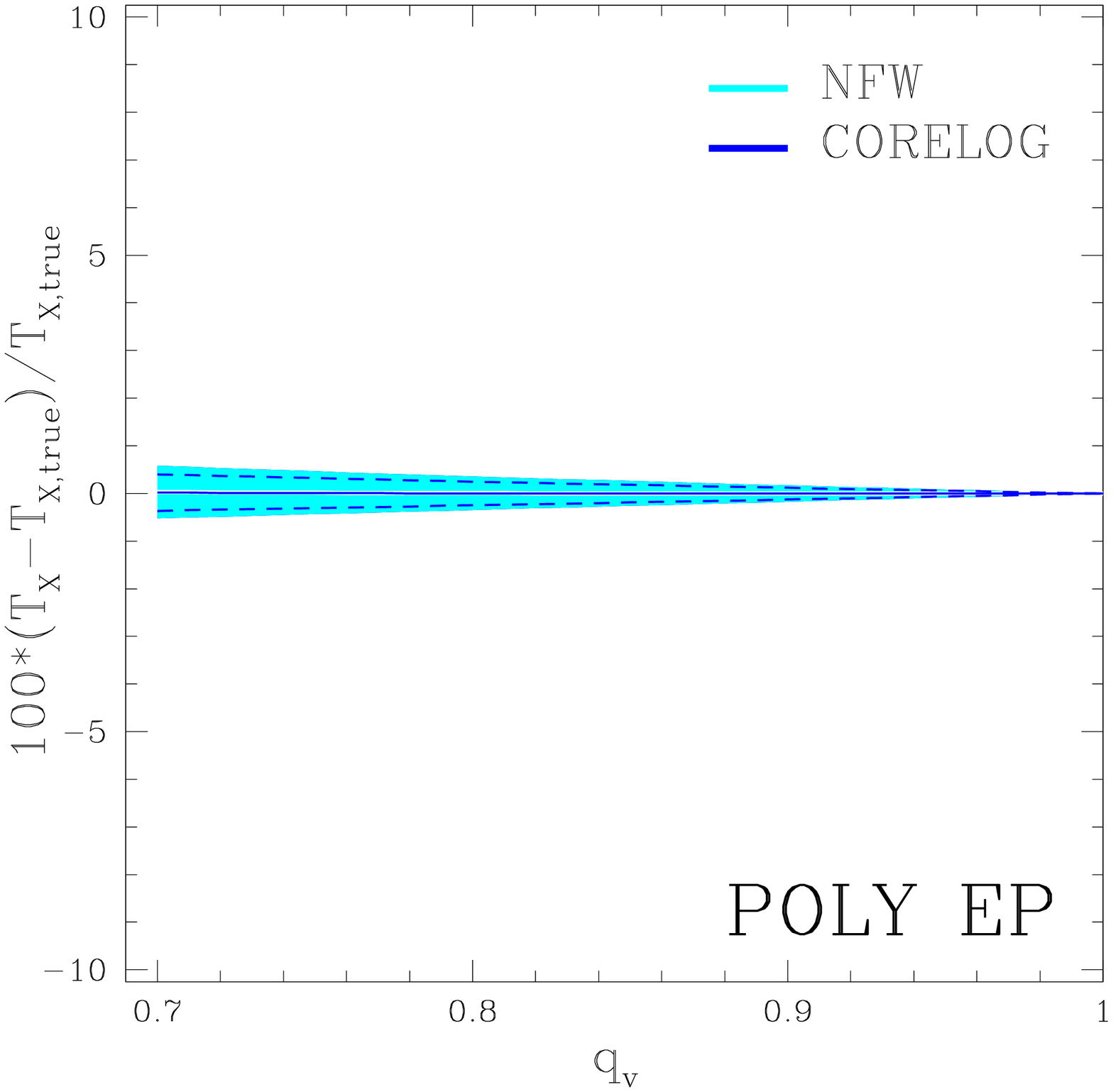}}}
\parbox{0.49\textwidth}{
\centerline{\includegraphics[scale=0.29,angle=0]{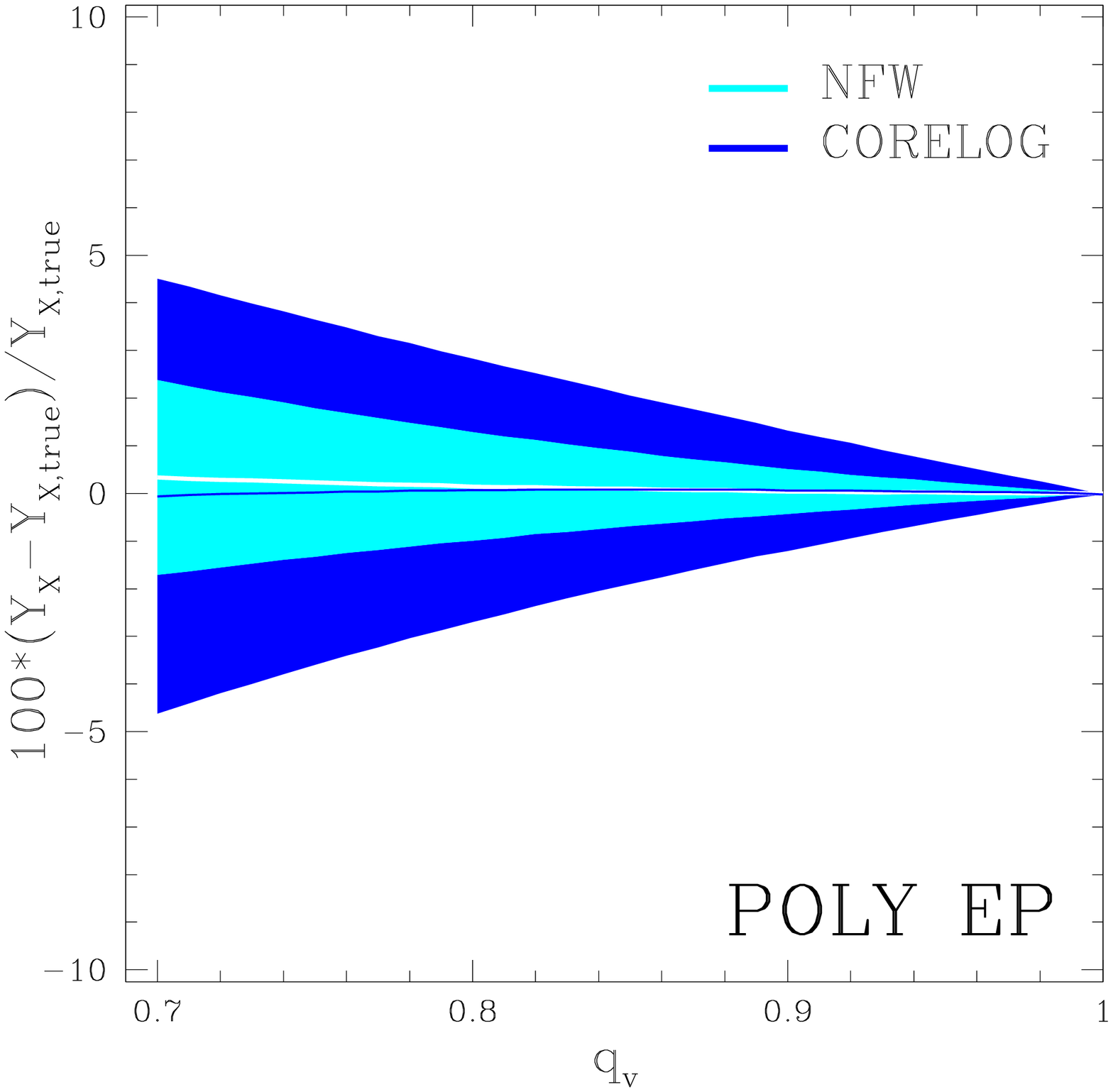}}}

\vskip 0.05cm

\parbox{0.49\textwidth}{
\centerline{\includegraphics[scale=0.29,angle=0]{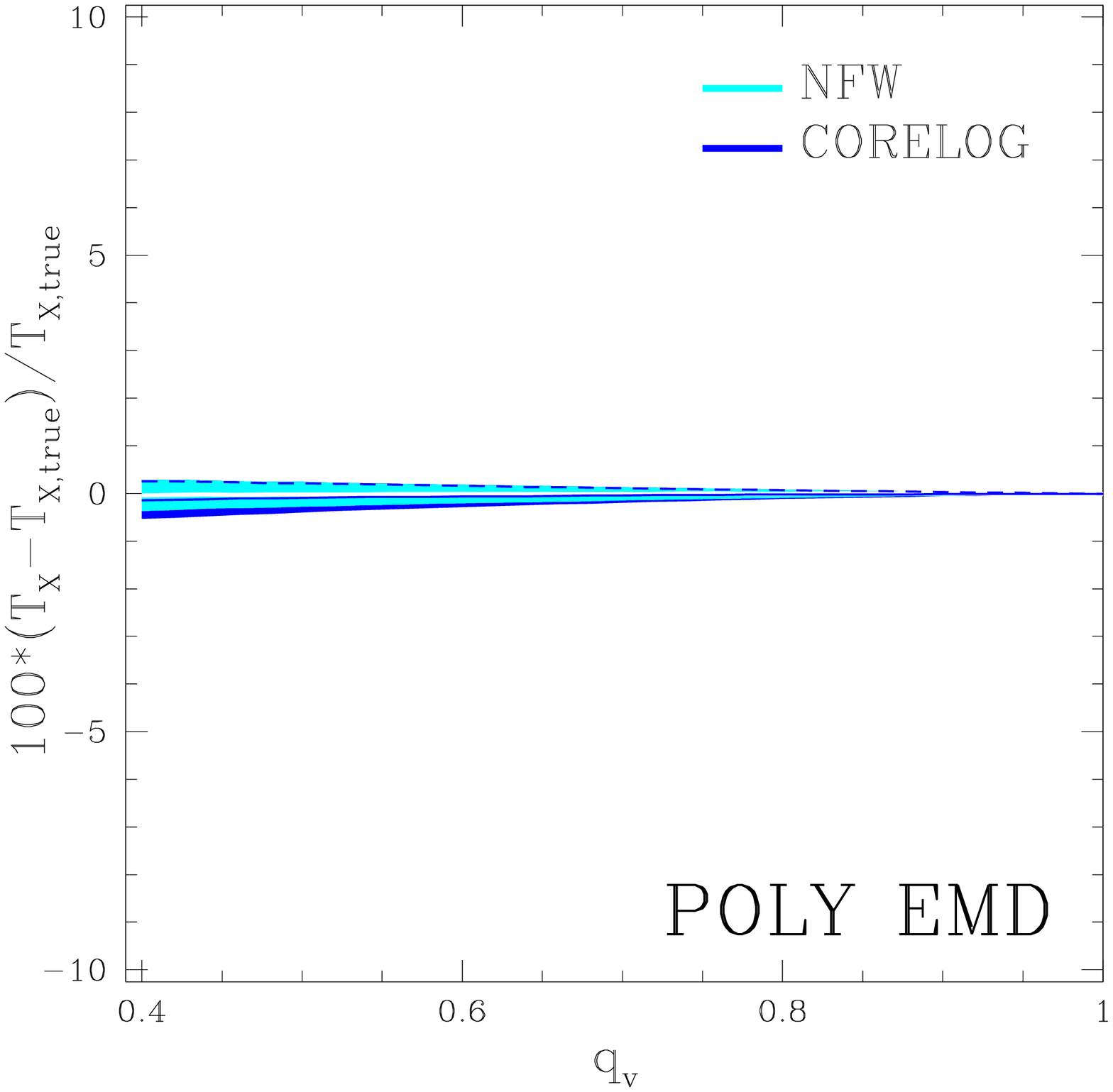}}}
\parbox{0.49\textwidth}{
\centerline{\includegraphics[scale=0.29,angle=0]{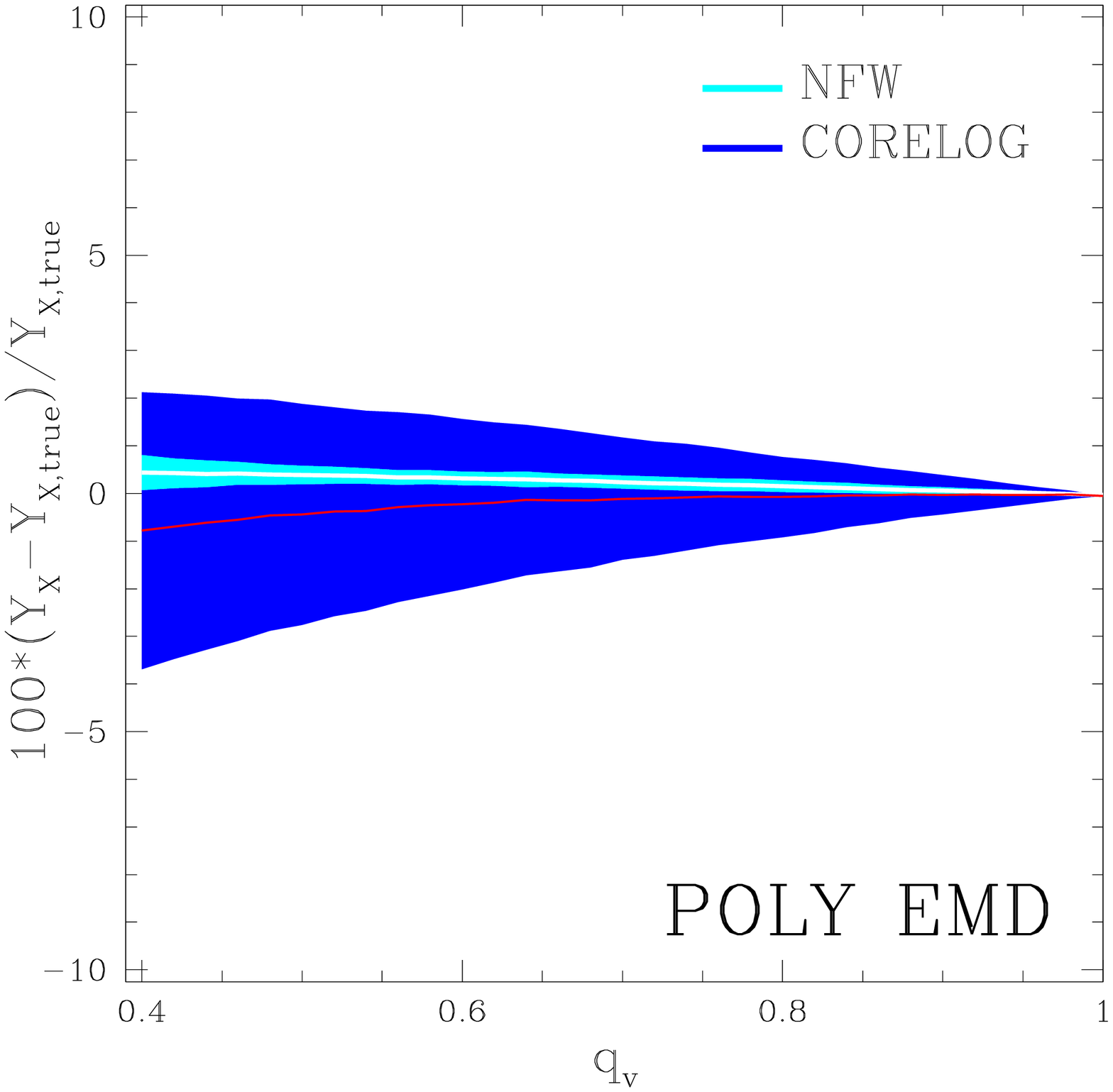}}}
\caption{\label{fig.error.yx}
Similar to Figure \ref{fig.error.c} except now the biases on $T_{\rm
X}$ and $Y_{\rm X}$ are shown for the ({\sl Top Panels}) polytropic EP
models and ({\sl Bottom Panels}) polytropic EMD models. Results for
$Y_{\rm X}$ for isothermal models are identical to those of the gas
mass (Figure \ref{fig.error.mgas}). For clarity, the blue dashed lines
also represent the $1\sigma$ CORELOG region when it is contained
within the $1\sigma$ NFW region, and the average bias for CORELOG-EMD is
denoted by the red solid line.}
\end{figure*}

\begin{figure*}
\parbox{0.32\textwidth}{
\centerline{\includegraphics[scale=0.29,angle=0]{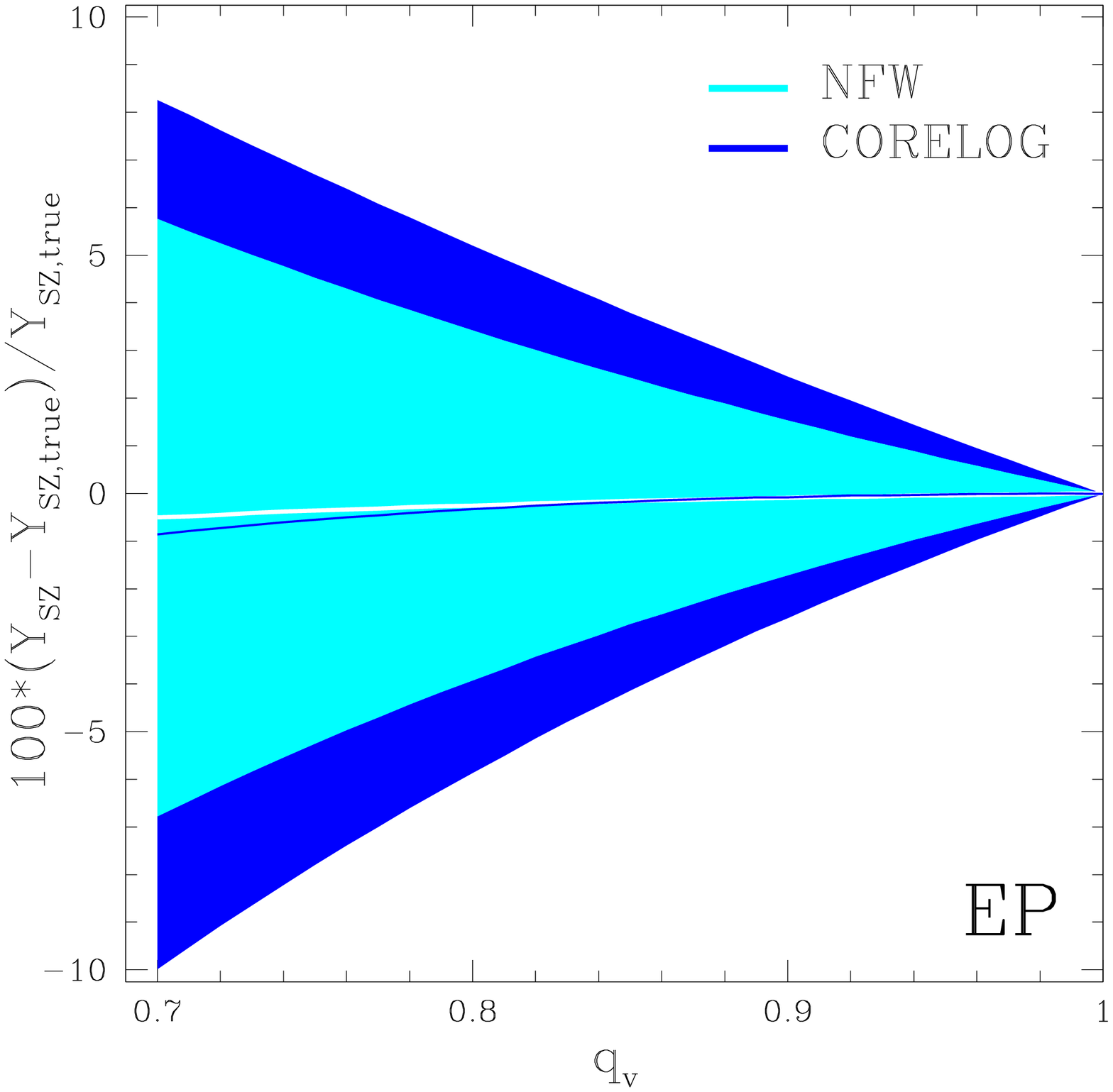}}}
\parbox{0.32\textwidth}{
\centerline{\includegraphics[scale=0.29,angle=0]{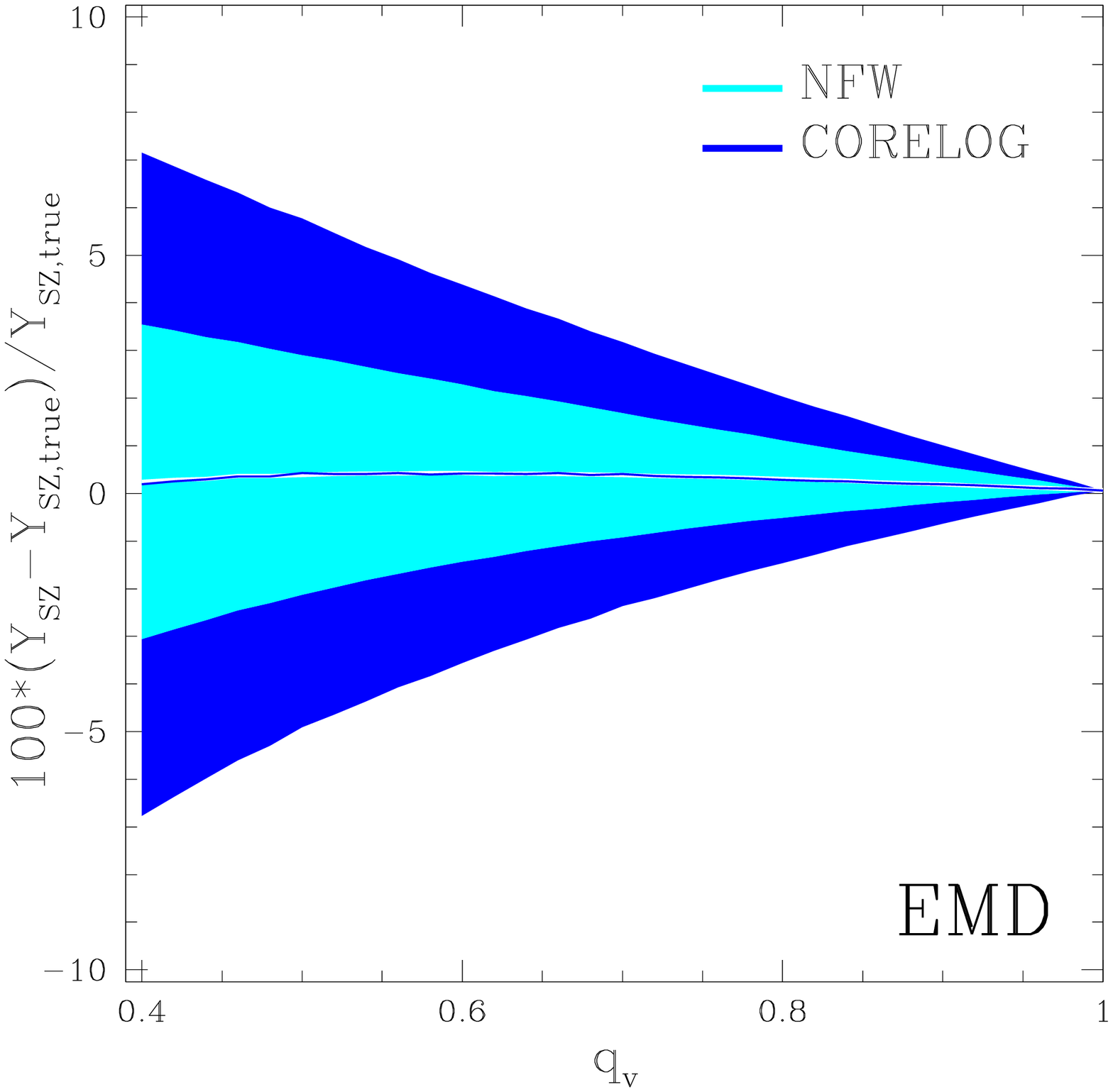}}}
\parbox{0.32\textwidth}{
\centerline{\includegraphics[scale=0.29,angle=0]{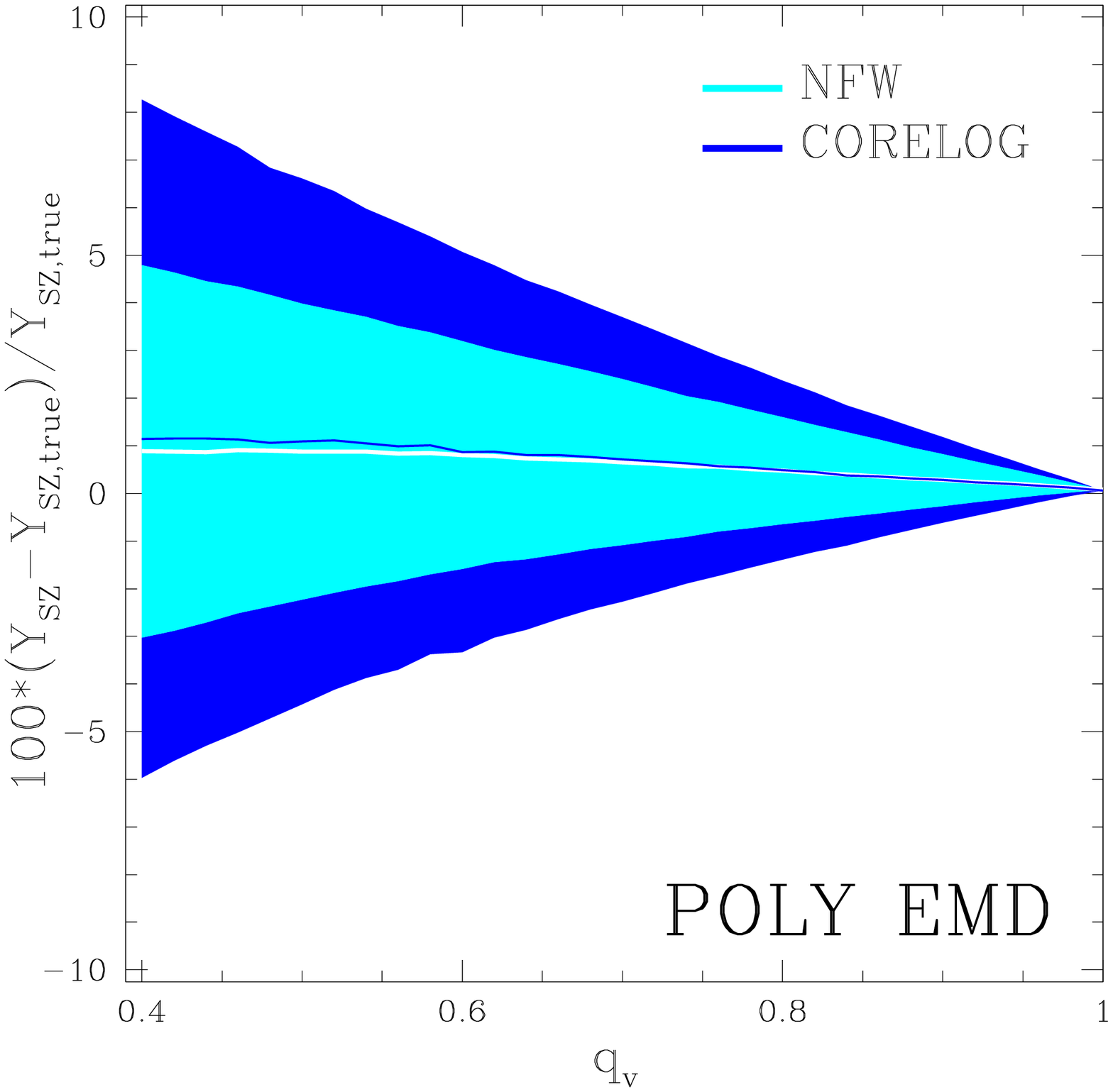}}}
\caption{\label{fig.error.ysz} Same as Figure \ref{fig.error.c} except
now the bias on the $Y_{\rm SZ}$ is shown.}
\end{figure*}

\begin{figure*}
\parbox{0.32\textwidth}{
\centerline{\includegraphics[scale=0.29,angle=0]{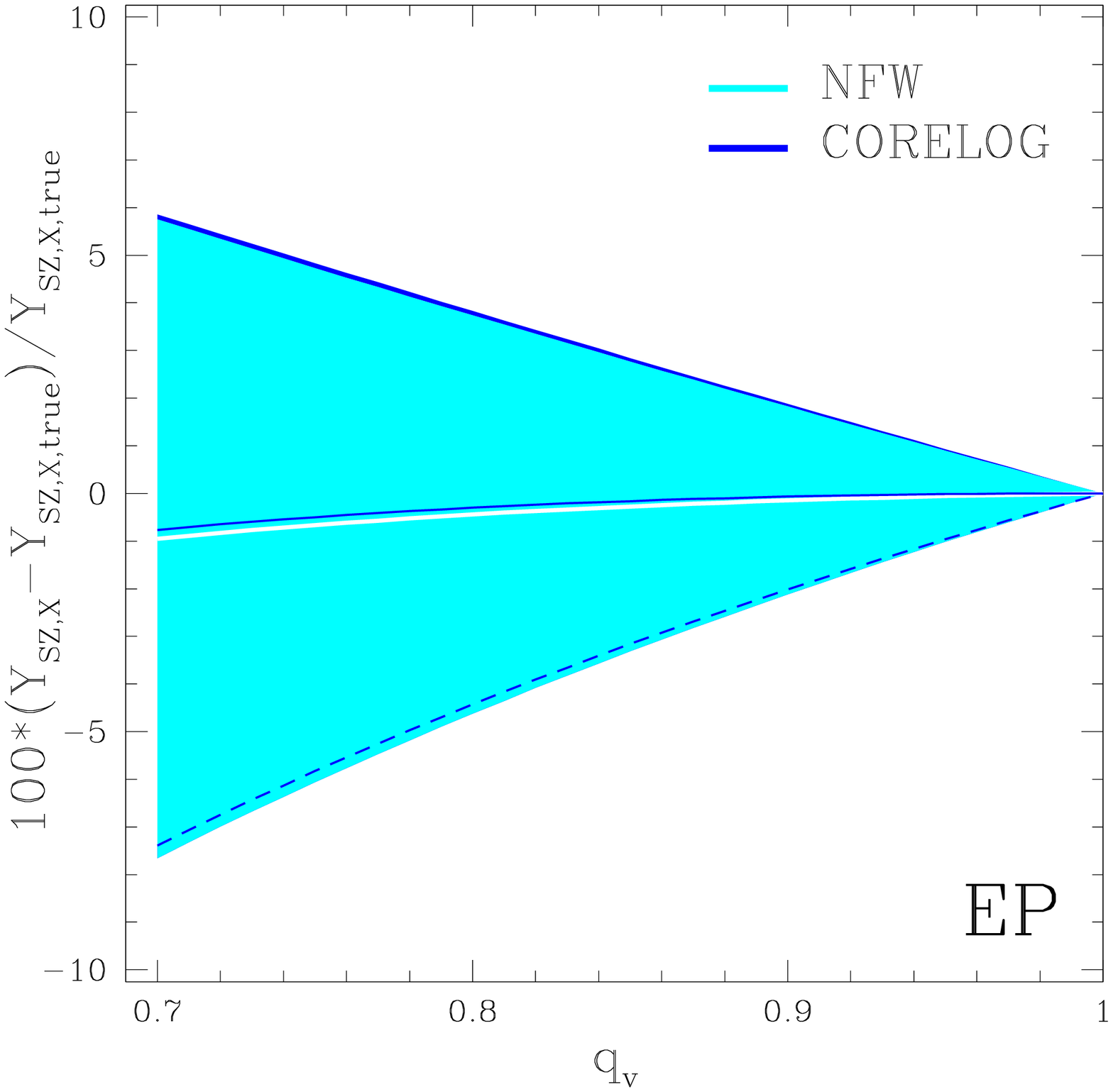}}}
\parbox{0.32\textwidth}{
\centerline{\includegraphics[scale=0.29,angle=0]{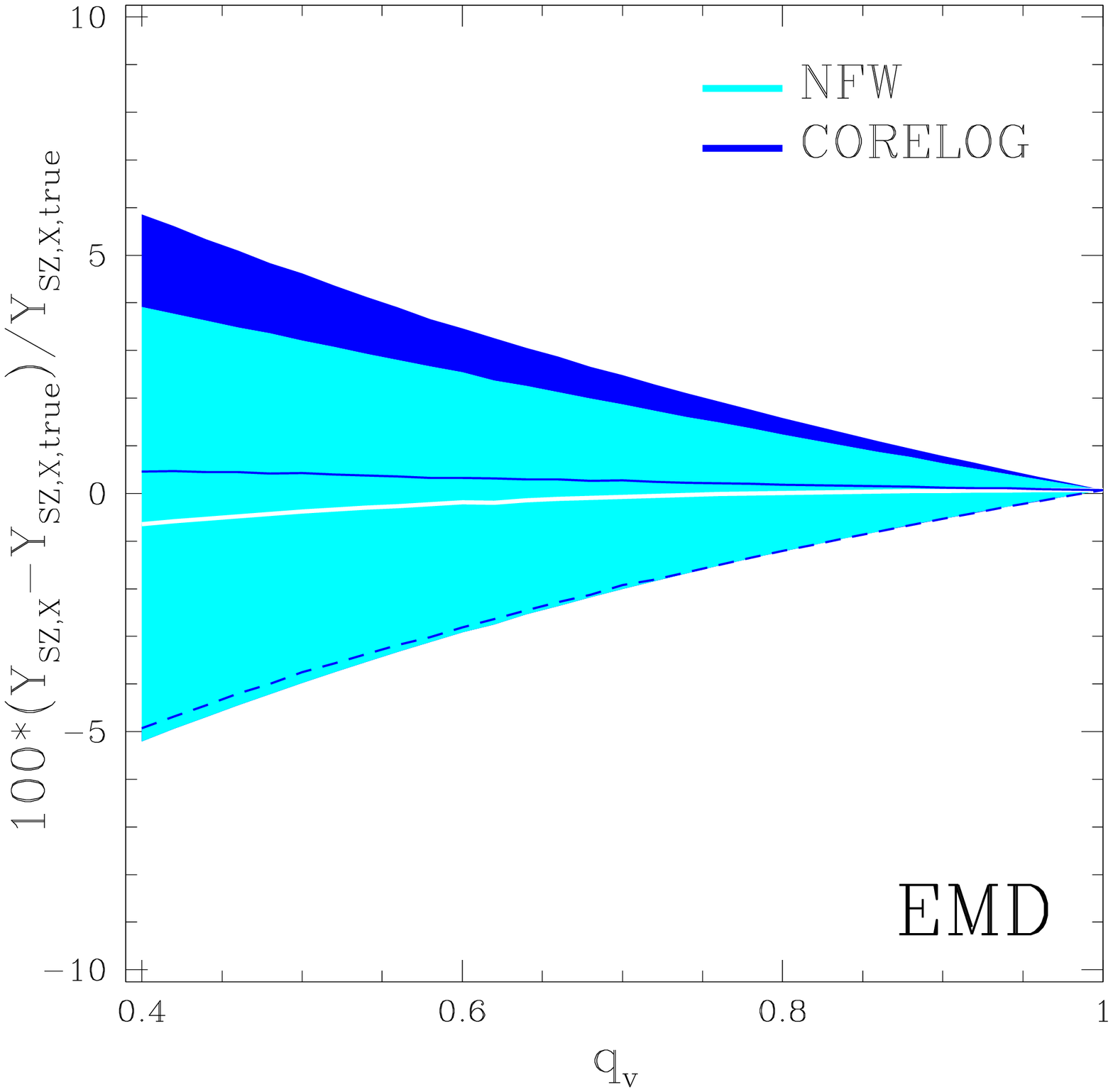}}}
\parbox{0.32\textwidth}{
\centerline{\includegraphics[scale=0.29,angle=0]{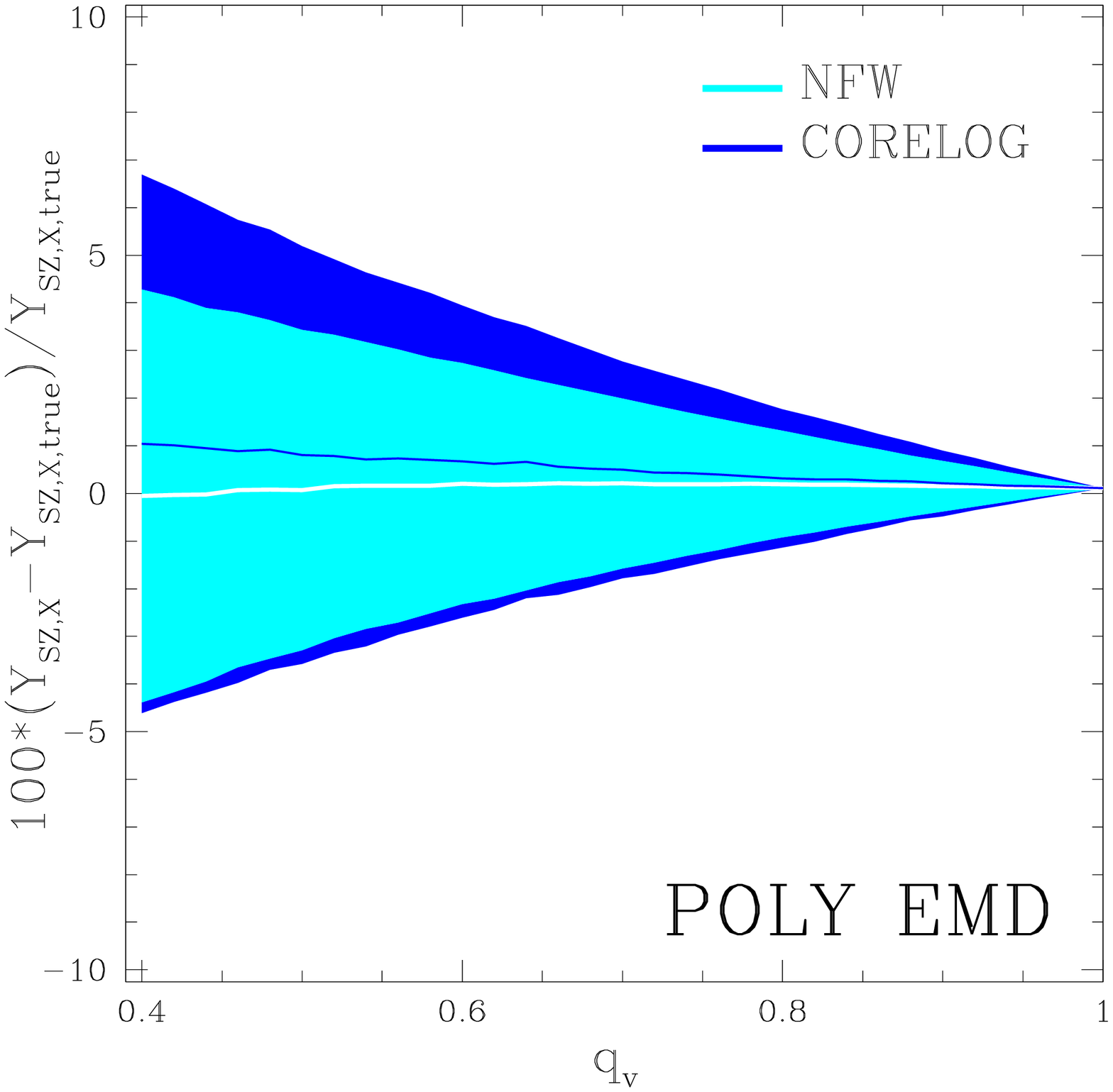}}}
\caption{\label{fig.error.yszx} Same as Figure \ref{fig.error.c} except
now the bias on the $Y_{\rm SZ,X}$ is shown.}
\end{figure*}

\begin{figure*}
\parbox{0.32\textwidth}{
\centerline{\includegraphics[scale=0.29,angle=0]{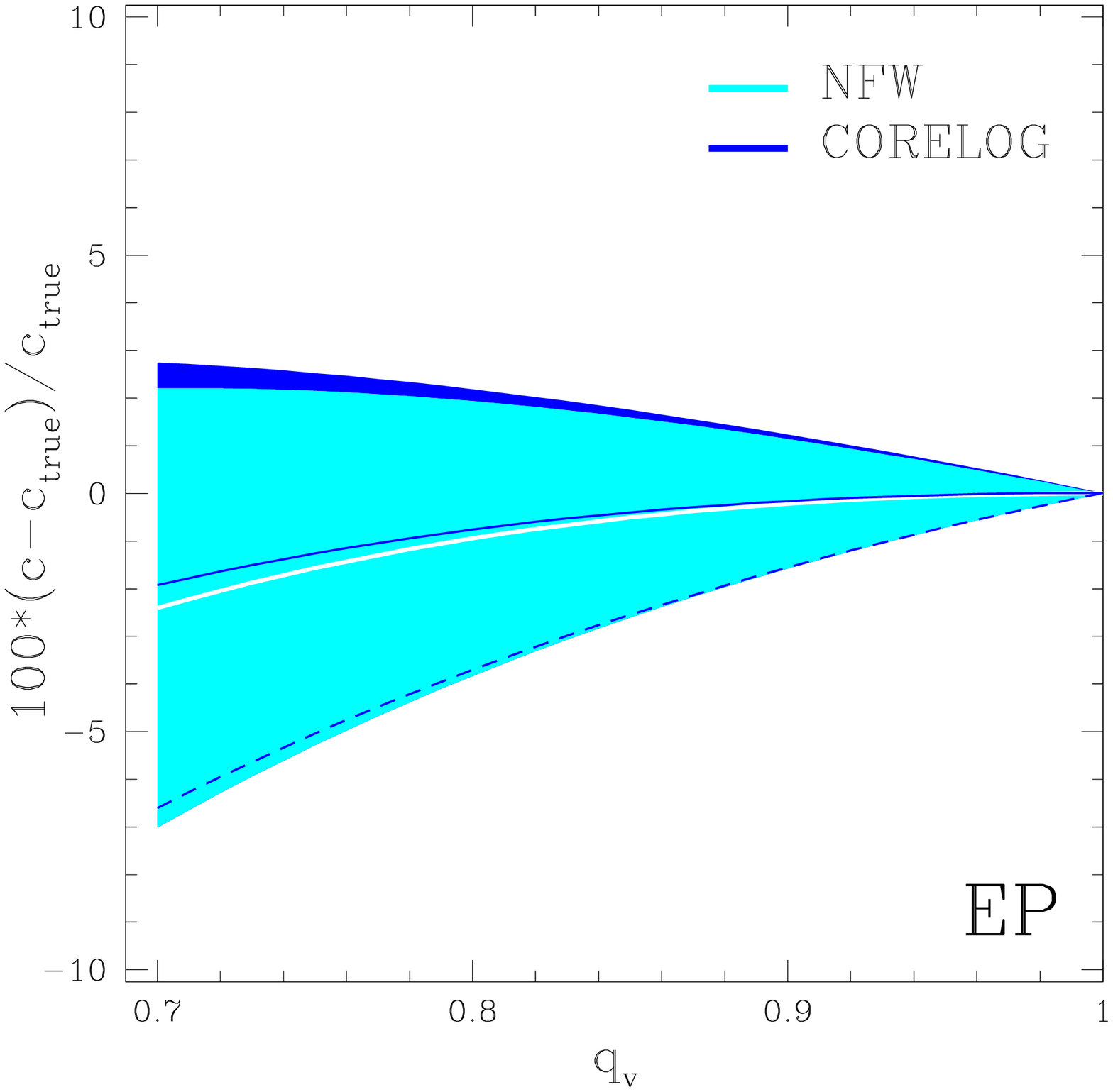}}}
\parbox{0.32\textwidth}{
\centerline{\includegraphics[scale=0.29,angle=0]{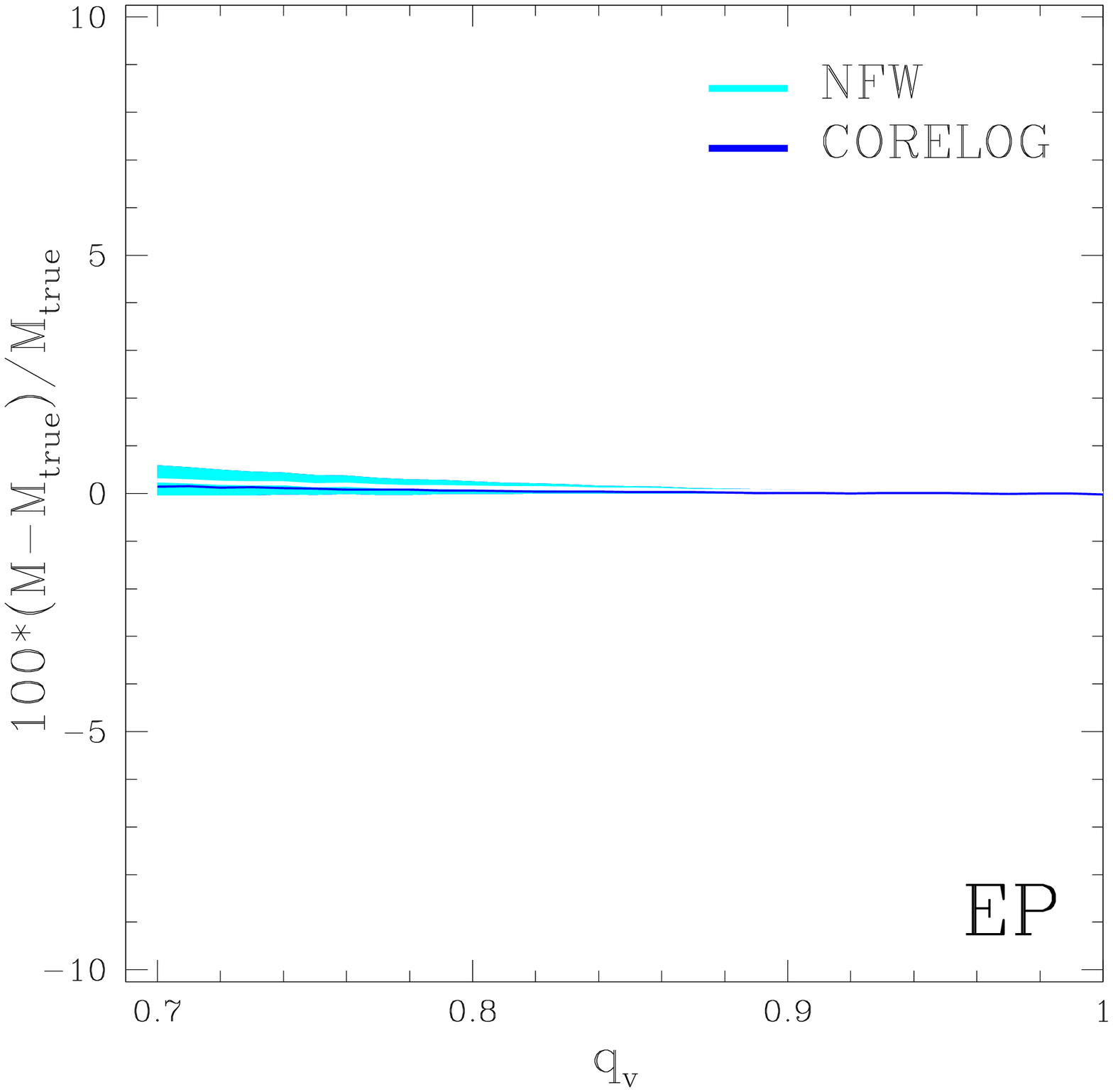}}}
\parbox{0.32\textwidth}{
\centerline{\includegraphics[scale=0.29,angle=0]{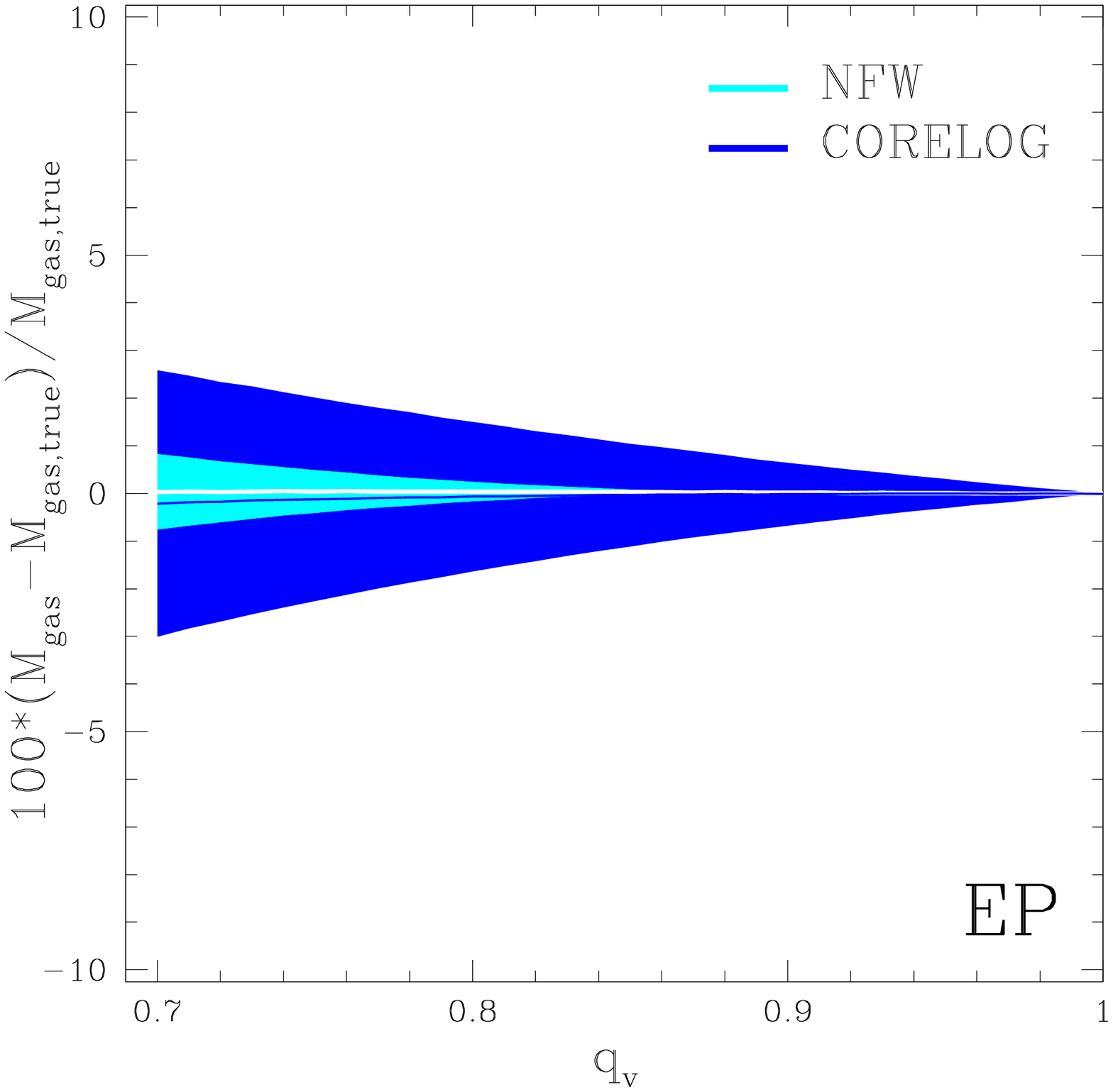}}}

\vskip 0.25cm

\parbox{0.32\textwidth}{
\centerline{\includegraphics[scale=0.29,angle=0]{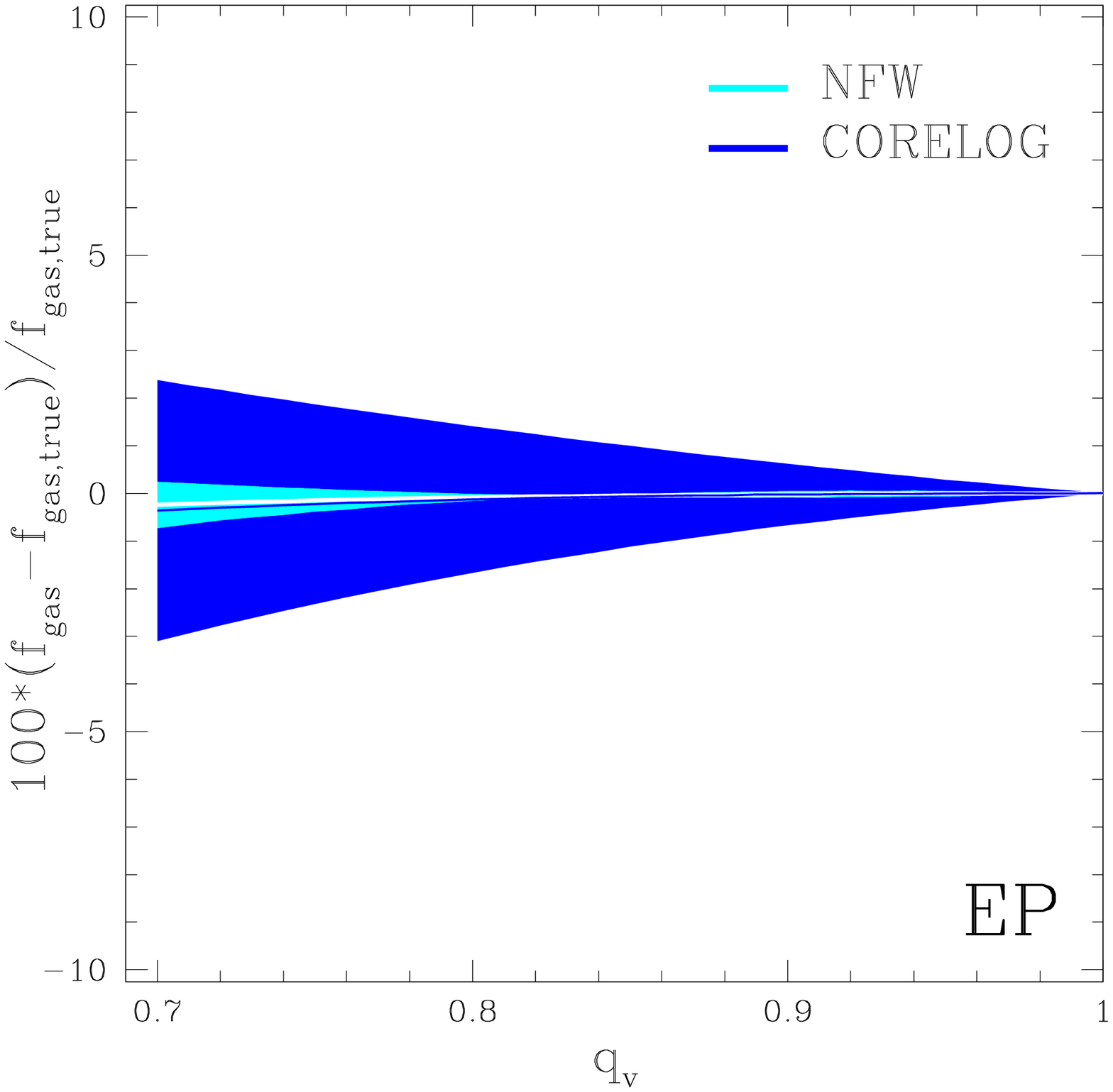}}}
\parbox{0.32\textwidth}{
\centerline{\includegraphics[scale=0.29,angle=0]{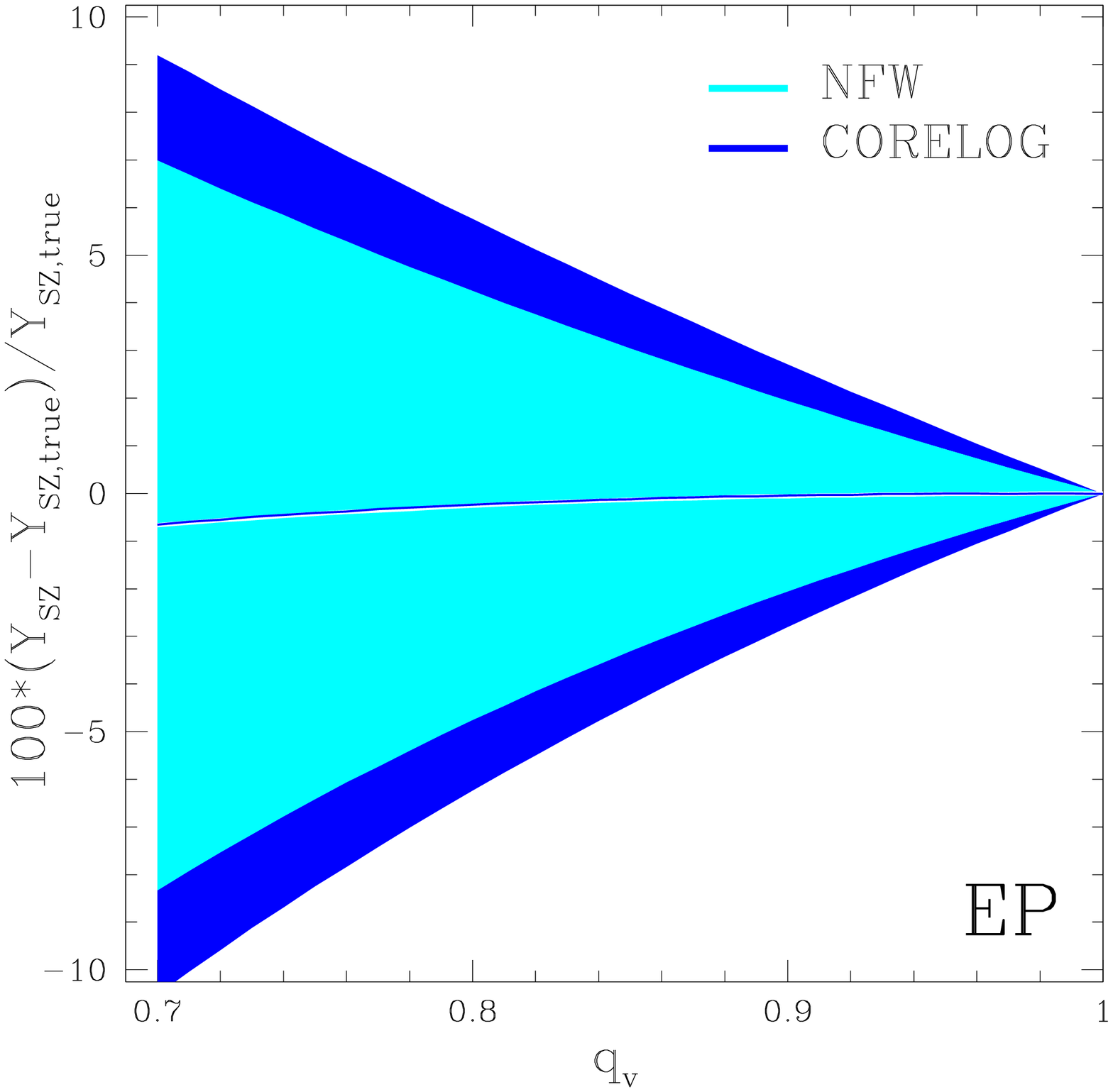}}}
\parbox{0.32\textwidth}{
\centerline{\includegraphics[scale=0.29,angle=0]{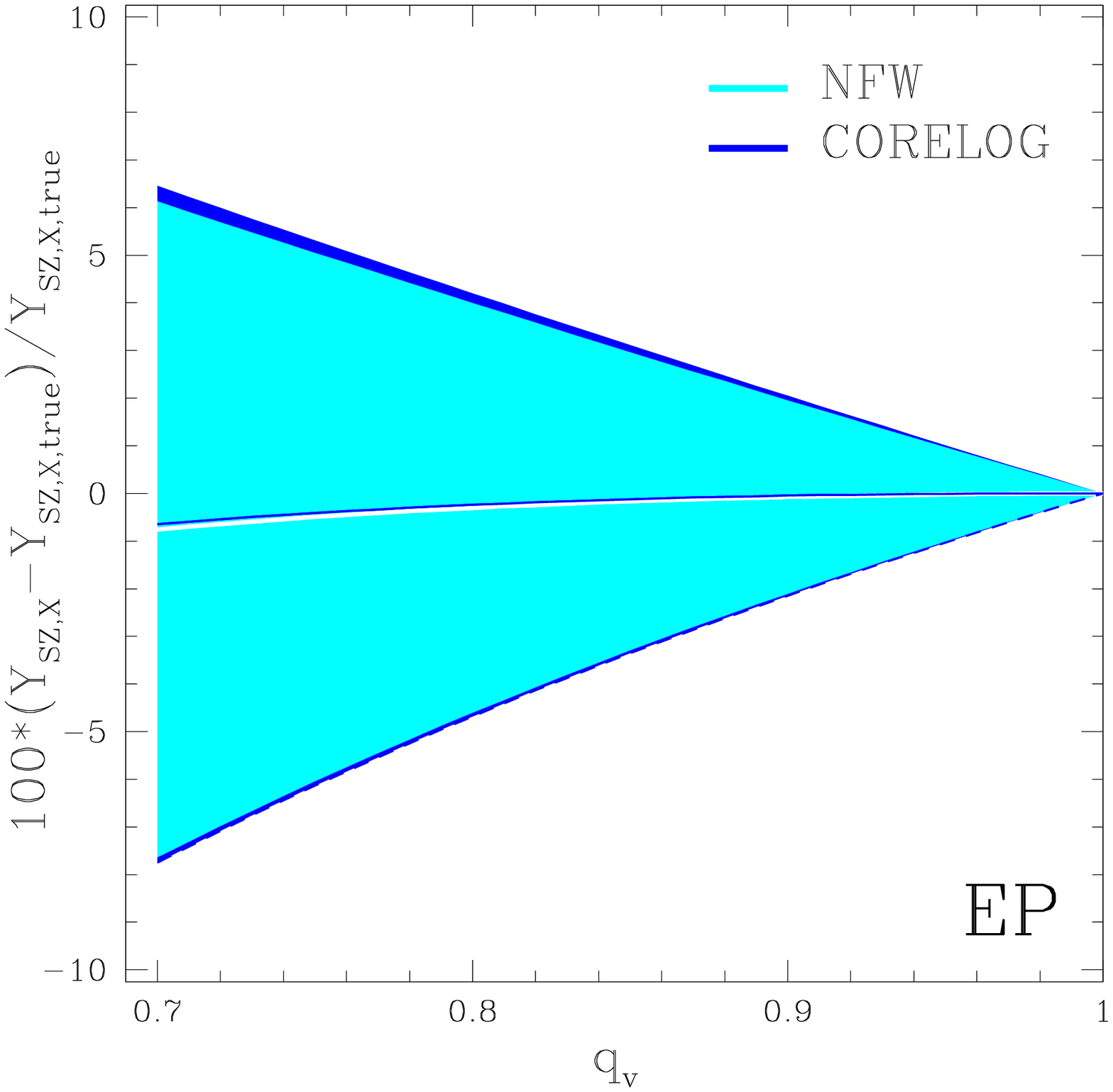}}}
\caption{\label{fig.error.2500}
Results computed within $r_{2500}$ for isothermal EP models.}
\end{figure*}

\begin{figure*}
\parbox{0.32\textwidth}{
\centerline{\includegraphics[scale=0.29,angle=0]{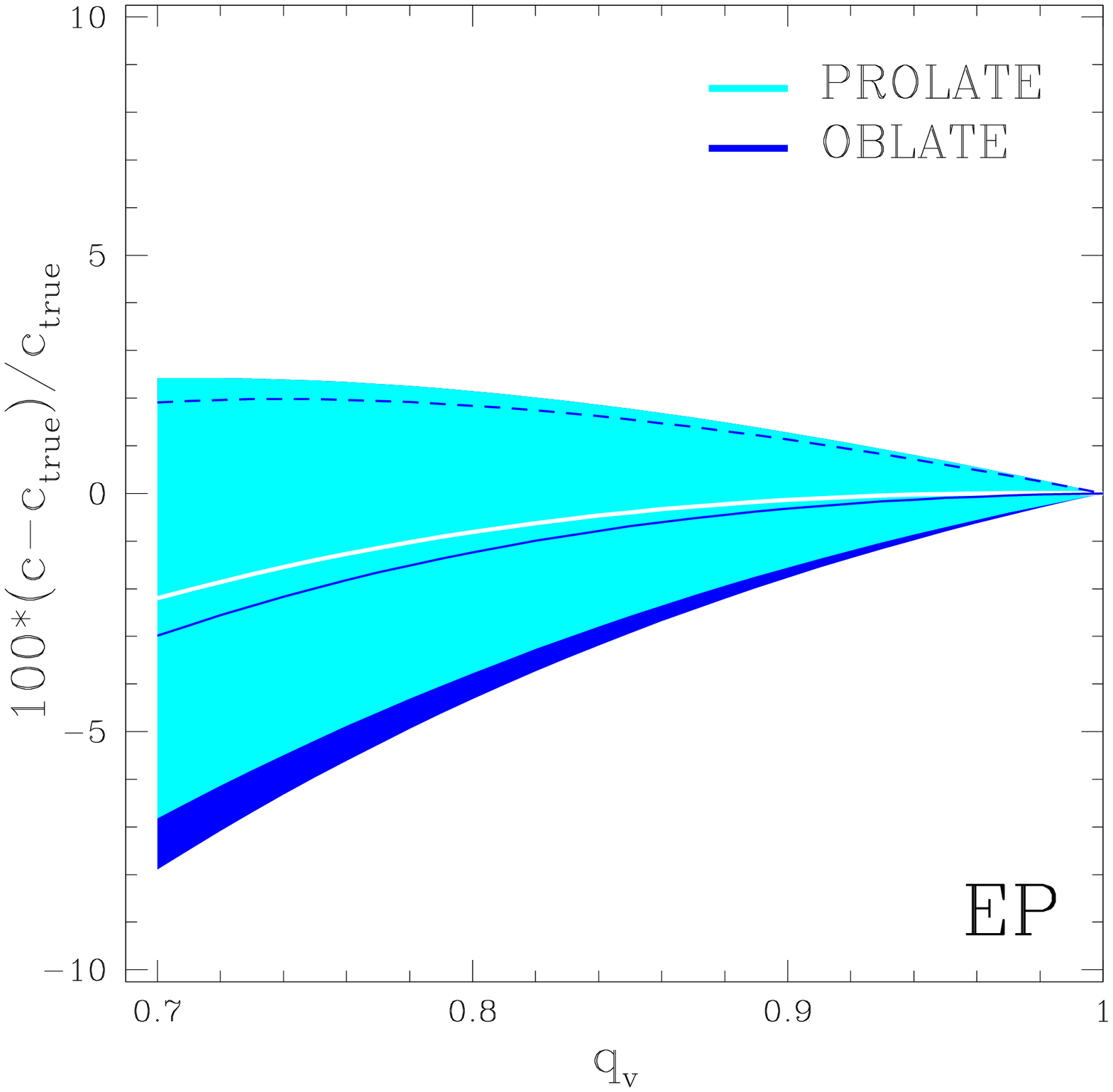}}}
\parbox{0.32\textwidth}{
\centerline{\includegraphics[scale=0.29,angle=0]{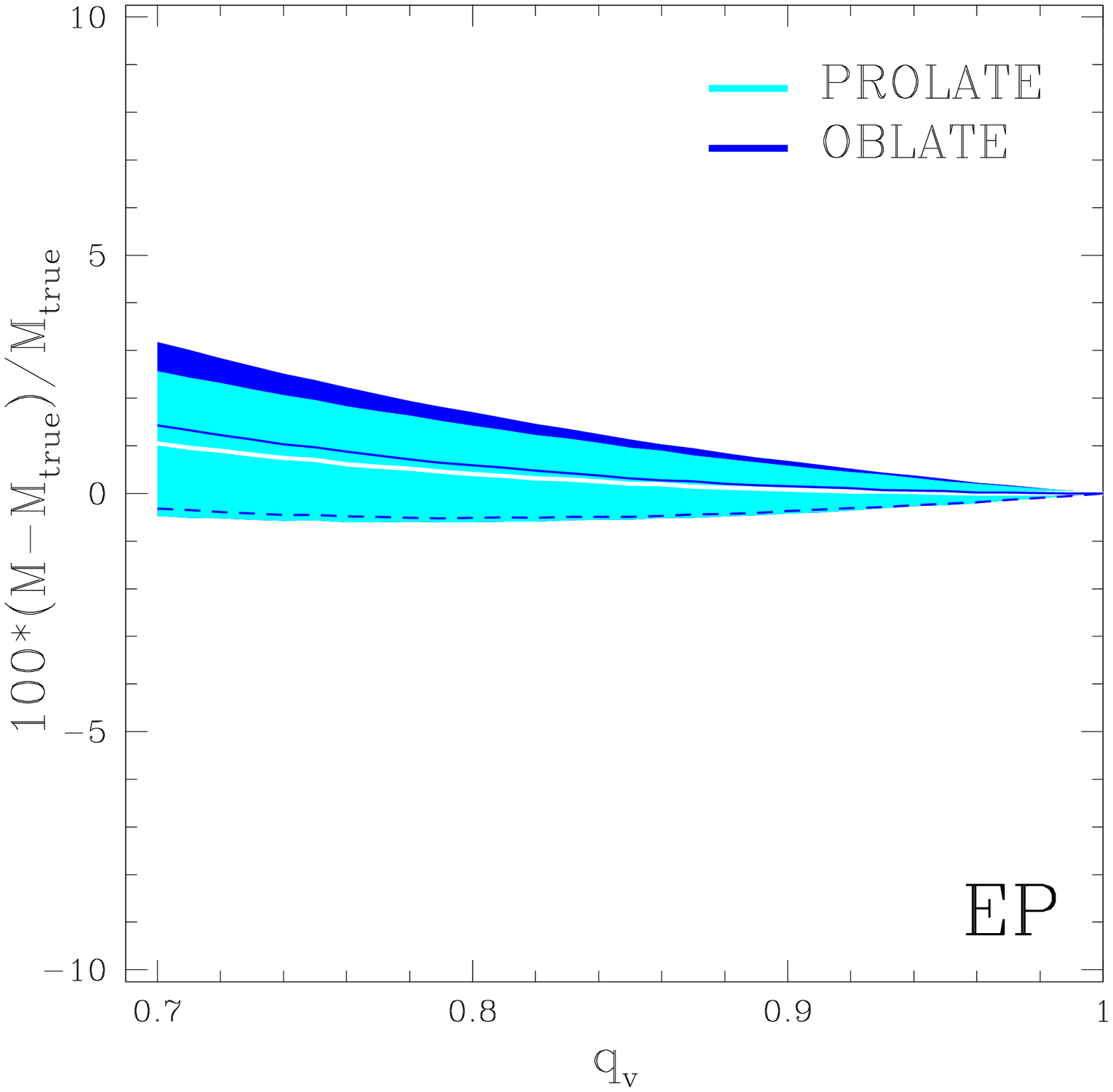}}}
\parbox{0.32\textwidth}{
\centerline{\includegraphics[scale=0.29,angle=0]{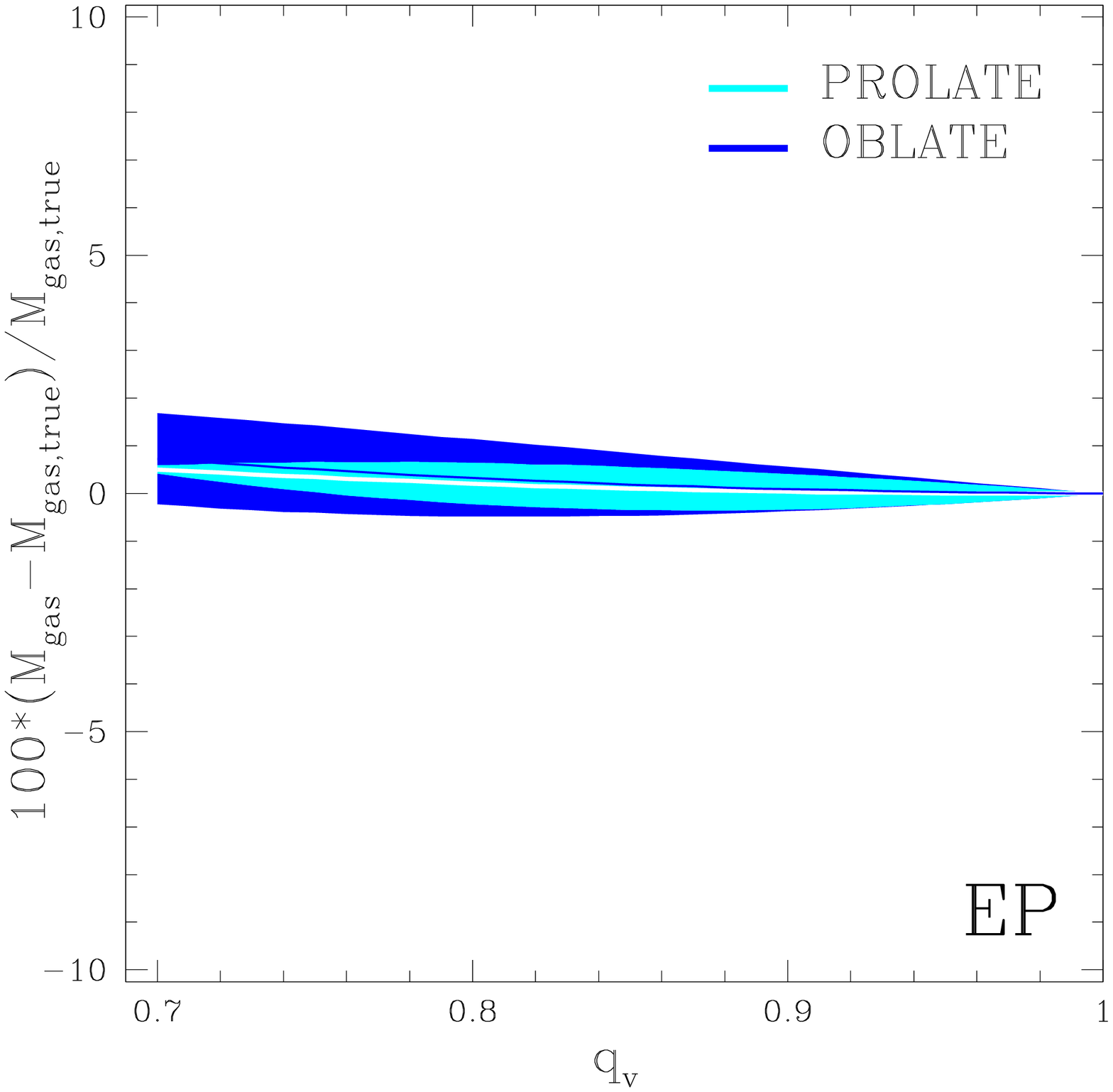}}}

\vskip 0.25cm

\parbox{0.32\textwidth}{
\centerline{\includegraphics[scale=0.29,angle=0]{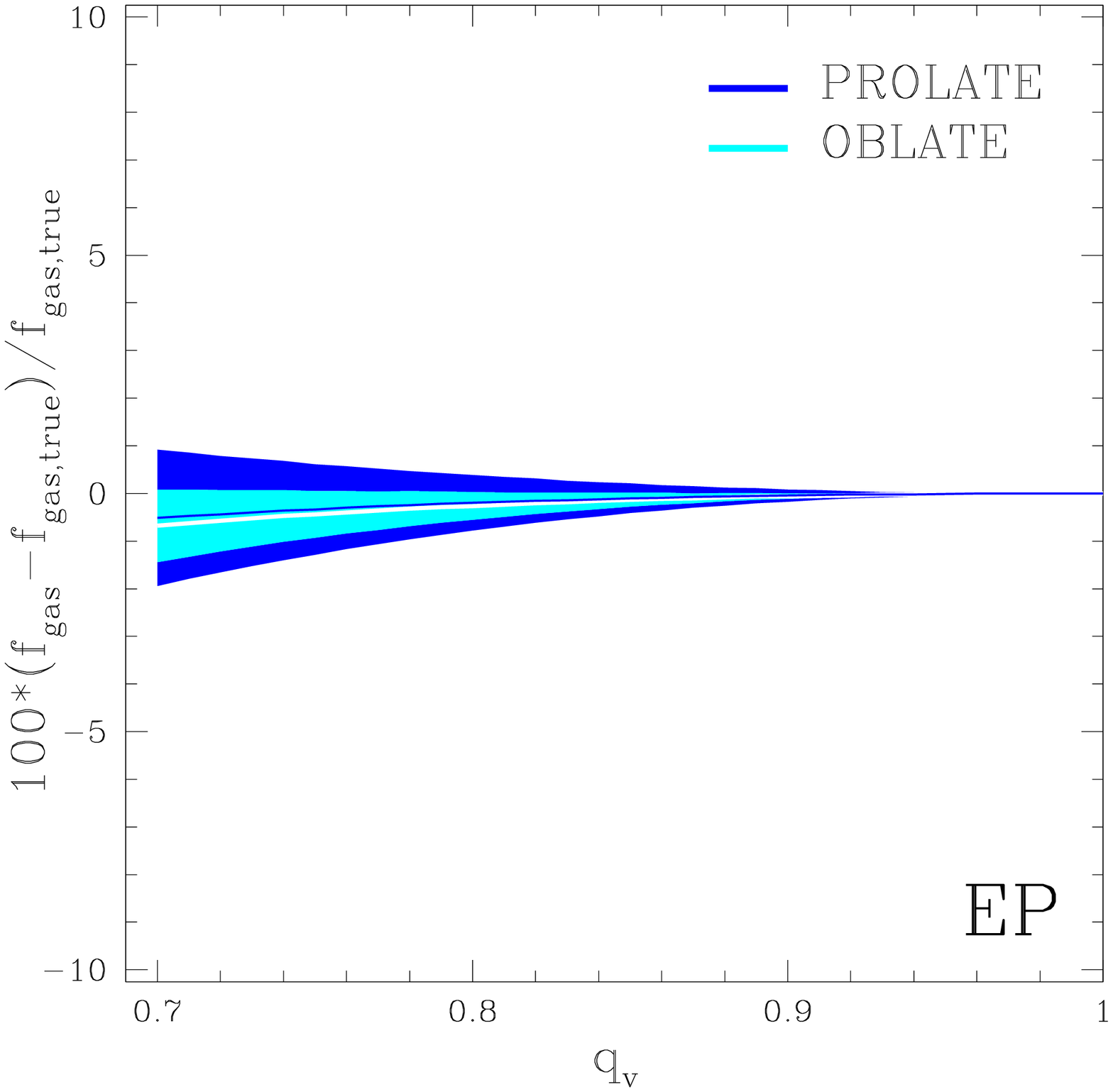}}}
\parbox{0.32\textwidth}{
\centerline{\includegraphics[scale=0.29,angle=0]{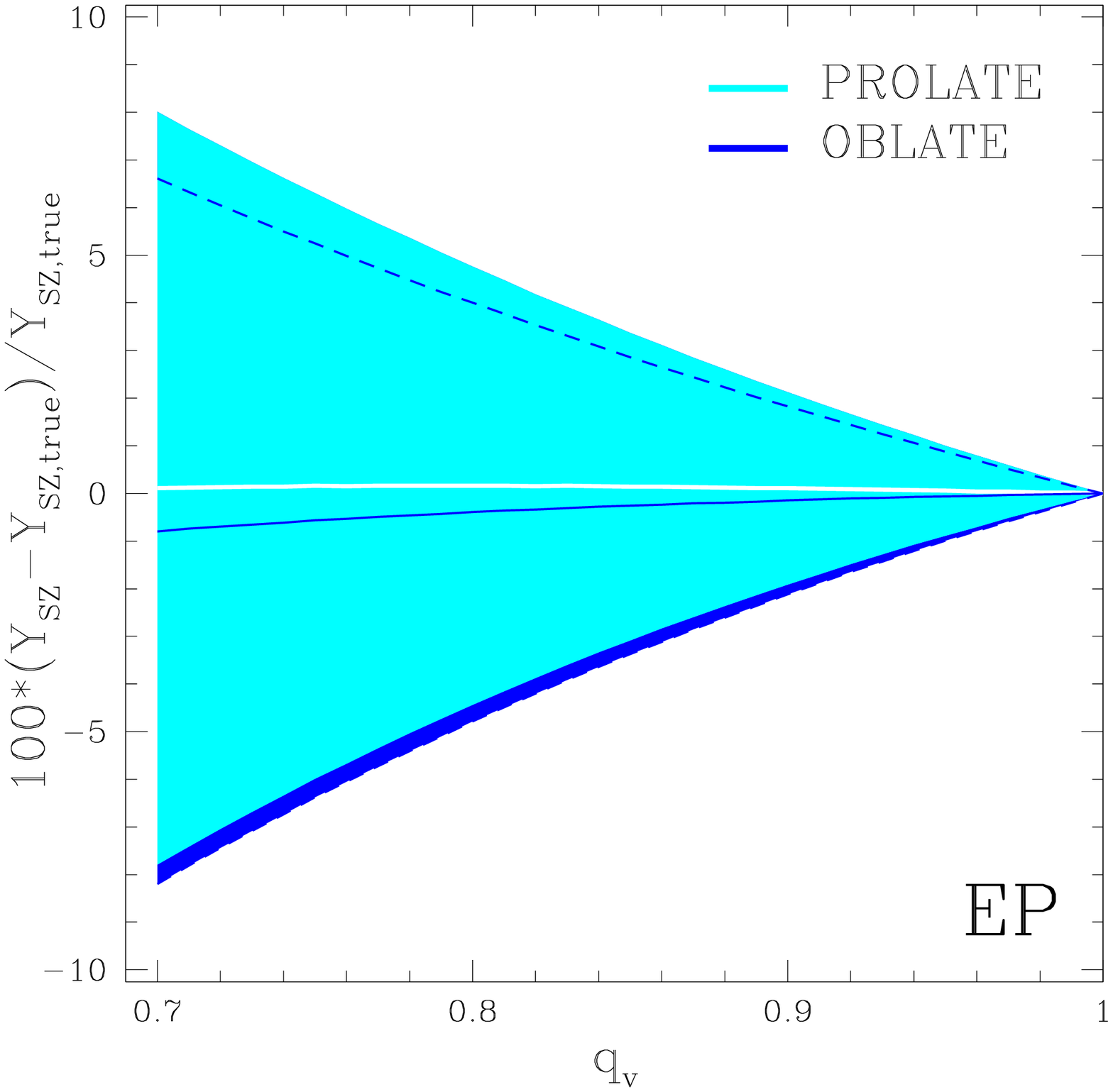}}}
\parbox{0.32\textwidth}{
\centerline{\includegraphics[scale=0.29,angle=0]{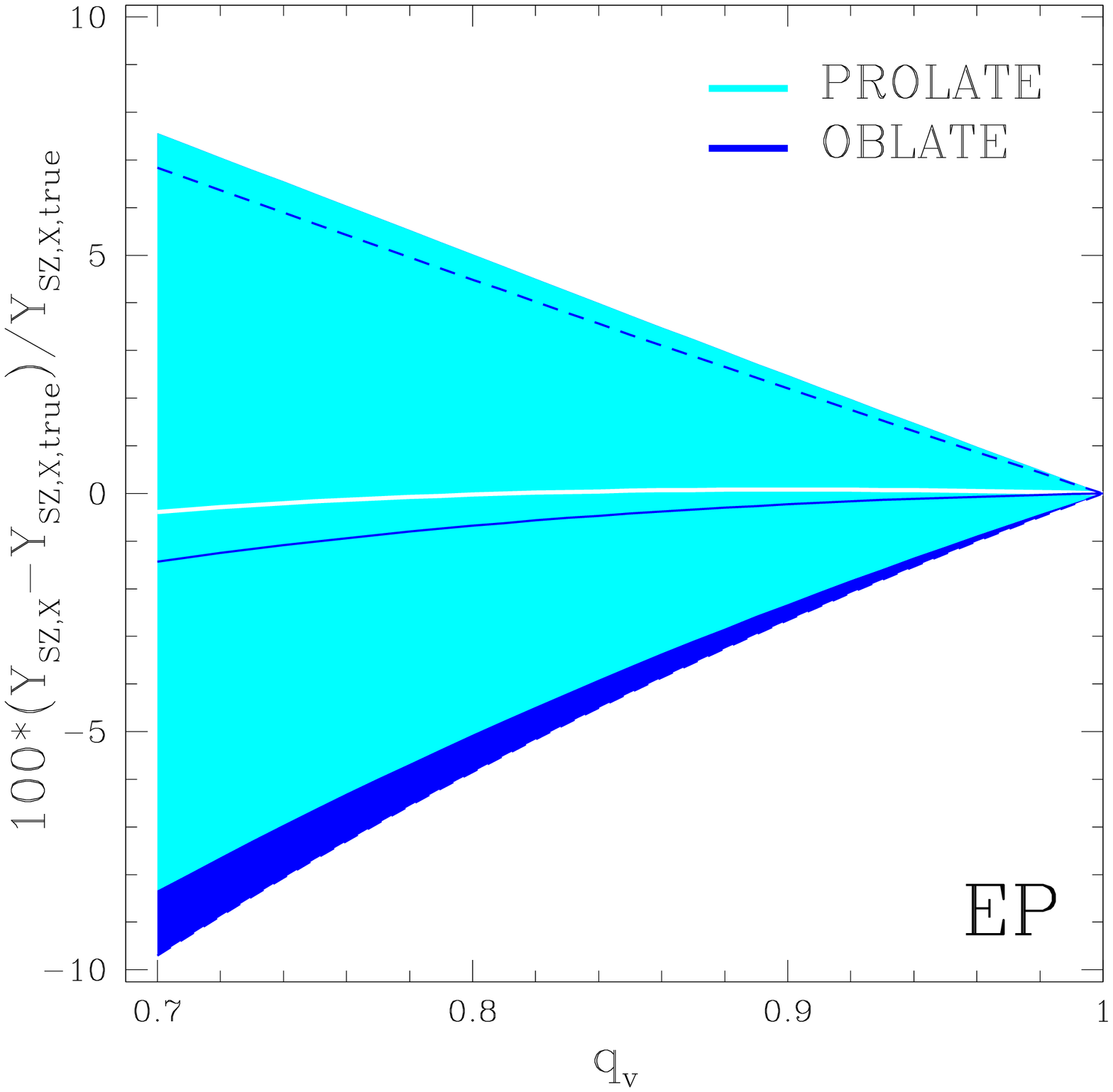}}}
\caption{\label{fig.error.spheroid}
Comparison of results for oblate and prolate isothermal NFW-EP models.
The solid lines are orientation angle-averaged values, while the
shaded regions are the $1\sigma$ ranges.  Note the color coding is
reversed for the gas fraction.}
\end{figure*}

\begin{table*}
\begin{minipage}{126mm}
\caption{Polynomial Approximations of the Average Biases as a Function of $q_v$}
\label{poly}
\begin{tabular}{lrrrrrrrr}  
\hline
& \multicolumn{4}{c}{NFW-EP} & \multicolumn{4}{c}{NFW-EMD}\\ 
Bias or $\sigma$ & $c_0$ & $c_1$ & $c_2$ & $c_3$ & $c_0$ & $c_1$ & $c_2$ & $c_3$\\
\hline
Concentration	& -42.93 & 114.87 & -101.68 & 29.74   & -9.32   &  24.93  &  -23.45  &  7.87  \\	
$\sigma$	& 21.82  & -37.60 & 20.81   & -5.04   & 14.64   & -32.60  &   27.54  &  -9.60  \\    
Mass		& 17.95  & -45.95 & 38.18   & -10.18  & 5.60    & -15.26  &   14.53  &  -4.89  \\    
$\sigma$	& 6.58   & -10.47 &  5.28   & -1.39   & 6.38    & -14.94  &   13.27  &  -4.71  \\    
Gas Mass	& 6.29   & -14.34 &  9.96   & -1.91   & 1.12    &   0.83  &   -3.80  &   1.85  \\    
$\sigma$	& -10.52 & 33.89  & -32.09  &  8.72   & 1.70    &   0.45  &   -3.96  &   1.82  \\    
Gas Fraction	& -11.07 & 29.86  & -26.52  &  7.73   & -4.25   &  15.24  &  -17.27  &   6.31  \\    
$\sigma$	& 16.03  & -40.87 &  33.56  & -8.72   & 4.29    & -11.68  &   11.17  &  -3.79  \\ 
$Y_{\rm SZ}$    & -7.67  &  19.97 & -17.58  &  5.30   & -1.74   &  8.67   & -10.80   &  3.95   \\ 
$\sigma$	& 50.80  & -108.78 &  79.78 & -21.80  & 7.97    & -15.47  &  10.75   & -3.26   \\
$Y_{\rm SZ,X}$  & -15.49 &  40.19 & -34.84  &  10.14  & -2.63   &  7.28   & -6.58    &  2.00   \\
$\sigma$	& 39.08  & -71.35 &  44.20  & -11.93  & 9.46    & -14.65  &  6.47    & -1.28 \\
\hline
\end{tabular}

\medskip

The orientation-angle-average bias and standard deviation as a function of $q_v$
approximated by a cubic polynomial: $c_0 + c_1q_v + c_2q_v^2 +
c_3q_v^3$. Results are given for the triaxial, isothermal NFW models
evaluated at $r_{500}$; c.f.\ Figures \ref{fig.error.c},
\ref{fig.error.m}, \ref{fig.error.mgas},
\ref{fig.error.fgas}.

\end{minipage}
\end{table*}

In Figures~\ref{fig.error.c}, \ref{fig.error.m}, \ref{fig.error.mgas},
\ref{fig.error.fgas}, \ref{fig.error.ysz}, and \ref{fig.error.yszx} 
we plot the mean and standard deviations of the bias distributions
versus $q_v$ for the concentration, mass, gas mass, gas fraction,
$Y_{\rm SZ}$ and $Y_{\rm SZ, X}$ computed within $r_{500}$ for the
isothermal models as well as the polytropic EMD models.  The results
for EP models reflect 500 random orientations while the more
computationally expensive EMD models reflect 200 random trials.  For
comparison, for the case of the isothermal EPs we also display results
for models computed within $r_{2500}$ (Figure~\ref{fig.error.2500})
and for spheroidal NFW-EP models
(Figure~\ref{fig.error.spheroid}). Results for $T_{\rm X}$ and $Y_{X}$
using polytropic models are displayed in
Figure~\ref{fig.error.yx}. All the figures plot the biases over the
same range to facilitate visual comparison.  Finally, in
Table~\ref{poly} we provide analytical approximations to the mean bias
and standard deviation profiles for the triaxial isothermal NFW models
computed within $r_{500}$, which we find are represented very well by
cubic polynomials.

\subsubsection{Concentration}
\label{c}

The concentration parameter displays the largest mean bias magnitude
of the parameters investigated. While still not large in absolute
terms, as seen in Figure~\ref{fig.error.c} the average bias is nearly
-2\% for the smallest axial ratios considered. In addition, next to
$Y_{\rm SZ}$ (\S \ref{ysz}) the concentration possesses the largest
scatter. The NFW and CORELOG models have nearly identical average
biases, though CORELOG has a slightly larger scatter. Since the
deprojected spherically averaged mass profile of an EP is independent
of the temperature profile (Theorem~7 of Paper~1), temperature
gradients do not affect the bias distribution for the concentration of
an EP. Comparing the isothermal and polytropic EMD models in
Figure~\ref{fig.error.c}, one sees that the EMDs are only a little
affected by having different temperature profiles; i.e., the
polytropic EMDs have slightly larger scatter and slightly smaller
average bias magnitudes. The concentration biases are essentially the
same when the results are computed within the smaller aperture
$r_{2500}$ (Fig.\ \ref{fig.error.2500}) or when spheroids are adopted
instead of the default maximally triaxial ellipsoid (Fig.\
\ref{fig.error.spheroid}). Perhaps the most notable difference is the
slightly more negative average bias for the oblate NFW spheroids.
Note also that because our implementation of the CORELOG model with a
large concentration results in essentially a scale-free model, all the
CORELOG parameter biases (including the concentration) are essentially
unchanged when computed within $r_{2500}$ compared to the default
$r_{500}$.

Overall, the concentration parameter biases exhibit some general
behavior that apply to most of the other parameters. These trends are
usefully described with respect to the projections down each of the
three principal axes, which we summarize here. Of the three principal
axes, the smallest mean bias magnitude occurs for the intermediate-axis
projection, while the projections along the short and long axes
bracket the full range of biases for all orientations. The short-axis
projection underestimates the true value, and the long-axis projection
overestimates it. For the specific case of the concentration parameter,
these trends can be explained because the concentration is defined to
be inversely proportional to the scale radius of the mass profile.  It
is reasonable to expect that the scale radius tends to be
overestimated when the ellipsoid is projected along the short axis
because the two longest axes then lie in the sky plane. Conversely,
the scale radius tends to be underestimated when the ellipsoid is
projected down the long axis because then the two shortest axes lie
in the sky plane.

\subsubsection{Mass}
\label{m}

The NFW and CORELOG models exhibit interesting differences in their
mass biases. The average bias of the NFW models is small but
significant, increasing to $\sim 1\%$ for the smallest axial ratios
(Fig.\ \ref{fig.error.m}). The $1\sigma$ scatter for NFW models is
similarly small: $\la 1\%$ for EP and about twice as large for EMD.
Since the mass bias of an EP does not depend on the temperature
profile, the results for the biases for the isothermal and polytropic
EPs are identical (not shown). The strong similarity observed between
the isothermal and polytropic NFW-EMD models in Fig.\
\ref{fig.error.m} indicates that the mass bias of EMDs is also mostly
insensitive to the temperature profile. We obtain very similar results
for the spheroidal NFW models (Fig.\
\ref{fig.error.spheroid}).

In contrast, the isothermal CORELOG-EP model possesses essentially
zero bias for all projections. This is a consequence of our adopting a
large concentration $(c=100)$ so that CORELOG-EP approximates the
scale-free logarithmic potential, which we demonstrated in Theorem~9
of Paper~1~\citep[see also][]{chur08a} always has zero mass bias when
viewed from any orientation -- independent of the temperature profile;
i.e., the polytropic CORELOG-EP (not shown) also has zero mass bias
for any projection. While not aways identically zero, CORELOG-EMD
displays a very small range of biases $(\sigma < 1\%)$, with the
isothermal case having essentially zero average bias. The polytropic
CORELOG-EMD model attains an average mass bias of about $-0.4\%$ for
the smallest axial ratio studied ($q_v=0.40$), though we caution that
the polytropic EMD models are the least accurate of all the models
investigated (see end of \S
\ref{prelim}).

For our fiducial models (excluding the zero-bias CORELOG-EP) the sign
of the mass bias is opposite to that found for the concentration
parameter (\S \ref{c}); i.e., projecting down the short axis results
in a positive mass bias while the opposite happens when projecting
down the long axis. We note, however, that if instead we adopt a
smaller value for the concentration (say 10 instead of 100) for the
CORELOG-EP model, the behavior reverses for the short and long axis
projections. We conclude that the sign of the bias for the mass has a
more complex origin than for the concentration.

Our results broadly agree with those obtained by \citet{piff03a} and
\citet{gava05a} who have previously performed the most extensive
examinations of the effects of spherical averaging on the the masses
of galaxy clusters inferred from X-ray observations. \citet{gava05a}
computed the mass bias as a function of radius for spheroidal,
isothermal, NFW-EMD models viewed down their symmetry axes. Visual
inspection of Figure 6 of \citet{gava05a} reveals that the mass bias
for such face-on projections of spheroidal models is generally only a
few percent for a radius $\sim 4r_s\sim r_{500}$, which agrees
reasonably well with our calculations. \citet{piff03a} instead fitted
\rosat\ X-ray imaging data of ten clusters with an isothermal triaxial
$\beta$ model (i.e., isothermal CORELOG-EP) projected down the
intermediate principal axis (i.e., edge-on). They obtained mass biases
of magnitude $\la 4\%$ broadly similar to those of our isothermal
NFW-EP model, which provides a more appropriate comparison because our
fiducial CORELOG-EP model has a much larger concentration than used
by~\citet{piff03a}.

When the NFW-EP model is computed within the smaller radius
$r_{2500}\sim 0.46r_{500}$ (Fig.\ \ref{fig.error.2500}), both the mean
bias $(<0.3\%)$ and scatter $(<0.3\%)$ are strikingly smaller than
obtained for $r_{500}$. Such large changes justify a more detailed
examination of the radial dependence of the mass bias for the NFW
models. (We do not include CORELOG in this discussion because for the
large adopted concentration there are negligible changes in the mass
bias with radius.) In Figure \ref{fig.radpro.ep} we show the radial
profile of the mass bias for NFW-EP models with $q_v=0.7,0.9$. While
the projection of the intermediate axis shows little radial variation,
the other two projections have larger variations, each of similar
magnitude, but of opposite direction. Interestingly, both the short-
and long-axis projections cross the line of zero bias at nearly the
same radius $\sim 0.4\,r_{500}\sim 1.7r_s$. Although we have shown
profiles only for two values of $q_v$, we find similar results for all
$q_v$ examined. 

For comparison, in Figure \ref{fig.radpro.emd} we show the radial
profiles for the mass bias of the NFW-EMD models with $q_v=0.4,0.6$,
which have mass axial ratios at $r_{500}$ similar to the EPs shown in
Figure \ref{fig.radpro.ep}. Although the intermediate-axis projection
of the EMDs shows more radial variation than for the EPs, the
characteristics of the profiles are very similar. In particular, the
bias profiles of the short- and long-axis projections cross the zero
bias line at nearly the same radius, this time near $\sim
0.36\,r_{500}\sim 1.5r_s$, which also applies to all the EMD models
with different $q_v$ that we investigated. These results are little
changed for the polytropic models, consistent with the insensitivity
of the global mass bias to temperature gradients discussed above. The
largest effects are observed for the $q_v=0.4$ EMD model where the
crossing of the zero-bias line for the polytropic model occurs closer
to $1.8r_s$, but the differences in the biases between the short- and
long-axis projections at $1.8r_s$ are almost the same as at $\sim
1.6r_s$. 

Very similar results are also obtained for spheroidal models, where we
have examined short- and long-axis projections, as well as projections
along an intermediate direction defined by ($\theta=45^{\circ}$,
$\phi=45^{\circ}$). We conclude that a useful, general procedure to
minimize the mass bias for a galaxy or cluster expected to follow the
NFW profile is to measure the mass at a radius $\approx
0.4\,r_{500}\approx 1.7r_s$, which should be accurate to within
$\approx 0.5\%$ for any intrinsic axial ratio (and triaxiality),
projection orientation, and (isothermal or $\gamma=1.2$ polytropic)
temperature profile. Since for our models $r_{2500}\approx 0.46
r_{500}$, the use of $r_{2500}$ should also be a very suitable and
convenient choice to minimize the orientation mass bias.

As with the global mass bias, we find that our results for the radial
variation of the mass bias broadly agree with the face-on projections
of spheroidal isothermal NFW-EMD models examined by~\citet{gava05a}.
From visual inspection of Figure 6 of \citet{gava05a}, his
symmetry-axis projections cross the zero-bias line near a radius of
$2r_s$, similar to the radius we find minimizes the mass bias for NFW
models. Figure 6 of \citet{gava05a} also shows that the face-on
prolate models tend to overestimate the mass within $\sim 2r_s$ and
underestimate the mass for radii larger than $\sim 3r_s$. The opposite
is true for the oblate models. We confirm a similar trend for
spheroidal models (not shown).

\subsubsection{Gas Mass}
\label{mgas}

The mean bias for the gas mass is very similar to that obtained for
the total mass. As shown in Fig.\ \ref{fig.error.mgas}, the largest
bias is $\approx 1\%$ which occurs for the isothermal NFW-EMD model at
$q_v=0.40$, while all the CORELOG models have nearly zero average
bias. (We obtain very similar results for the isothermal spheroidal
NFW models shown in Fig.\ \ref{fig.error.spheroid}.)  The bias scatter
for the gas mass is generally larger for the CORELOG models, and it
exhibits a diversity of model-dependent behavior. For example, the
presence of temperature gradients affects the scatter of NFW-EMD and
CORELOG-EMD in the opposite manner; i.e., the scatter is larger for
the polytropic model compared to the isothermal model for CORELOG and
smaller for NFW. In contrast, for EPs the scatter is larger for the
polytropic models for both the NFW and CORELOG (not shown). Finally,
the sign of the bias in the gas mass for principal-axis projections is
also model-dependent; e.g., for isothermal models CORELOG behaves as
the concentration while NFW has the opposite behavior (like the total
mass).

When the isothermal EPs are computed within the smaller radius
$r_{2500}$, the mean bias and scatter (Fig.\ \ref{fig.error.2500}) do
not change by very much. The most notable difference is that the mean
bias for NFW-EP is nearly zero for $r_{2500}$. It is also instructive
to examine in more detail the radial dependence of the gas mass bias
to compare to the total mass. In Figure \ref{fig.radpro.mgas} we
display the radial profile of the gas mass bias for projections down
the principal axes for the isothermal and polytropic NFW-EP models
with $q_v=0.7$.  The bias in the gas mass for the isothermal model
displays similarities to the bias in the mass, including crossings of
the zero-bias line for the short-axis and long-axis projections near
radius $\sim 0.4r_{500}$. However, the polytropic model has
qualitatively different behavior, showing larger biases that do not
cross the zero-bias bias line. In fact, the polytropic NFW-EP profile
is more similar to profiles of the CORELOG models (right panel of
Figure \ref{fig.radpro.mgas}) which display very little radial
dependence as noted in \S \ref{c}.

\subsubsection{Gas Fraction}
\label{fgas}

The behavior of the mean and scatter in the bias of the gas fraction
is extremely similar to the gas mass. Generally the magnitude of the
mean bias of the gas fraction is the same as, or slightly smaller
than, for the gas mass.

\subsubsection{$T_{\rm X}$}
\label{tx}

In Fig.\ \ref{fig.error.yx} we display results for the
emission-weighted temperature integrated over the volume of default
radius $r_{500}$ for polytropic models. Not only do the models all
have nearly zero average bias, but overall the bias scatter is the
smallest for all the parameters considered. We note that the sign of
the bias (for principal-axis projections) for $T_{\rm X}$ behaves the
same as for the mass.

\subsubsection{$Y_{\rm X}$}
\label{yx}

For isothermal models the biases of $Y_{\rm X}$ are identical to
the gas mass since $Y_{\rm X} = M_{\rm gas}T_{\rm X}$.
Given the extremely small bias and scatter for $T_{\rm X}$ (\S
\ref{tx}), it is unsurprising that even for polytropic models the
biases of $Y_{\rm X}$ are extremely similar to the gas mass (Fig.\
\ref{fig.error.yx}). Hence, with regard only to the orientation bias
of the type considered in our study, we find no significant advantage
in terms of either the mean bias or scatter of using $Y_{\rm X}$
instead of the gas mass alone.

\subsubsection{$Y_{\rm SZ}$ and $Y_{\rm SZ,X}$}
\label{ysz}

The integrated Compton-y parameter has a very small $(\la 1\%)$ mean
bias for all the models considered (Figs.\ \ref{fig.error.ysz},
\ref{fig.error.2500}, \ref{fig.error.spheroid}). However, $Y_{\rm SZ}$
does exhibit the largest bias scatter; e.g., for isothermal models,
$\sigma=6\%$ for NFW-EP and $\sigma=9\%$ for CORELOG-EP. Although
differing in detail, results for $Y_{\rm SZ,X}= Y_{\rm SZ}/Y_{\rm X}$,
are fairly similar to those of $Y_{\rm SZ}$. Probably the most
noticeable difference is the smaller bias scatter for $Y_{\rm SZ,X}$
for CORELOG models.  These results for $Y_{\rm SZ,X}$ suggest that for
small cluster samples attempts to calibrate the relationship between
$Y_{\rm SZ}$ and $Y_{\rm X}$ will encounter significant intrinsic
scatter owing to orientation bias. Fortunately, our models indicate
that by averaging over a large number of randomly oriented clusters
this source of systematic error can be reduced to less than one
percent.  Current observations~\citep[e.g.,][]{ande11a,plan11a} are
consistent with larger intrinsic scatter than that associated with the
orientation bias on $Y_{\rm SZ,X}$ that we have computed.

\subsection{Comment on the Assumption of Random Orientations}
\label{random}

Our analysis assumes that clusters are randomly oriented to the
observer's line of sight. However, in an X-ray flux-limited sample there may
be a bias toward selecting clusters viewed down their major axis,
since this orientation gives rise to the highest central surface
brightness in an ellipsoidal cluster. As mentioned in \S \ref{intro},
we have assumed ``ideal measurements'' of galaxy clusters, which in
this context means we have assumed a volume-limited cluster sample
free of such observational biases.  Similarly, the gravitational
fields generated by large-scale structures likely establish weak
cluster alignments on $\sim 40$~Mpc scales~\citep[e.g.,][and
references therein]{paz11a}.  By assuming a sample volume much larger
than this scale, we expect that small deviations from random
orientations due to this effect impact our calculations less than the
systematic differences observed between the different models explored
in our paper; e.g., EP vs.\ EMD, NFW vs.\ CORELOG, isothermal vs.\
polytropic. Cosmological N-body simulations are required to
investigate this further. 

\section{Conclusions}
\label{conc}

This is the second of two papers investigating the spherical averaging
of ellipsoidal galaxy clusters (and massive elliptical galaxies) in
the context of X-ray and SZ studies. In Paper~1 we present analytical
formulas describing both the deprojection and spherical averaging of
EPs. In the present study we quantify the mean bias and scatter in
cluster observables (e.g., mass) resulting from spherical averaging in
the following sense. We associate ``true'' spherically averaged values
with those computed from direct spherical averaging of the true
three-dimensional cluster, as is done typically in theoretical
studies. To obtain the observer's perspective, we initially fill up
the ellipsoidal cluster's potential well with hot ICM by assuming
hydrostatic equilibrium. The X-ray emission from this ellipsoidal
model is projected onto the sky for a particular orientation. From
this point onward the hypothetical observer treats the cluster as
spherical; i.e., the observer circularly averages the X-ray image and
temperature map and deprojects each of them assuming spherical
symmetry. We use these profiles to compute ``observed'' spherically
averaged quantities.  By comparing the ``true'' and ``observed''
spherically averaged quantities we compute the bias arising from the
inconsistent spherical averaging procedures.

Our study extends previous work~\citep[e.g.,][]{piff03a,gava05a} on
this topic in a few key respects. Of most importance, unlike previous
studies that examined biases using only a single projection direction,
we compute statistical orientation bias distributions for each cluster
parameter. This allows us to present the first calculations of the
orientation-averaged bias and scatter in cluster properties derived
from X-ray and SZ studies. (Note that although we find that the
parameter bias distributions are non-gaussian, because they are also
centrally peaked it is nevertheless useful for our present
investigation to describe them using the mean and standard deviation.)
We also employ a more diverse set of cluster models than considered in
previous individual studies. First, we consider clusters having a
constant shape or one that changes with radius -- in either the mass
or ICM; i.e., both EPs and EMDs. Second, our fiducial models are
maximally triaxial (triaxiality parameter, $T=0.5$), though we also
examine both oblate ($T=0$) and prolate ($T=1$) spheroids for
comparison.  Second, we examine gravitational potentials corresponding
either to the NFW profile or a (nearly scale-free) logarithmic
potential. Third, for the ICM we consider both isothermal models and
(polytropic) models with temperature gradients. Finally, we examine
the biases for several parameters: concentration, (total) mass, gas
mass, gas fraction, emission-weighted temperature ($T_{\rm X}$),
$Y_{\rm X}$, integrated Compton-y parameter ($Y_{\rm SZ}$), and the
ratio $Y_{\rm SZ,X}=Y_{\rm SZ}/Y_{\rm X}$.

We find that substantial biases can result from different viewing
orientations, where $Y_{\rm SZ}$ and the concentration have the
largest scatter (as large as $\sigma\sim 10\%$ for $Y_{\rm SZ}$) and
$T_{\rm X}$ has the smallest $(\sigma\la 0.5\%)$. As expected, the
biases of largest magnitude occur for the flattest models; i.e.,
$q_v=0.7$ for EPs and $q_v=0.4$ for EMDs. Because the orientation bias
of $T_{\rm X}$ displays such a small scatter, we find that the
scatters for $Y_{\rm X}$ and $M_{\rm gas}$ are virtually the same (as
are their mean biases). Hence, spherical averaging of ellipsoidal
clusters has essentially the same effect on $Y_{\rm X}$ as it does on
$M_{\rm gas}$. We conclude in light of this that the smaller scatter
for $Y_{\rm X}$ obtained from cosmological hydrodynamical
simulations~\citep{krav06a} might be attributed to effects not
considered in our study, especially deviations from ellipsoidal
geometry and hydrostatic equilibrium. 
%Finally, our results for the concentration and mass inferred for X-ray
%studies are completely different from the very large orientation
%biases for the mass (up to $\sim 40\%$) and the negligible biases for
%the concentration recently inferred for weak lensing studies of
%clusters~\citep{fero11a}.

Although the mean biases are small and almost always $<1\%$, the
flattest models generally have a significant non-zero bias; i.e.,
typically orientation angle averaging does not completely eliminate
projection biases.  The mean biases of largest magnitude occur for the
concentration parameter $\sim -3\%$, while $T_{\rm X}$ always has
nearly zero mean bias. Despite $Y_{\rm SZ}$ having the largest
scatter, we find that its mean bias is always $\la 1\%$.  Finally, the
masses of the nearly scale-free CORELOG models -- both EP and EMD --
exhibit very small (or zero) mean bias and scatter.

Orientation bias can generate significant scatter (up to $\sigma\sim
8\%$ for the flattest systems) in the relationship between $Y_{\rm
SZ}$ and $Y_{\rm X}$ (as expressed by $Y_{\rm SZ, X}$). This will be a
significant source of uncertainty for measurements of the $Y_{\rm
SZ}-Y_{\rm X}$ relation using small cluster samples. Fortunately,
since the mean bias for $Y_{\rm SZ, X}$ is small ($\sigma\la 1\%$),
for large enough cluster samples the orientation bias can be reduced
below one percent.

Our results do not change significantly over a wide range in halo mass
$(10^{12}-10^{15}\, \msun)$ spanning massive elliptical galaxies to
galaxy clusters. However, we do observe an interesting trend in the
radial profile of the orientation bias for the mass for a cluster that
follows the NFW profile.  The scatter in the mass bias attains a
minimum near the radius $0.4\,r_{500}\approx 1.7r_s\approx
r_{2500}$. Consequently, considering the effect of orientation bias,
the mass computed at this radius is always within $\approx 0.5\%$ of
the true value for any orientation for all of our models. For
comparisons between a small number of clusters, it is useful to quote
masses at $r_{2500}$ to minimize the orientation bias.

It is important to recognize that even though the $1\sigma$ scatter of
the orientation bias for any parameter we have investigated is $\la
10\%$, the scatter is not small relative to other sources of
systematic error. Take for example the gas fraction of the fossil
group/cluster RXJ~1159+5531 for which we provide a detailed systematic
error budget from a combined \chandra\ and \suzaku\
analysis~\citep[see Table~2 of][]{hump11b}. The systematic errors
reported on the gas fraction computed within $r_{500}$ are $\sim 3\%$
which is very similar to the $1\sigma$ scatter in the orientation bias
for the flattest ellipsoids (Fig.\ \ref{fig.error.fgas}). Hence,
orientation bias is not negligible compared to other effects, and this
is especially so in a fossil system like RXJ~1159+5531 where there is
evidence that the hydrostatic equilibrium approximation is accurate.

To facilitate the accounting for orientation bias in X-ray and SZ
cluster studies, we provide cubic polynomial approximations to the
mean bias and $1\sigma$ scatter as a function of axial ratio $q_v$ for
each parameter for the (maximally triaxial) isothermal NFW models
(Table~\ref{poly}). These can be consulted to estimate the typical
range of orientation bias for a given parameter and to construct a
prior probability distribution for Bayesian studies. In the latter
case, to average over $q_v$ for a given halo mass the probability
distribution for $\Lambda$CDM halos by~\citet{jing02a} can be employed
directly for the NFW-EMD model.

\section*{Acknowledgments}
We gratefully acknowledge partial support from the National
Aeronautics and Space Administration under Grant No.\ NNX10AD07G
issued through the Astrophysics Data Analysis Program.

%XXX bibtex bibliography \\
\bibliographystyle{mn2e}
\bibliography{dabrefs}

\begin{thebibliography}{}

\bibitem[\protect\citeauthoryear{{Allen}, {Evrard} \& {Mantz}}{{Allen}
  et~al.}{2011}]{alle11a}
{Allen} S.~W.,  {Evrard} A.~E.,    {Mantz} A.~B.,  2011, \araa, 49, 409

\bibitem[\protect\citeauthoryear{{Allgood}, {Flores}, {Primack}, {Kravtsov},
  {Wechsler}, {Faltenbacher} \& {Bullock}}{{Allgood} et~al.}{2006}]{allg06a}
{Allgood} B.,  {Flores} R.~A.,  {Primack} J.~R.,  {Kravtsov} A.~V.,  {Wechsler}
  R.~H.,  {Faltenbacher} A.,    {Bullock} J.~S.,  2006, \mnras, 367, 1781

\bibitem[\protect\citeauthoryear{{Andersson}, {Benson}, {Ade}, {Aird},
  {Armstrong}, {Bautz}, {Bleem}, {Brodwin}, {Carlstrom}, {Chang}, {Crawford},
  {Crites} \& et al.}{{Andersson} et~al.}{2011}]{ande11a}
{Andersson} K.,  {Benson} B.~A.,  {Ade} P.~A.~R.,  {Aird} K.~A.,  {Armstrong}
  B.,  {Bautz} M.,  {Bleem} L.~E.,  {Brodwin} M.,  {Carlstrom} J.~E.,  {Chang}
  C.~L.,  {Crawford} T.~M.,  {Crites} A.~T.,    et al. 2011, \apj, 738, 48

\bibitem[\protect\citeauthoryear{{Arnaud}}{{Arnaud}}{1996}]{xspec}
{Arnaud} K.~A.,  1996, in {Jacoby} G.~H.,  {Barnes} J.,  eds, Astronomical Data
  Analysis Software and Systems V Vol.~101 of Astronomical Society of the
  Pacific Conference Series, {XSPEC: The First Ten Years}.
p.~17

\bibitem[\protect\citeauthoryear{{Arnaud}}{{Arnaud}}{2005}]{arna05b}
{Arnaud} M.,  2005, in {F.~Melchiorri \& Y.~Rephaeli} ed., Background Microwave
  Radiation and Intracluster Cosmology {X-ray observations of clusters of
  galaxies}.
p.~77

\bibitem[\protect\citeauthoryear{{Arnaud}, {Pointecouteau} \& {Pratt}}{{Arnaud}
  et~al.}{2005}]{arna05a}
{Arnaud} M.,  {Pointecouteau} E.,    {Pratt} G.~W.,  2005, \aap, 441, 893

\bibitem[\protect\citeauthoryear{{Arnaud}, {Pointecouteau} \& {Pratt}}{{Arnaud}
  et~al.}{2007}]{arna07a}
{Arnaud} M.,  {Pointecouteau} E.,    {Pratt} G.~W.,  2007, \aap, 474, L37

\bibitem[\protect\citeauthoryear{{Bailin} \& {Steinmetz}}{{Bailin} \&
  {Steinmetz}}{2005}]{bail05a}
{Bailin} J.,  {Steinmetz} M.,  2005, \apj, 627, 647

\bibitem[\protect\citeauthoryear{{Binney}}{{Binney}}{1981}]{binn81a}
{Binney} J.,  1981, \mnras, 196, 455

\bibitem[\protect\citeauthoryear{{Binney} \& {Strimpel}}{{Binney} \&
  {Strimpel}}{1978}]{binn78a}
{Binney} J.,  {Strimpel} O.,  1978, \mnras, 185, 473

\bibitem[\protect\citeauthoryear{{Binney} \& {Tremaine}}{{Binney} \&
  {Tremaine}}{2008}]{bt}
{Binney} J.,  {Tremaine} S.,  2008, {Galactic Dynamics: Second Edition}.
Princeton University Press

\bibitem[\protect\citeauthoryear{{Borgani}, {Finoguenov}, {Kay}, {Ponman},
  {Springel}, {Tozzi} \& {Voit}}{{Borgani} et~al.}{2005}]{borg05a}
{Borgani} S.,  {Finoguenov} A.,  {Kay} S.~T.,  {Ponman} T.~J.,  {Springel} V.,
  {Tozzi} P.,    {Voit} G.~M.,  2005, \mnras, 361, 233

\bibitem[\protect\citeauthoryear{{Brighenti} \& {Mathews}}{{Brighenti} \&
  {Mathews}}{2001}]{brig01a}
{Brighenti} F.,  {Mathews} W.~G.,  2001, \apj, 553, 103

\bibitem[\protect\citeauthoryear{{Buote} \& {Canizares}}{{Buote} \&
  {Canizares}}{1994}]{buot94}
{Buote} D.~A.,  {Canizares} C.~R.,  1994, \apj, 427, 86

\bibitem[\protect\citeauthoryear{{Buote} \& {Canizares}}{{Buote} \&
  {Canizares}}{1996a}]{buot96a}
{Buote} D.~A.,  {Canizares} C.~R.,  1996a, \apj, 457, 177

\bibitem[\protect\citeauthoryear{{Buote} \& {Canizares}}{{Buote} \&
  {Canizares}}{1996b}]{buot96c}
{Buote} D.~A.,  {Canizares} C.~R.,  1996b, \apj, 457, 565

\bibitem[\protect\citeauthoryear{{Buote}, {Gastaldello}, {Humphrey},
  {Zappacosta}, {Bullock}, {Brighenti} \& {Mathews}}{{Buote}
  et~al.}{2007}]{buot07a}
{Buote} D.~A.,  {Gastaldello} F.,  {Humphrey} P.~J.,  {Zappacosta} L.,
  {Bullock} J.~S.,  {Brighenti} F.,    {Mathews} W.~G.,  2007, \apj, 664, 123

\bibitem[\protect\citeauthoryear{{Buote} \& {Humphrey}}{{Buote} \&
  {Humphrey}}{2011}]{buot11c}
{Buote} D.~A.,  {Humphrey} P.~J.,  2011, \mnras, in press (arXiv:1109.6921)

\bibitem[\protect\citeauthoryear{{Buote} \& {Humphrey}}{{Buote} \&
  {Humphrey}}{2012}]{buot12a}
{Buote} D.~A.,  {Humphrey} P.~J.,  2012, in {D.-W.~Kim \& S.~Pellegrini} ed.,
  Astrophysics and Space Science Library Vol.~378 of Astrophysics and Space
  Science Library, {Dark Matter in Elliptical Galaxies}.
p.~235

\bibitem[\protect\citeauthoryear{{Cavaliere} \& {Fusco-Femiano}}{{Cavaliere} \&
  {Fusco-Femiano}}{1976}]{cava76a}
{Cavaliere} A.,  {Fusco-Femiano} R.,  1976, \aap, 49, 137

\bibitem[\protect\citeauthoryear{{Chandrasekhar}}{{Chandrasekhar}}{1987}]{chan87}
{Chandrasekhar} S.,  1987, {Ellipsoidal figures of equilibrium}

\bibitem[\protect\citeauthoryear{{Churazov}, {Forman}, {Vikhlinin}, {Tremaine},
  {Gerhard} \& {Jones}}{{Churazov} et~al.}{2008}]{chur08a}
{Churazov} E.,  {Forman} W.,  {Vikhlinin} A.,  {Tremaine} S.,  {Gerhard} O.,
  {Jones} C.,  2008, \mnras, 388, 1062

\bibitem[\protect\citeauthoryear{{Debattista}, {Moore}, {Quinn}, {Kazantzidis},
  {Maas}, {Mayer}, {Read} \& {Stadel}}{{Debattista} et~al.}{2008}]{deba08a}
{Debattista} V.~P.,  {Moore} B.,  {Quinn} T.,  {Kazantzidis} S.,  {Maas} R.,
  {Mayer} L.,  {Read} J.,    {Stadel} J.,  2008, \apj, 681, 1076

\bibitem[\protect\citeauthoryear{{Evans}}{{Evans}}{1993}]{evan93a}
{Evans} N.~W.,  1993, \mnras, 260, 191

\bibitem[\protect\citeauthoryear{{Evrard}, {Metzler} \& {Navarro}}{{Evrard}
  et~al.}{1996}]{evra96a}
{Evrard} A.~E.,  {Metzler} C.~A.,    {Navarro} J.~F.,  1996, \apj, 469, 494

\bibitem[\protect\citeauthoryear{{Fabian}, {Hu}, {Cowie} \&
  {Grindlay}}{{Fabian} et~al.}{1981}]{deproj}
{Fabian} A.~C.,  {Hu} E.~M.,  {Cowie} L.~L.,    {Grindlay} J.,  1981, \apj,
  248, 47

\bibitem[\protect\citeauthoryear{{Fabricant}, {Rybicki} \&
  {Gorenstein}}{{Fabricant} et~al.}{1984}]{frg}
{Fabricant} D.,  {Rybicki} G.,    {Gorenstein} P.,  1984, \apj, 286, 186

\bibitem[\protect\citeauthoryear{{Franx}, {Illingworth} \& {de Zeeuw}}{{Franx}
  et~al.}{1991}]{fran91a}
{Franx} M.,  {Illingworth} G.,    {de Zeeuw} T.,  1991, \apj, 383, 112

\bibitem[\protect\citeauthoryear{{Gavazzi}}{{Gavazzi}}{2005}]{gava05a}
{Gavazzi} R.,  2005, \aap, 443, 793

\bibitem[\protect\citeauthoryear{{Grego}, {Carlstrom}, {Joy}, {Reese},
  {Holder}, {Patel}, {Cooray} \& {Holzapfel}}{{Grego} et~al.}{2000}]{greg00a}
{Grego} L.,  {Carlstrom} J.~E.,  {Joy} M.~K.,  {Reese} E.~D.,  {Holder} G.~P.,
  {Patel} S.,  {Cooray} A.~R.,    {Holzapfel} W.~L.,  2000, \apj, 539, 39

\bibitem[\protect\citeauthoryear{{Henry}}{{Henry}}{2003}]{henr03a}
{Henry} J.~P.,  2003, in {S.~Bowyer \& C.-Y.~Hwang} ed., Astronomical Society
  of the Pacific Conference Series Vol.~301 of Astronomical Society of the
  Pacific Conference Series, {Evolution of the X-Ray Properties of Clusters of
  Galaxies}.
p.~5

\bibitem[\protect\citeauthoryear{{Humphrey}, {Buote}, {Brighenti}, {Flohic},
  {Gastaldello} \& {Mathews}}{{Humphrey} et~al.}{2011}]{hump11b}
{Humphrey} P.~J.,  {Buote} D.~A.,  {Brighenti} F.,  {Flohic} H.~M.~L.~G.,
  {Gastaldello} F.,    {Mathews} W.~G.,  2011, ArXiv e-prints (1106.3322)

\bibitem[\protect\citeauthoryear{{Jing} \& {Suto}}{{Jing} \&
  {Suto}}{2002}]{jing02a}
{Jing} Y.~P.,  {Suto} Y.,  2002, \apj, 574, 538

\bibitem[\protect\citeauthoryear{{Kassiola} \& {Kovner}}{{Kassiola} \&
  {Kovner}}{1993}]{kass93a}
{Kassiola} A.,  {Kovner} I.,  1993, \apj, 417, 450

\bibitem[\protect\citeauthoryear{{Kazantzidis}, {Kravtsov}, {Zentner},
  {Allgood}, {Nagai} \& {Moore}}{{Kazantzidis} et~al.}{2004}]{kaza04a}
{Kazantzidis} S.,  {Kravtsov} A.~V.,  {Zentner} A.~R.,  {Allgood} B.,  {Nagai}
  D.,    {Moore} B.,  2004, \apjl, 611, L73

\bibitem[\protect\citeauthoryear{{Krause}, {Pierpaoli}, {Dolag} \&
  {Borgani}}{{Krause} et~al.}{2011}]{krau11a}
{Krause} E.,  {Pierpaoli} E.,  {Dolag} K.,    {Borgani} S.,  2011, ArXiv
  e-prints

\bibitem[\protect\citeauthoryear{{Kravtsov}, {Vikhlinin} \& {Nagai}}{{Kravtsov}
  et~al.}{2006}]{krav06a}
{Kravtsov} A.~V.,  {Vikhlinin} A.,    {Nagai} D.,  2006, \apj, 650, 128

\bibitem[\protect\citeauthoryear{{Kriss}, {Cioffi} \& {Canizares}}{{Kriss}
  et~al.}{1983}]{kris83}
{Kriss} G.~A.,  {Cioffi} D.~F.,    {Canizares} C.~R.,  1983, \apj, 272, 439

\bibitem[\protect\citeauthoryear{{Lewis}, {Buote} \& {Stocke}}{{Lewis}
  et~al.}{2003}]{lewi03a}
{Lewis} A.~D.,  {Buote} D.~A.,    {Stocke} J.~T.,  2003, \apj, 586, 135

\bibitem[\protect\citeauthoryear{{Macci{\`o}}, {Dutton} \& {van den
  Bosch}}{{Macci{\`o}} et~al.}{2008}]{macc08a}
{Macci{\`o}} A.~V.,  {Dutton} A.~A.,    {van den Bosch} F.~C.,  2008, \mnras,
  391, 1940

\bibitem[\protect\citeauthoryear{{McCarthy}, {Babul}, {Bower} \&
  {Balogh}}{{McCarthy} et~al.}{2008}]{mcca08a}
{McCarthy} I.~G.,  {Babul} A.,  {Bower} R.~G.,    {Balogh} M.~L.,  2008,
  \mnras, 386, 1309

\bibitem[\protect\citeauthoryear{{McCarthy}, {Schaye}, {Ponman}, {Bower},
  {Booth}, {Dalla Vecchia}, {Crain}, {Springel}, {Theuns} \&
  {Wiersma}}{{McCarthy} et~al.}{2010}]{mcca10a}
{McCarthy} I.~G.,  {Schaye} J.,  {Ponman} T.~J.,  {Bower} R.~G.,  {Booth}
  C.~M.,  {Dalla Vecchia} C.,  {Crain} R.~A.,  {Springel} V.,  {Theuns} T.,
  {Wiersma} R.~P.~C.,  2010, \mnras, 406, 822

\bibitem[\protect\citeauthoryear{{McLaughlin}}{{McLaughlin}}{1999}]{mcla99a}
{McLaughlin} D.~E.,  1999, \aj, 117, 2398

\bibitem[\protect\citeauthoryear{{Merritt} \& {Fridman}}{{Merritt} \&
  {Fridman}}{1996}]{merr96a}
{Merritt} D.,  {Fridman} T.,  1996, \apj, 460, 136

\bibitem[\protect\citeauthoryear{{Merritt}, {Navarro}, {Ludlow} \&
  {Jenkins}}{{Merritt} et~al.}{2005}]{merr05a}
{Merritt} D.,  {Navarro} J.~F.,  {Ludlow} A.,    {Jenkins} A.,  2005, \apjl,
  624, L85

\bibitem[\protect\citeauthoryear{{Mu{\~n}oz-Cuartas}, {Macci{\`o}},
  {Gottl{\"o}ber} \& {Dutton}}{{Mu{\~n}oz-Cuartas} et~al.}{2011}]{muno11a}
{Mu{\~n}oz-Cuartas} J.~C.,  {Macci{\`o}} A.~V.,  {Gottl{\"o}ber} S.,
  {Dutton} A.~A.,  2011, \mnras, 411, 584

\bibitem[\protect\citeauthoryear{{Nagai} \& {Lau}}{{Nagai} \&
  {Lau}}{2011}]{naga11a}
{Nagai} D.,  {Lau} E.~T.,  2011, \apjl, 731, L10

\bibitem[\protect\citeauthoryear{{Nagai}, {Vikhlinin} \& {Kravtsov}}{{Nagai}
  et~al.}{2007}]{naga07a}
{Nagai} D.,  {Vikhlinin} A.,    {Kravtsov} A.~V.,  2007, \apj, 655, 98

\bibitem[\protect\citeauthoryear{{Navarro}, {Frenk} \& {White}}{{Navarro}
  et~al.}{1997}]{nfw}
{Navarro} J.~F.,  {Frenk} C.~S.,    {White} S.~D.~M.,  1997, \apj, 490, 493

\bibitem[\protect\citeauthoryear{{Navarro}, {Hayashi}, {Power}, {Jenkins},
  {Frenk}, {White}, {Springel}, {Stadel} \& {Quinn}}{{Navarro}
  et~al.}{2004}]{nava04a}
{Navarro} J.~F.,  {Hayashi} E.,  {Power} C.,  {Jenkins} A.~R.,  {Frenk} C.~S.,
  {White} S.~D.~M.,  {Springel} V.,  {Stadel} J.,    {Quinn} T.~R.,  2004,
  \mnras, 349, 1039

\bibitem[\protect\citeauthoryear{{Ostriker}, {Bode} \& {Babul}}{{Ostriker}
  et~al.}{2005}]{ostr05a}
{Ostriker} J.~P.,  {Bode} P.,    {Babul} A.,  2005, \apj, 634, 964

\bibitem[\protect\citeauthoryear{{Paz}, {Sgr{\'o}}, {Merch{\'a}n} \&
  {Padilla}}{{Paz} et~al.}{2011}]{paz11a}
{Paz} D.~J.,  {Sgr{\'o}} M.~A.,  {Merch{\'a}n} M.,    {Padilla} N.,  2011,
  \mnras, 414, 2029

\bibitem[\protect\citeauthoryear{{Piffaretti}, {Jetzer} \&
  {Schindler}}{{Piffaretti} et~al.}{2003}]{piff03a}
{Piffaretti} R.,  {Jetzer} P.,    {Schindler} S.,  2003, \aap, 398, 41

\bibitem[\protect\citeauthoryear{{Piffaretti} \& {Valdarnini}}{{Piffaretti} \&
  {Valdarnini}}{2008}]{piff08a}
{Piffaretti} R.,  {Valdarnini} R.,  2008, \aap, 491, 71

\bibitem[\protect\citeauthoryear{{Planck Collaboration}, {Ade}, {Aghanim},
  {Arnaud}, {Ashdown}, {Aumont}, {Baccigalupi}, {Balbi}, {Banday}, {Barreiro}
  \& et al.}{{Planck Collaboration} et~al.}{2011}]{plan11a}
{Planck Collaboration} {Ade} P.~A.~R.,  {Aghanim} N.,  {Arnaud} M.,  {Ashdown}
  M.,  {Aumont} J.,  {Baccigalupi} C.,  {Balbi} A.,  {Banday} A.~J.,
  {Barreiro} R.~B.,    et al. 2011, \aap, 536, A11

\bibitem[\protect\citeauthoryear{{Pratt}, {Arnaud}, {Piffaretti},
  {B{\"o}hringer}, {Ponman}, {Croston}, {Voit}, {Borgani} \& {Bower}}{{Pratt}
  et~al.}{2010}]{prat10a}
{Pratt} G.~W.,  {Arnaud} M.,  {Piffaretti} R.,  {B{\"o}hringer} H.,  {Ponman}
  T.~J.,  {Croston} J.~H.,  {Voit} G.~M.,  {Borgani} S.,    {Bower} R.~G.,
  2010, \aap, 511, A85

\bibitem[\protect\citeauthoryear{{Rasia}, {Ettori}, {Moscardini}, {Mazzotta},
  {Borgani}, {Dolag}, {Tormen}, {Cheng} \& {Diaferio}}{{Rasia}
  et~al.}{2006}]{rasi06a}
{Rasia} E.,  {Ettori} S.,  {Moscardini} L.,  {Mazzotta} P.,  {Borgani} S.,
  {Dolag} K.,  {Tormen} G.,  {Cheng} L.~M.,    {Diaferio} A.,  2006, \mnras,
  369, 2013

\bibitem[\protect\citeauthoryear{{Romeo}, {Sommer-Larsen}, {Portinari} \&
  {Antonuccio-Delogu}}{{Romeo} et~al.}{2006}]{rome06a}
{Romeo} A.~D.,  {Sommer-Larsen} J.,  {Portinari} L.,    {Antonuccio-Delogu} V.,
   2006, \mnras, 371, 548

\bibitem[\protect\citeauthoryear{{Sanders}, {Fabian} \& {Smith}}{{Sanders}
  et~al.}{2011}]{sand11a}
{Sanders} J.~S.,  {Fabian} A.~C.,    {Smith} R.~K.,  2011, \mnras, 410, 1797

\bibitem[\protect\citeauthoryear{{Sarazin}}{{Sarazin}}{1986}]{sara86a}
{Sarazin} C.~L.,  1986, Reviews of Modern Physics, 58, 1

\bibitem[\protect\citeauthoryear{{Schuecker}}{{Schuecker}}{2005}]{schu05a}
{Schuecker} P.,  2005, in {S.~R{\"o}ser} ed., Reviews in Modern Astronomy
  Vol.~18 of Reviews in Modern Astronomy, {New Cosmology with Clusters of
  Galaxies}.
pp 76--105

\bibitem[\protect\citeauthoryear{{Schuecker}, {Finoguenov}, {Miniati},
  {B{\"o}hringer} \& {Briel}}{{Schuecker} et~al.}{2004}]{schu04a}
{Schuecker} P.,  {Finoguenov} A.,  {Miniati} F.,  {B{\"o}hringer} H.,
  {Briel} U.~G.,  2004, \aap, 426, 387

\bibitem[\protect\citeauthoryear{{Shaw}, {Holder} \& {Bode}}{{Shaw}
  et~al.}{2008}]{shaw08a}
{Shaw} L.~D.,  {Holder} G.~P.,    {Bode} P.,  2008, \apj, 686, 206

\bibitem[\protect\citeauthoryear{{Simionescu}, {Allen}, {Mantz}, {Werner},
  {Takei}, {Morris}, {Fabian}, {Sanders}, {Nulsen}, {George} \&
  {Taylor}}{{Simionescu} et~al.}{2011}]{simi11a}
{Simionescu} A.,  {Allen} S.~W.,  {Mantz} A.,  {Werner} N.,  {Takei} Y.,
  {Morris} R.~G.,  {Fabian} A.~C.,  {Sanders} J.~S.,  {Nulsen} P.~E.~J.,
  {George} M.~R.,    {Taylor} G.~B.,  2011, Science, 331, 1576

\bibitem[\protect\citeauthoryear{{Smith}, {Brickhouse}, {Liedahl} \&
  {Raymond}}{{Smith} et~al.}{2001}]{apec}
{Smith} R.~K.,  {Brickhouse} N.~S.,  {Liedahl} D.~A.,    {Raymond} J.~C.,
  2001, \apjl, 556, L91

\bibitem[\protect\citeauthoryear{{Takahashi}}{{Takahashi}}{2011}]{taka11a}
{Takahashi} T.,  2011, in {J.-U.~Ness \& M.~Ehle} ed., The X-ray Universe 2011,
  Presentations of the Conference held in Berlin, Germany, 27-30 June 2011.
  Available online at:
  http://xmm.esac.esa.int/external/xmm\_science/workshops/2011symposium/, p.433
  {The ASTRO-H Mission}.
p.~433

\bibitem[\protect\citeauthoryear{{Tozzi}}{{Tozzi}}{2007}]{tozz07a}
{Tozzi} P.,  2007, in {L.~Papantonopoulos} ed., The Invisible Universe: Dark
  Matter and Dark Energy Vol.~720 of Lecture Notes in Physics, Berlin Springer
  Verlag, {Cosmological Parameters from Galaxy Clusters: An Introduction}.
p.~125

\bibitem[\protect\citeauthoryear{{Tozzi} \& {Norman}}{{Tozzi} \&
  {Norman}}{2001}]{tozz01a}
{Tozzi} P.,  {Norman} C.,  2001, \apj, 546, 63

\bibitem[\protect\citeauthoryear{{Trinchieri}, {Fabbiano} \&
  {Canizares}}{{Trinchieri} et~al.}{1986}]{trin86}
{Trinchieri} G.,  {Fabbiano} G.,    {Canizares} C.~R.,  1986, \apj, 310, 637

\bibitem[\protect\citeauthoryear{{Tsai}, {Katz} \& {Bertschinger}}{{Tsai}
  et~al.}{1994}]{tsai94a}
{Tsai} J.~C.,  {Katz} N.,    {Bertschinger} E.,  1994, \apj, 423, 553

\bibitem[\protect\citeauthoryear{{Urban}, {Werner}, {Simionescu}, {Allen} \&
  {B{\"o}hringer}}{{Urban} et~al.}{2011}]{urba11a}
{Urban} O.,  {Werner} N.,  {Simionescu} A.,  {Allen} S.~W.,    {B{\"o}hringer}
  H.,  2011, \mnras, 414, 2101

\bibitem[\protect\citeauthoryear{{Vazza}, {Brunetti}, {Gheller}, {Brunino} \&
  {Br{\"u}ggen}}{{Vazza} et~al.}{2011}]{vazz11a}
{Vazza} F.,  {Brunetti} G.,  {Gheller} C.,  {Brunino} R.,    {Br{\"u}ggen} M.,
  2011, \aap, 529, A17

\bibitem[\protect\citeauthoryear{{Voit}}{{Voit}}{2005}]{voit05b}
{Voit} G.~M.,  2005, Reviews of Modern Physics, 77, 207

\bibitem[\protect\citeauthoryear{{Voit}, {Bryan}, {Balogh} \& {Bower}}{{Voit}
  et~al.}{2002}]{voit02a}
{Voit} G.~M.,  {Bryan} G.~L.,  {Balogh} M.~L.,    {Bower} R.~G.,  2002, \apj,
  576, 601

\bibitem[\protect\citeauthoryear{{Voit}, {Kay} \& {Bryan}}{{Voit}
  et~al.}{2005}]{voit05a}
{Voit} G.~M.,  {Kay} S.~T.,    {Bryan} G.~L.,  2005, \mnras, 364, 909

\bibitem[\protect\citeauthoryear{{White}, {Fabian}, {Allen}, {Edge},
  {Crawford}, {Johnstone}, {Stewart} \& {Voges}}{{White} et~al.}{1994}]{daw94}
{White} D.~A.,  {Fabian} A.~C.,  {Allen} S.~W.,  {Edge} A.~C.,  {Crawford}
  C.~S.,  {Johnstone} R.~M.,  {Stewart} G.~C.,    {Voges} W.,  1994, \mnras,
  269, 589

\bibitem[\protect\citeauthoryear{{White}, {Hernquist} \& {Springel}}{{White}
  et~al.}{2002}]{whit02a}
{White} M.,  {Hernquist} L.,    {Springel} V.,  2002, \apj, 579, 16

\bibitem[\protect\citeauthoryear{{Younger} \& {Bryan}}{{Younger} \&
  {Bryan}}{2007}]{youn07a}
{Younger} J.~D.,  {Bryan} G.~L.,  2007, \apj, 666, 647

\bibitem[\protect\citeauthoryear{{Zaroubi}, {Squires}, {Hoffman} \&
  {Silk}}{{Zaroubi} et~al.}{1998}]{zaro98a}
{Zaroubi} S.,  {Squires} G.,  {Hoffman} Y.,    {Silk} J.,  1998, \apjl, 500,
  L87

\end{thebibliography}

\end{document}